\documentclass[11pt,a4paper]{article}
\usepackage[utf8]{inputenc}
\usepackage{geometry}
\usepackage{authblk}
\usepackage{color}
\usepackage{graphicx}
\usepackage{amsmath}
\usepackage{amssymb}
\usepackage{subcaption}
\usepackage{booktabs}
\usepackage{commath}
\usepackage[colorlinks=true, pdfstartview=FitV, linkcolor=blue, citecolor=blue, urlcolor=blue]{hyperref}
\usepackage{slashed}
\usepackage{tikz}
\usepackage{makecell}
\usepackage[square,numbers,comma,sort&compress]{natbib}
\usepackage{tablefootnote}
\allowdisplaybreaks

\newcommand{\braket}[1]{\ensuremath{\left\langle #1 \right\rangle}}
\renewcommand{\Re}{\ensuremath{\mathrm{Re}}}
\title{A Tale of Invisibility: Constraints on New Physics in $b\to s\nu\nu$}
\author{Tobias Felkl\thanks{t.felkl@unsw.edu.au}}
\author{Sze Lok Li\thanks{szell0305@gmail.com}}
\author{Michael A.~Schmidt\thanks{m.schmidt@unsw.edu.au}}
\affil{Sydney Consortium for Particle Physics and Cosmology,
School of Physics, The University of New South Wales, Sydney, NSW 2052, Australia}

\begin{document}
\maketitle
\begin{abstract}
    The Belle II experiment will measure the rare decays $B\to K\nu\nu$ and $B\to K^* \nu\nu$ with increased sensitivity which can hence be expected to serve as a very efficient probe of new physics. 
    We calculate the relevant branching ratios in low-energy effective field theory (LEFT) including an arbitrary number of massive sterile neutrinos and discuss the expected sensitivity to the different operators. We also take into account the longitudinal polarisation fraction $F_L$ and the inclusive decay rate $B\to X_s\nu\nu$.
    In our investigation we consider new physics dominantly contributing to one and two operators both for massless and massive (sterile) neutrinos. Our results show a powerful interplay of the exclusive decay rates $B\to K\nu\nu$ and $B\to K^*\nu\nu$, and a surprisingly large sensitivity of the inclusive decay mode to vector operators even under conservative assumptions about its uncertainty. Furthermore, the sensitivity of $F_L$ is competitive with the branching ratio of $B\to K^* \nu\nu$ in the search for new physics contributing to scalar operators and thus also complementary to $B\to K\nu\nu$ and $B\to X_s\nu\nu$. 
\end{abstract}

\begin{tikzpicture}[remember picture,overlay]
\node[below left,xshift=-2cm,yshift=-1cm] at (current page.north east) {CPPC-2021-13};
\end{tikzpicture}

\newpage
\tableofcontents
\newpage
\section{Introduction}

As of today, we know for sure that the flavour structure of nature is more complicated than what is implied by the Standard Model (SM) of particle physics. This has first become manifest with the measurement of neutrino oscillations which provide conclusive evidence that lepton flavour is not exactly conserved. Currently observed anomalies such as the long-standing tensions between the SM predictions for and measurements of the magnetic dipole moment of the muon~\cite{Muong-2:2021ojo} as well as in observables related to lepton-flavour universality in semi-leptonic $B$-meson decays like $R(D^{(*)})$ and $R(K^{(*)})$ (see for instance Ref.~\cite{HFLAV:2019otj}) suggest the existence of new physics.

A particularly promising avenue to probe and constrain extensions of the SM is via the investigation of rare processes. In the search for light and weakly-interacting particles, rare processes with missing energy are particularly interesting, because they may not only be enhanced via new intermediate states, but also via exotic sterile final states which escape undetected.

Furthermore, the amplitudes for $b\to s\nu\nu$ transitions completely factorise into a hadronic and a leptonic part and are therefore under very good theoretical control. Indeed, quantum chromodynamics (QCD) involved in exclusive decays is entirely captured via an appropriate set of form factors, whereas the inclusive decay mode is at leading order given by the underlying parton-level process which is calculable in perturbation theory and receives corrections only at quadratic order in the heavy-quark effective theory (HQET) expansion. Processes like $b\to s\nu\nu$ are mediated by flavour-changing neutral currents (FCNCs) which in the SM are suppressed in a rather accidental manner via the Glashow-Iliopoulos-Maiani (GIM) mechanism, which however generically does not hold anymore if flavour-sensitive new physics is introduced. In this paper, we exploit this feature and study the constraining power of measurements of several observables related to the $b\to s\nu\nu$ transition in the light of the expected sensitivity of Belle II~\cite{Kou:2018nap}.

Indeed, due to the large suppression of $b\to s$ transitions as predicted by the SM, currently only experimental upper bounds on the decay channels $B\to K^{(*)}\nu\nu$ and $B\to X_s\nu\nu$ exist.
Most recently, the Belle-II collaboration presented a new analysis for $B^+\to K^+ \nu\nu$~\cite{Belle-II:2021rof} and reported an upper bound Br($B^+\to K^+\nu\nu)<4.1\times 10^{-5}$ at the 90\% confidence level. 
A simple weighted average of their result with earlier results~\cite{Belle:2013tnz,BaBar:2013npw,Belle:2017oht} leads to Br$(B^+\to K^+ \nu\nu)=(1.1\pm0.4)\times 10^{-5}$~\cite{Dattola:2021cmw,Belle-II:2021rof}. If substantiated further, this would imply an enhancement on top of the SM expectation Br$(B^+\to K^+\nu\nu)=(4.4\pm0.7)\times 10^{-6}$~\cite{Straub:2018kue}, which has been interpreted in terms of leptoquarks and $Z^\prime$ bosons~\cite{Browder:2021hbl,He:2021yoz}.

Moreover, if the aforementioned observed tensions in $b\to s\ell\ell$ processes are confirmed as being induced by new physics, this may intriguingly also imply effects in the decay channels with neutrinos, since the latter are part of the same weak-isospin doublets as left-handed charged leptons.
Recently, the interplay of the observed anomalies in $b\to s\mu^+\mu^-$ and rare decays such as $B\to K^{(*)}\nu\nu$ and $K\to\pi\nu\nu$ was studied in \cite{Descotes-Genon:2020buf,Calibbi:2015kma}, and \cite{Bause:2021ply} provides a more general analysis of the interplay of di-neutrino and di-charged-lepton modes based on the relevant four-fermion vector operators.
Finally, an observation of $b\to s\nu\nu$ may place constraints on semi-leptonic $B$-meson decays with $\tau$ leptons in the final state which are currently less precisely determined by 
experimental data. 

There are several earlier model-independent studies of semi-leptonic $B$-meson decays with final-state neutrinos in terms of effective field theory for different classes of operators. Vector operators with left-handed massless neutrinos have been studied in~\cite{Colangelo:1996ay,Melikhov:1998ug,Altmannshofer:2009ma,Altmannshofer:2011gn,Buras:2014fpa,Bause:2021ply}. 
 Contributions from scalar and tensor operators were taken into account in \cite{Kim:1999waa,Aliev:2001in}, but no dependence on (sterile) neutrino mass and consequently neither any interference between scalar, vector and tensor operators.
 The inclusive mode $B\to X_s\nu\nu$ was studied earlier in \cite{Hou:1986ug,Grossman:1995gt,Buchalla:1993bv,Altmannshofer:2009ma} where only vector operators were taken into account. Reference \cite{Fukae:1998qy} contains an investigation of the process $B\to X_s\ell^+\ell^-$ including contributions from scalar and tensor operators which can be applied to $B\to X_s \nu\nu$. 
 
 We go beyond previous work by considering the full set of dimension-6 operators in low-energy effective theory (LEFT) which contribute to $b\to s\nu\nu$~\cite{Aebischer:2017gaw,Jenkins:2017jig} for an arbitrary number of generations to account for the possible existence of massive sterile neutrinos. Right-handed sterile neutrinos $\nu_R$ are included as left-handed fields $\nu_R^c \equiv C \overline{\nu_R}^T$.
There are only five operators at dimension 6, i.e. vector and scalar operators with left-handed and right-handed quark bilinears, respectively, and tensor operators with left-handed quark bilinears. The dimension-5 dipole operators are already strongly constrained from searches for neutrino magnetic dipole moments~\cite{Beda:2012zz,Borexino:2017fbd} (see \cite{Giunti:2014ixa} for a recent review) and are thus not considered.

In this work, we investigate
the current constraints on the dimension-6 LEFT operators and their improvement in the light of
the future sensitivity of Belle II.
We discuss
 the implications of an interpretation of the aforementioned simple weighted average of Br($B^+\to K^+\nu\nu$) in terms of an additional sterile neutrino. We also provide the leading-order result for the inclusive decay mode $B\to X_s\nu\nu$ with all contributing operators including interference terms and arbitrary masses for both final-state neutrinos. 
Our results are entirely general and can be matched onto any specific new-physics model yielding non-zero contributions to one or several of the considered operators.

The paper is organised as follows. The effective field theory framework is explained in Sec.~\ref{sec:EFT}. In Sec.~\ref{sec:observables} we introduce the considered observables and present compact expressions for massless neutrinos. In Sec.~\ref{sec:results} we discuss the results of our phenomenological study and conclude in Sec.~\ref{sec:conclusions}. Expressions for the observables in the case of massive neutrinos as well as further technical details are summarised in the appendices. 

\section{Effective Field Theory Framework}
\label{sec:EFT}

We consider the Standard Model extended by an arbitrary number of sterile neutrinos and work entirely within LEFT~\cite{Jenkins:2017jig}. The matching of the LEFT operators to SM effective field theory (SMEFT) operators is presented in App.~\ref{sec:Matching}. Throughout the paper, we assume massless SM neutrinos $\nu_{1,2,3}$, i.e.~they refer both to flavour eigenstates and to mass eigenstates. We
neglect mixing between active $\nu_{1,2,3}$ and sterile neutrinos $\nu_{4,\dots}$, thus we also treat the latter as well-defined mass and flavour eigenstates. The relevant interactions for $b\to s \nu \nu$ processes are described by the Lagrangian~\cite{Aebischer:2017gaw,Jenkins:2017jig}
\begin{align}
    \mathcal{L} &= \sum_{X=L,R}  
    C^{\text{VLX}}_{\nu d} \mathcal{O}^{\text{VLX}}_{\nu d}
    + \left( \sum_{X=L,R} C_{\nu d}^{\text{SLX}} \mathcal{O}_{\nu d}^{\text{SLX}}
    +C_{\nu d}^{\text{TLL}} \mathcal{O}_{\nu d}^{\text{TLL}}
    +\mathrm{h.c.}
    \right)
\end{align}
with the effective operators
\begin{equation}
\begin{aligned}
    \mathcal{O}_{\nu d}^{\text{VLL}}  & = ( \overline{\nu_L}\gamma_\mu  \nu_L)(\overline{d_L} \gamma^\mu  d_L) 
    &
    \mathcal{O}_{\nu d}^{\text{VLR}}  & =
    ( \overline{\nu_L}\gamma_\mu  \nu_L)(\overline{d_R} \gamma^\mu  d_R) 
    \\
    \mathcal{O}_{\nu d}^{\text{SLL}} 
    & =  ( \overline{\nu^c_L}  \nu_L)(\overline{d_R}  d_L)
    &
    \mathcal{O}_{\nu d}^{\text{SLR}}  & 
    = ( \overline{\nu^c_L}  \nu_L) (\overline{d_L}  d_R)
     \\
    \mathcal{O}_{\nu d}^{\text{TLL}} & 
    = (\overline{\nu^c_L} \sigma_{\mu\nu}  \nu_L)  (\overline{d_R} \sigma^{\mu\nu}  d_L) \;,
\end{aligned}
\end{equation}
where the superscripts indicate the chirality and $\nu_L^c \equiv C \overline{\nu_L}^T$ with the charge conjugation matrix $C=i\gamma^2\gamma^0$.
Note that the scalar operators $\mathcal{O}_{\nu d}^{\text{SLL}}$, $\mathcal{O}_{\nu d}^{\text{SLR}}$ are symmetric in the neutrino flavours and the tensor operator $\mathcal{O}_{\nu d}^{\text{TLL}}$ is antisymmetric in the neutrino flavours as shown in Eq.~\eqref{eq:symmetry-properties}.
The vector operators $\mathcal{O}_{\nu d}^{\text{VLL}}$, $\mathcal{O}_{\nu d}^{\text{VLR}}$ do not exhibit any manifest symmetry properties. 
As the dimension-5 neutrino dipole operator $\overline{\nu_L^c}\sigma^{\mu\nu}\nu_L F_{\mu\nu}$ only contributes together with the 
dipole operator $\overline{d_L}\sigma^{\mu\nu}d_R F_{\mu\nu}$ for down-type quarks, it effectively contributes at the same order in the LEFT expansion. Still, both dipole operators are only generated at loop-level and thus further suppressed.
Moreover, the neutrino dipole operator is strongly constrained by searches for magnetic dipole moments of neutrinos~\cite{Beda:2012zz,Borexino:2017fbd} (see \cite{Giunti:2014ixa} for a recent review). Hence we do not include contributions from the neutrino dipole operator in this study. 

 The Weyl fermions $\nu_L$ for the neutrino fields and their respective charge conjugate can be combined to form Majorana neutrino fields $\nu \equiv \nu_L + \nu_L^c$, e.g. the scalar operator $\mathcal{O}^{\text{SLL}}_{\nu d}$ can be rewritten as $\mathcal{O}^{\text{SLL}}_{\nu d} = (\overline{\nu} P_L \nu) (\overline{d_R} d_L)$ where we explicitly included the chiral projection operators $P_{L,R}=\tfrac12 (1\mp \gamma_5)$.
The vector and axial-vector Majorana neutrino bilinears are antisymmetric and symmetric in the neutrino flavours, respectively.
In App.~\ref{sec:SPVAT} we present the matching to a basis in terms of (pseudo)scalar, (axial)vector and tensor neutrino bilinears which we use for the exclusive decays following \cite{Gratrex:2015hna}.

Most of the relevant Wilson coefficients (WCs) are zero in the SM. The only sizeable non-vanishing WC which contributes to $b\to s\nu_\alpha\overline{\nu_\alpha}$ is 
\begin{equation}\label{eq:SM}
    C_{\nu d,\alpha\alpha sb}^{\text{VLL},\text{SM}} = -\frac{4G_F}{\sqrt{2}} \frac{\alpha}{2\pi} V_{ts}^* V_{tb} \left(\frac{X}{\sin^2\theta_W}\right) \;,
\end{equation}
including two-loop electroweak corrections induced by top quarks as captured by the function $X$. 
The latter has been calculated in~\cite{Brod:2010hi} and is numerically given by~\cite{Straub:2018kue}
$X = 6.402 \sin^2\theta_W$.

In LEFT the dominant quantum corrections originate from QCD running. 
The vector (and axial-vector) current operators do not run at one-loop order because of the 
Ward identity. However, the scalar and tensor currents do exhibit renormalisation group (RG) running and their one-loop RG equations for the corresponding Wilson coefficients are well-known (see e.g.~\cite{Aebischer:2017gaw,Jenkins:2017dyc})
\begin{equation}
\begin{aligned}
\mu\frac{d}{d\mu}C^{\text{SLL}}_{\nu d} &=-
3C_F\,\frac{\alpha_s}{2\pi}\,
C^{\text{SLL}}_{\nu d}, 
&
\mu\frac{d}{d\mu}C^{\text{TLL}}_{\nu d}&=
C_F\,\frac{\alpha_s}{2\pi}\,
C^{\text{TLL}}_{\nu d}
\end{aligned}
\end{equation}
where $C_F=(N^2_c-1)/2N_c=4/3$ and $N_c=3$ is the second Casimir invariant of the colour group $SU(3)_c$ and $\alpha_s=g_s^2/(4\pi)$ is the strong fine structure constant. Here, one may exchange $\text{SLL}\leftrightarrow\text{SLR}$. The solutions to the above equations are given by
\begin{equation}
\begin{aligned}
C^{\text{SLL}}_{\nu d}(\mu_1)&=\left( \frac{\alpha_s(\mu_2)}{\alpha_s(\mu_1)} \right)^{3C_F/b}
C^{\text{SLL}}_{\nu d}(\mu_2)
\;, \quad
&
C^{\text{TLL}}_{\nu d}(\mu_1)&=\left( \frac{ \alpha_s(\mu_2)}{\alpha_s(\mu_1)} \right)^{-C_F/ b}
C^{\text{TLL}}_{\nu d}(\mu_2)
\end{aligned}
\end{equation}
between two scales $\mu_1$ and $\mu_2$. Here $b=-11+\tfrac23n_f$ with $n_f$ being the number of active quark flavors between $\mu_1$ and $\mu_2$, and one may exchange $C^{\text{SLL}}_{\nu d} \leftrightarrow C^{\text{SLR}}_{\nu d}$. We use RunDec~\cite{Chetyrkin:2000yt} to obtain precise values for the strong fine structure constant at the different scales. Numerically, we find for the Wilson coefficients at the hadronic scale 
$\mu=4.8$ GeV as a function of the Wilson coefficients at the scale $\mu=m_Z$
 \begin{equation}
     \begin{aligned}
     C_{\nu d}^{\text{SLL}}(4.8 \mathrm{GeV}) & = 1.370\, C_{\nu d}^{\text{SLL}}(m_Z)\;, 
     &
     C_{\nu d}^{\text{TLL}}(4.8 \mathrm{GeV}) & = 0.900\, C_{\nu d}^{\text{TLL}}(m_Z)\;,
     \\
     C_{\nu d}^{\text{SLR}}(4.8 \mathrm{GeV}) & = 1.370\, C_{\nu d}^{\text{SLR}}(m_Z)\;. 
 \end{aligned}
 \end{equation}


\mathversion{bold}
\section{Observables $b\to s \nu\nu$}
\label{sec:observables}
\mathversion{normal}

In our analysis we consider the two exclusive decays $B\to K^{(*)} \nu\nu$ and the inclusive decay $B\to X_s\nu\nu$ decay. While the only observable for $B\to K\nu\nu$ is the differential branching ratio because the final-state neutrinos escape the detector unobserved, the decay to a vector meson $B\to K^*(\to K\pi)\nu\nu$ provides additional angular information which is contained in the $K^*$ longitudinal polarisation fraction $F_L$~\cite{Altmannshofer:2009ma,Buras:2014fpa}. Belle II is anticipated to measure the different branching ratios for $B\to K^{(*)}\nu\nu$ at the level of 10\% with the full integrated luminosity and will also be sensitive to $F_L$~\cite{Kou:2018nap}.
Throughout this work we use the $B\to K$ form factors in \cite{Gubernari:2018wyi} and the $B\to K^*$ form factors in \cite{Bharucha:2015bzk}, the analytical expressions for which are summarised in App~\ref{sec:FF}. Both of them are based on a combined fit to data extracted from light-cone sum rules (LCSR) and lattice QCD (LQCD). 
We summarise the SM predictions\footnote{
We used \texttt{flavio}~\cite{Straub:2018kue,david_straub_2021_5543714} to determine the SM uncertainties of the exclusive decays. Our results for the central values of the SM prediction are the same. Using $B\to K^*$ form factors based on the LCSR+LQCD in~\cite{Gubernari:2018wyi} yields slightly smaller values for the branching ratio and the longitudinal polarisation fraction $F_L$, but within the quoted theoretical errors.
Our result for the inclusive decay slightly overestimates the branching ratio by about 20\%, because it does not take into account QCD and subleading HQET corrections, and hence we refer to~\cite{Altmannshofer:2009ma} for the SM prediction.}, current constraints and future  sensitivities in Tab.~\ref{tab:BelleII}. There is no projection for the inclusive decay $B\to X_s \nu\nu$~\cite{Kou:2018nap}. 

Recently, the Belle II collaboration presented a new analysis with a new upper bound Br($B^+ \to K^+ \nu \nu)<4.1 \times 10^{-5}$~\cite{Dattola:2021cmw,Belle-II:2021rof}. A simple weighted average of the result with previous analyses~\cite{Belle:2013tnz,BaBar:2013npw,Belle:2017oht} results in Br$(B^+\to K^+ \nu\nu)=(1.1\pm0.4)\times 10^{-5}$\cite{Dattola:2021cmw,Belle-II:2021rof} which suggests an enhancement over the SM expectation. We discuss its implications in terms of new physics in Sec.~\ref{sec:average}.
\begin{table}[t]
    \centering
    \begin{tabular}{lcc cc}\toprule
        Observable & SM prediction & current constraint & \multicolumn{2}{c}{Belle II \cite{Kou:2018nap}}  \\ 
         & LQCD+LCSR  & & 5 ab$^{-1}$ & 50 ab$^{-1}$  \\ \midrule
         Br($B^0\to K^0 \nu\nu$) & $(4.1\pm0.5) \times 10^{-6}$~\cite{david_straub_2021_5543714} &
         $<2.6\times 10^{-5}$~\cite{Belle:2017oht}\tablefootnote{Reference \cite{Belle:2017oht} quotes the upper bound on the branching ratio for $B^0\to K_S^0\nu\nu$ which we translated to $B^0\to K^0\nu\nu$.}
         &  & \\
         Br($B^+\to K^+ \nu\nu$) & $(4.4\pm0.7)\times 10^{-6}$~\cite{david_straub_2021_5543714} & $<1.6\times 10^{-5}$~\cite{BaBar:2013npw} & 30\% & 11\% \\
         Br($B^0\to K^{*0} \nu\nu$) & $(11.6\pm 1.1)\times 10^{-6}$~\cite{david_straub_2021_5543714} &$<1.8\times 10^{-5}$~\cite{Belle:2017oht} & 26\% & 9.6\% \\
         Br($B^+\to K^{*+} \nu\nu$) & $(12.4\pm 1.2)\times 10^{-6}$~\cite{david_straub_2021_5543714} &$<4.0\times 10^{-5}$~\cite{Belle:2013tnz} & 25\% & 9.3\% \\
         $F_L(B^0\to K^{*0} \nu\nu$) & $0.49\pm0.04$~\cite{david_straub_2021_5543714} & & & 0.079 \\
         $F_L(B^+\to K^{*+} \nu\nu$) & $0.49\pm0.04$~\cite{david_straub_2021_5543714} & & & 0.077 \\\midrule
         Br$(B\to X_s\nu\nu$) & $(2.7\pm 0.2)\times 10^{-5}$~\cite{Altmannshofer:2009ma} &$<6.4\times 10^{-4}$~\cite{ALEPH:2000vvi} & &  \\\bottomrule
    \end{tabular}
    \caption{Observables for $b\to s \nu\nu$. The SM predictions for the exclusive decays and their uncertainties are based on light-cone sum rules (LCSR) and lattice QCD and are taken from \cite{Gubernari:2018wyi} for $B\to K\nu\nu$ and from \cite{Bharucha:2015bzk} for $B\to K^*\nu\nu$ including a 10\% increase of the $B\to K^*$ form factors due to finite-width effects~\cite{Descotes-Genon:2019bud}. The last two columns list the Belle-II sensitivities to exclusive $B$-meson decays to a $K^{(*)}$ meson and active neutrinos~\cite{Kou:2018nap} if the respective SM predictions are assumed.
   } 
    \label{tab:BelleII}
\end{table}

For the discussion of the exclusive decays $B\to K^{(*)}\nu\nu$ we employ the helicity formalism~\cite{Jacob:1959at} and make use of the general discussion in \cite{Gratrex:2015hna} which employs the narrow-width approximation. Finite-width effects have been considered for $B\to K^*$ form factors in \cite{Das:2017ebx,Descotes-Genon:2019bud}. Following~\cite{Descotes-Genon:2019bud} we increase all $B\to K^*$ form factors by 10\% to take these effects into account.
In order to check our results, we performed independent
calculations for $B\to K\nu\nu$ without the use of helicity amplitudes~\cite{Li:2020Honours} and for $B\to K^*\nu\nu$ using transversity amplitudes~\cite{Altmannshofer:2008dz}. Finally, we find agreement when comparing our results to a calculation of exclusive decays in the SM extended with vector operators with \texttt{flavio}~\cite{david_straub_2021_5543714} and the calculation of the  inclusive decay in~\cite{Altmannshofer:2009ma}.

\mathversion{bold}
\subsection{$B\to  K\nu\nu$}
\mathversion{normal}
After integrating over the phase space of the final state neutrinos which escape the detector unobserved, the differential decay width reads~\cite{Gratrex:2015hna}
\begin{align}
\frac{d \Gamma (B\to  K \nu_\alpha\nu_\beta)}{d q^2}= \frac{1}{4} \bar G^{(0)}(q^2)\;,
\end{align}
where $q^2$ denotes the square of the 4-momentum of the neutrino pair and
$\bar G^{(0)}(q^2)$ is the coefficient of the Wigner-$D$ function $D^0_{0,0}(\Omega_\nu)=1$\footnote{$\Omega_\nu$ denotes the solid angle of $\nu_\alpha$ in the centre of mass frame of the neutrino pair.}. In App.~\ref{sec:BKnunu} we report the function $G^{(0)}(q^2)$ which describes the CP-conjugate process $\bar B\to \bar K \nu_\alpha\nu_\beta$. It is related to $\bar G^{(0)}$ via replacing all Wilson coefficients by their complex conjugates.

We refer the reader to App.~\ref{sec:BKnunu} for the full expression with massive neutrinos as it is lengthy, and only quote the differential decay rate for massless neutrinos
\begin{equation}
\begin{aligned}
\frac{d\Gamma(B\to K\nu_\alpha\nu_\beta)}{dq^2} 
=&\frac{\sqrt{\lambda_{BK}} q^2}{(4\pi)^3 m_B^3 (1+\delta_{\alpha\beta})}  \Bigg[
\frac{\lambda_{BK}}{24 q^2} |f_+|^2  \left|C^{\text{VLL}}_{\nu d,\alpha\beta sb} +C^{\text{VLR}}_{\nu d,\alpha\beta sb}\right|^2
\\ &
+ \frac{(m_B^2-m_K^2)^2}{8(m_b-m_s)^2} |f_0|^2
\Big( \left| C^{\text{SLL}}_{\nu d,\alpha\beta sb}  +C^{\text{SLR}}_{\nu d,\alpha\beta sb} \right|^2 + \left|C^{\text{SLL}}_{\nu d,\alpha\beta bs}+ C^{\text{SLR}}_{\nu d,\alpha\beta bs}\right|^2
\Big) 
\\&
+\frac{2\lambda_{BK}}{3(m_B+m_K)^2} |f_T|^2\left(  
 \left| C^{\text{TLL}}_{\nu d,\alpha\beta sb} \right|^2 +\left| C^{\text{TLL}}_{\nu d,\alpha\beta bs} \right|^2 
\right)
+(\alpha\leftrightarrow \beta)\Bigg]
\end{aligned}
\end{equation}
where $\lambda_{BK}$ is an abbreviation for the K\"all\'en function evaluated as $\lambda_{BK}\equiv\lambda(m_B^2,m_K^2,q^2)$.

Note that there is no interference between scalar, vector and tensor operators for massless neutrinos due to the different chiralities of the final-state neutrinos and the symmetry properties of the scalar and tensor operators. As expected, the differential decay rate is symmetric under exchange of the final-state neutrinos and also under exchange of the quark-flavour indices $sb\leftrightarrow bs$ for the scalar and tensor operators. The same exchange symmetries hold for massive neutrinos. 


\mathversion{bold}
\subsection{$ B\to K^*(\to K\pi)\nu\nu$}
\mathversion{normal}
As the final-state neutrinos escape unobserved from the detector, there are two independent  observables which can be parameterised in terms of the coefficients $\bar G^{0,0}_0$ and $\bar G^{2,0}_0$ of the Wigner-$D$ functions in the differential decay rate~\cite{Gratrex:2015hna}
\begin{equation}
    \frac{d\Gamma(B\to K^*\nu_\alpha\nu_\beta)}{dq^2 d\cos\theta_{K} } 
    = \frac{3}{8}\left[ 
    \bar G_0^{0,0}(q^2) D^0_{0,0}(\Omega_{K}) + \bar G_0^{2,0}(q^2) D^2_{0,0}(\Omega_{K})
    \right]
\end{equation}
where $q^2$ denotes the square of the 4-momentum of the neutrino pair.
The relevant Wigner-$D$ functions are 
$D^{0}_{0,0}(\Omega_K) = 1$ and $D^{2}_{0,0}(\Omega_K) = \frac{1}{2} \left( 3 \cos^2 \theta_K -1 \right)$\footnote{$\Omega_K$ denotes the solid angle of the final-state $K$ meson in the $K^*$ rest frame.} and their coefficients $\bar G$ are 
given in App.~\ref{sec:BKstarnunu}. The CP conjugate process $\bar B \to \bar K^* \nu_\alpha\nu_\beta$ is obtained by replacing the $\bar G$ functions with the corresponding $G$ functions $G^{0,0}_0$ and $G^{2,0}_0$ for which all Wilson coefficients are replaced by their complex conjugates.

As there are two observable final-state particles $K$ and $\pi$ in addition to the missing energy of the neutrino pair, there are two independent observables, the differential decay rate $d\Gamma/dq^2$ and the longitudinal polarisation fraction $F_L(q^2)$~\cite{Bobeth:2012vn}, 
\begin{align}
    \frac{d\Gamma}{dq^2}
    & = \frac34 
    \bar G^{0,0}_0(q^2)
    \;,
    &
    F_L(q^2)  & = \frac{\bar G^{0,0}_0(q^2) + \bar G^{2,0}_0(q^2) }{3 \bar G^{0,0}_0(q^2)}\;.
\end{align}
The corresponding transverse polarisation fraction $F_T$ is related to the longitudinal polarisation fraction by $F_L+F_T=1$.
Experiments measure the integrated longitudinal polarisation fraction
\begin{align}\label{eq:def_FL}
    F_L  & = \frac{\left\langle \bar G^{0,0}_0(q^2)\right\rangle+ \left\langle \bar G^{2,0}_0(q^2) \right\rangle}{3 \left\langle \bar G^{0,0}_0(q^2)\right\rangle}\;,
\end{align}
where angle brackets denote the binning over $q^2$ including a summation over the final-state neutrino flavours\footnote{If no endpoints $q_{0,1}$ are specified, the full kinematic range is integrated over.}
\begin{equation}
    \braket{X} \equiv \sum_{\alpha,\beta}\frac{1}{(q_1^2-q_0^2)} \int_{q_{0}^2}^{q_1^2} dq^2 X
    \;. 
\end{equation}
The analytic expressions in the general case of massive neutrinos are lengthy and reported in App.~\ref{sec:BKstarnunu}, but there are compact expressions for massless neutrinos. In this case, the differential decay rate
is given by
\begin{equation}
\begin{aligned}
\frac{d\Gamma(B\to K^*\nu_\alpha\nu_\beta)}{dq^2} 
=  &\frac{\sqrt{\lambda_{BK^*}} q^2}{(4\pi)^3 m_B^3 (1+\delta_{\alpha\beta})}\Bigg[ 
\frac{\lambda_{BK^*}|V|^2}{12 (m_B+m_{K^*})^2} \left| C^{\text{VLL}}_{\nu d,\alpha\beta sb} +C^{\text{VLR}}_{\nu d,\alpha\beta sb} \right|^2 
\\&
+\left(\frac{8 m_B^2 m_{K^*}^2}{3q^2} |A_{12}|^2 
+\frac{(m_B+m_{K^*})^2|A_1|^2}{12}
\right)
\left|C^{\text{VLL}}_{\nu d,\alpha\beta sb}-C^{\text{VLR}}_{\nu d,\alpha\beta sb}\right|^2
\\&
+\frac{\lambda_{BK^*}}{8 (m_b+m_s)^2}|A_0|^2\Big( 
 \left|C^{\text{SLR}}_{\nu d,\alpha\beta sb} -C^{\text{SLL}}_{\nu d,\alpha\beta sb}\right|^2 + \left|C^{\text{SLR}}_{\nu d,\alpha\beta bs}-C^{\text{SLL}}_{\nu d,\alpha\beta bs}   \right|^2
\Big)
\\&
+ \left(\frac{32m_B^2m_{K^*}^2 |T_{23}|^2 }{3(m_B+m_{K^*})^2}+\frac{4\lambda_{BK^*} |T_1|^2 +4(m_B^2-m_{K^*}^2)^2 |T_2|^2}{3q^2}\right)
\\&\qquad\qquad\qquad\times
\left(\left| C^{\text{TLL}}_{\nu d,\alpha\beta bs}\right|^2 +\left|C^{\text{TLL}}_{\nu d,\alpha\beta sb} \right|^2
\right)
+(\alpha\leftrightarrow\beta)
\Bigg]
\end{aligned}
\end{equation}
where $\lambda_{BK^*}=\lambda(m_B^2,m_{K^*}^2,q^2)$. The longitudinal polarisation fraction reads
\begin{equation}
    \begin{aligned}
    F_L  
    =1-&\sum_{\alpha,\beta}\frac{1}{3(4\pi)^3 m_B^3 (1+\delta_{\alpha\beta})\Gamma(B\to K^*\nu\nu)}
    \int dq^2 \sqrt{ \lambda_{BK^*}}  
    \\&
    \times
    \Bigg(
    \frac{\lambda_{BK^*} |V|^2 q^2}{4(m_B +m_{K^*})^2} 
    \left|C^{\text{VLL}}_{\nu d,\alpha\beta sb} +C^{\text{VLR}}_{\nu d,\alpha\beta sb}\right|^2 
     + \frac{(m_B+m_{K^*})^2q^2 |A_1|^2}{4} 
    \left|C^{\text{VLL}}_{\nu d,\alpha\beta sb} -C^{\text{VLR}}_{\nu d,\alpha\beta sb}\right|^2 
     \\&\qquad
     +\left(2\lambda_{BK^*}|T_1|^2  + 2(m_B^2-m_{K^*}^2)^2|T_2|^2 \right)
     \left( \left|C^{\text{TLL}}_{\nu d,\alpha\beta sb}\right|^2 +\left|C^{\text{TLL}}_{\nu d,\alpha\beta bs}\right|^2
    \right)
    +(\alpha\leftrightarrow\beta)
    \Bigg)
   \;,
    \end{aligned}
\end{equation}
where we integrate over the full kinematic range in $q^2$.
As it is the case for $B\to K\nu\nu$ there is no interference between the scalar, vector, and tensor operators because of the different chiralities and the symmetry properties of the scalar and tensor Wilson coefficients. 


\mathversion{bold}
\subsection{$B\to X_s \nu\nu$}
\mathversion{normal}

Formally, the inclusive decay rate of a hadron $H$ is related to its full propagator in the relevant effective theory described by a Hamiltonian $\mathcal{H}_{\text{eff}}$ via the optical theorem~\cite{ParticleDataGroup:2020ssz}
\begin{equation}
    \Gamma(H) = \frac{1}{m_H}\mathrm{Im}\left\langle H\left| i\int d^4x\;\mathrm{T}\big\{\mathcal{H}_{\text{eff}}(x)\mathcal{H}_{\text{eff}}(0)\big\}\right|H\right\rangle
    \;.
\end{equation}
In the case of $B$ hadrons, the comparatively large $b$-quark mass allows for an efficient expansion of the time-ordered product in terms of local operators defined in heavy-quark effective theory (HQET). The leading term in this expansion is determined by the decay width of a free $b$ quark, and corrections only appear at $\mathcal{O}(\Lambda^2_{\text{QCD}}/m^2_b)$ in the heavy-quark limit as it was first demonstrated in \cite{CHAY1990399}.
The fact that, at leading order, inclusive hadron decays are equivalent to the underlying partonic processes is seen as a manifestation of the notion of ``quark-hadron duality" (see for instance Ref.~\cite{Grossman:2017thq} for a review). For semi-leptonic inclusive decays, this enforces integrating out the phase space to a sufficiently large extent which is often referred to as a ``smearing procedure".

The rate of the inclusive decay $B\to X_s\nu\nu$ is very sensitive to $m_b$.
Besides the HQET corrections, there are also radiative QCD corrections to the leading-order result, the size of which  depends on the scheme one chooses for the $b$-quark mass. It has been argued that a so-called ``threshold mass" definition is favourable \cite{Hoang:1998ng,Bauer:2004ve} as this avoids renormalon ambiguities associated with the pole mass which does not directly correspond to a measurable quantity, and the QCD corrections are smaller and exhibit better convergence behaviour compared to the case of the $\overline{\text{MS}}$ mass. We will employ the 1S mass as originally proposed in \cite{Hoang:1998ng} and use the more recently determined value $m^{1S}_b = 4.75 \pm 0.04$ GeV \cite{Bernlochner:2020jlt}.

The leading-order result for the differential decay rate of the inclusive decay  $B\to X_s\nu_\alpha\nu_\beta$ reads
\begin{equation}
\begin{aligned}
    \frac{d\Gamma(B\to X_s\nu_\alpha\nu_\beta)}{dq^2} & = \frac{\sqrt{\lambda(m_b^2,m_s^2,q^2)\lambda(m_
    \alpha^2,m_\beta^2,q^2)}}{384\pi^3q^4(1 + \delta_{\alpha\beta})} \\
    & \quad \times \left(\frac{d\Gamma_{\text{incl,V}}^{\nu_\alpha\nu_\beta}}{dq^2} + \frac{d\Gamma_{\text{incl,S}}^{\nu_\alpha\nu_\beta}}{dq^2} + \frac{d\Gamma_{\text{incl,T}}^{\nu_\alpha\nu_\beta}}{dq^2} + \frac{d\Gamma_{\text{incl,VS}}^{\nu_\alpha\nu_\beta}}{dq^2} + \frac{d\Gamma_{\text{incl,VT}}^{\nu_\alpha\nu_\beta}}{dq^2}\right)
\end{aligned}
\end{equation}
with the different terms given in Appendix \ref{sec:inclusive} for arbitrary neutrino masses. 
Subleading HQET contributions will lead to a slight suppression by $\mathcal{O}(10\%)$~\cite{Altmannshofer:2009ma} compared to the leading-order result presented here.
In the limit of massless neutrinos, the expression simplifies as follows:
\begin{eqnarray}
\begin{aligned}
    \frac{d\Gamma(B\to X_s\nu_\alpha\nu_\beta)}{dq^2} & = \frac{\sqrt{\lambda(m_b^2,m_s^2,q^2)}}{768\pi^3m_b(1 + \delta_{\alpha\beta})}\Bigg(\left(3\frac{q^2}{m_b^2}(m_b^2 + m_s^2 - q^2) + \frac{1}{m_b^2}\lambda(m_b^2,m_s^2,q^2)\right) \\
    & \quad \times
    \Big[\left|C^{\text{VLL}}_{\nu d,\alpha\beta sb}\right|^2 + \left|C^{\text{VLL}}_{\nu d,\beta\alpha sb}\right|^2 + \left|C^{\text{VLR}}_{\nu d,\alpha\beta sb}\right|^2 + \left|C^{\text{VLR}}_{\nu d,\beta\alpha sb}\right|^2\Big] \\
    & \quad - 12q^2\frac{m_s}{m_b} \text{Re}\big(C^{\text{VLL}}_{\nu d,\alpha\beta sb}C^{\text{VLR}*}_{\nu d,\alpha\beta sb} + C^{\text{VLL}}_{\nu d,\beta\alpha sb}C^{\text{VLR}*}_{\nu d,\beta\alpha sb}\big) \\
    & \quad + 2\Big[3\frac{q^2}{m_b^2}(m_b^2 + m_s^2 - q^2)\Big[\left|C^{\text{SLL}}_{\nu d,\alpha\beta sb}\right|^2 + \left|C^{\text{SLR}}_{\nu d,\alpha\beta sb}\right|^2 + \left|C^{\text{SLL}}_{\nu d,\alpha\beta bs}\right|^2 + \left|C^{\text{SLR}}_{\nu d,\alpha\beta bs}\right|^2\Big] \\
    & \quad + 12q^2\frac{m_s}{m_b}\text{Re}\big(C^{\text{SLL}}_{\nu d,\alpha\beta sb}C^{\text{SLR}*}_{\nu d,\alpha\beta sb} + C^{\text{SLL}}_{\nu d,\alpha\beta bs}C^{\text{SLR}*}_{\nu d,\alpha\beta bs}\big)\Big] \\
    & \quad + 32\left(3\frac{q^2}{m_b^2}(m_b^2 + m_s^2 - q^2) + \frac{2}{m_b^2}\lambda(m_b^2,m_s^2,q^2)\right)\Big[\left|C^{\text{TLL}}_{\nu d,\alpha\beta sb}\right|^2 + \left|C^{\text{TLL}}_{\nu d,\alpha\beta bs}\right|^2\Big]
    \Bigg)
    \;.
\end{aligned}
\end{eqnarray}
Note that the result does not include QCD corrections and subleading HQET corrections which generally lead to a suppression of the differential decay rate. For the SM prediction it amounts to a suppression of $\mathcal{O}(20\%)$. As there are currently no projected sensitivities for the inclusive decay mode at Belle II, QCD and subleading HQET corrections are left for future work.

\section{Results}
\label{sec:results}

In this section we present our results, of which the discussion is split in four parts. In the first three subsections, we
demonstrate the reach for new physics in $b\to s\nu\nu$ processes at Belle II under the assumption of no experimental evidence of an enhancement or suppression of the SM expectation. In the fourth subsection, we consider the recently
reported simple weighted average~\cite{Dattola:2021cmw,Belle-II:2021rof} Br($B^+\to K^+ \nu\nu)=(1.1\pm 0.4)\times 10^{-5}$ and discuss how it could be explained in terms of a sterile neutrino.

We generally use \cite{Gubernari:2018wyi} for the $B\to K$ form factors and \cite{Bharucha:2015bzk} for the $B\to K^*$ form factors which are both based on a combined fit to LCSR and LQCD data. We increase the $B\to K^*$ form factors by 10\% to include finite-width effects following~\cite{Descotes-Genon:2019bud}. Note that only the leading-order contribution to the inclusive decay is taken into account in the following, which in particular overestimates the contributions to vector operators (and thus the SM contribution) by $\mathcal{O}(20\%)$.
All results are presented as constraints on real Wilson coefficients evaluated at the electroweak scale $\mu = m_Z$. Note that the scalar (tensor) Wilson coefficients are (anti)symmetric in the neutrino flavours, and thus the presence of a Wilson coefficient with neutrino flavours $\alpha\beta$ always implies the simultaneous presence of the Wilson coefficient with neutrino flavours $\beta\alpha$ in these cases.  

We typically only refer to $B\to K\nu\nu$ and $B\to K^*\nu\nu$ in the main text, but we generally imply $B^+\to K^+\nu\nu$ and $B^0\to K^{*0}\nu\nu$ for the current bounds as they are the most stringent ones, and the charged modes for the future sensitivity due to a slightly
better new-physics reach, unless differently specified. 
The results for the neutral mode would be essentially the same in the latter case, since any discrepancy is only due to the slightly different lifetimes and masses. Furthermore, as indicated in the caption in Figure~\ref{fig:belleII_1}, $\alpha$ refers to a fixed value $\in(1,2,3)$ in general, thus no summation is implied.

\subsection{One Operator with Massless Neutrinos}
\label{sec:massless_neutrinos_one operator}

\begin{table}[htb!]
    \centering
    \begin{tabular}{ccccccc}\toprule
        & \multicolumn{3}{c}{Current Bound} & \multicolumn{3}{c}{Future Sensitivity (50 ab$^{-1}$)} \\
        Operator & \makecell{Value \\ [-0.25ex] [TeV$^{-2}$]} & \makecell{NP scale \\ [-0.25ex] [TeV]} & Observable & \makecell{Value \\ [-0.25ex] [TeV$^{-2}$]} & \makecell{NP scale \\ [-0.25ex] [TeV]} & Observable \\
        \midrule
        $\mathcal{O}^{\text{VLL},\text{NP}}_{\nu d,\alpha\alpha sb}$ &  0.028
        & 6 & $B\to K^*\nu\nu$ &  0.023
        & 7 &
        $B\to K^{(*)}\nu\nu$ \\
        $\mathcal{O}^{\text{VLR}}_{\nu d,\alpha\alpha sb}$ & 0.021 & 7 & $B\to K\nu\nu$ & 0.002 & 25 & 
        $B\to K^{(*)}\nu\nu$ \\
        $\mathcal{O}^{\text{VLL}}_{\nu d,\gamma\delta sb}$ & 0.014
        &  9
        & $B\to K^*\nu\nu$ & 0.006 & 13 &  $B\to K^{(*)}\nu\nu$ \\
        $\mathcal{O}^{\text{SLL}}_{\nu d,\gamma\gamma sb}$ & 0.012 & 10 & $B\to K^{(*)}\nu\nu$ & 0.002 & 25 & $B\to K\nu\nu$ \\
        $\mathcal{O}^{\text{SLL}}_{\nu d,\gamma\delta sb}$ & 0.009 & 10 & $B\to K^{(*)}\nu\nu$ & 0.002 & 25 & $B\to K\nu\nu$ \\
        $\mathcal{O}^{\text{TLL}}_{\nu d,\gamma\delta sb}$ & 0.002
        &  25
        & $B\to K^*\nu\nu$ & 0.0009 & 35 & $B\to K^*\nu\nu$ \\
        \bottomrule
    \end{tabular}
    \caption{Most competitive bounds imposed on the absolute value of the respective Wilson coefficients if only one of them gets (sizeable) contributions from new physics at a time, both for the current situation and for the projections for the 50 ab$^{-1}$ Belle-II data set under the assumption of a confirmation of the SM predictions. Here, $\alpha\in(1,2,3)$ and $\gamma$ and $\delta$ arbitrary, but $\gamma\neq\delta$ (only in the case of $\mathcal{O}^{\text{VLL}}_{\nu d,\gamma\delta sb}$, $\gamma$ and $\delta$ may be equal if larger than 3), and neutrino masses are set to zero both for active and sterile states. Generally, the most conservative constraint is provided, with the possibility of interference with the SM taken into account. We also provide rough estimates for the corresponding new-physics scale and the observable from which the respective bound arises. If $B\to K^{(*)}\nu\nu$ is indicated, $B\to K\nu\nu$ and $B\to K^*\nu\nu$ yield similar bounds.}
    \label{tab:single_operator_constraints}
\end{table}

In this section we discuss the current constraints on and future sensitivities to new physics under the assumption that it contributes (dominantly) only to one of the considered operators, as summarised in Table~\ref{tab:single_operator_constraints}. The first column contains a representative selection of relevant operators which are bounded in different ways. The operators $\mathcal{O}^{\text{VLL},\text{NP}}_{\nu d,\alpha\alpha sb}$ and $\mathcal{O}^{\text{VLR}}_{\nu d,\alpha\alpha sb}$ both interfere with the SM, but since $B\to K^*\nu\nu$ and $F_L$ depend on $|C^{\text{VLL}}_{\nu d,\alpha\alpha sb} + C^{\text{VLR}}_{\nu d,\alpha\alpha sb}|^2$ and $ |C^{\text{VLL}}_{\nu d,\alpha\alpha sb} - C^{\text{VLR}}_{\nu d,\alpha\alpha sb}|^2$ with different $q^2$ dependencies each, contributions from $\mathcal{O}^{\text{VLR}}_{\nu d,\alpha\alpha sb}$ cannot efficiently cancel the SM contribution and thus it is subject to stronger bounds. The operator $\mathcal{O}^{\text{VLL}}_{\nu d,\gamma\delta sb}$ and the scalar operators could be replaced by the respective right-handed operators\footnote{In general, a ``{}left-handed (right-handed) operator" is to be understood as an operator which contains a left-handed (right-handed) projector in the quark bilinear.} without changing the constraints.

The second and fifth columns contain the current bounds on and future sensitivities to the Wilson coefficients in $\text{TeV}^{-2}$, respectively.
The values for the future sensitivities are obtained under the assumption that the central value of the Belle II measurement exactly coincides with the SM prediction. In each case, the given experimental uncertainty then translates into a constraint on the Wilson coefficient.
We generally provide the most conservative bound on the absolute value, with the possibility of interference with the SM contribution taken into account. Due to the latter, the current bounds on $\mathcal{O}^{\text{VLL},\text{NP}}_{\nu d,\alpha\alpha sb}$ and $\mathcal{O}^{\text{VLR}}_{\nu d,\alpha\alpha sb}$ are the least stringent ones. 

Scalar operators are more strongly constrained, both in the case of contributions to diagonal elements and those to off-diagonal elements, and tensor operators exhibit the tightest bounds. This general trend can be expected to remain so in the future as well, with the only exception given by $\mathcal{O}^{\text{VLR}}_{\nu d,\alpha\alpha sb}$. The bound on this operator is projected to outperform the one on $\mathcal{O}^{\text{VLL}}_{\nu d,\gamma\delta sb}$ due to interference with the SM, because of which there is a comparatively large contribution $\propto  C^{\text{VLL},\text{SM}}_{\nu d,\alpha\alpha sb}C^{\text{VLR}}_{\nu d,\alpha\alpha sb}$ to the relevant observables. 

Therefore, the future sensitivity to $\mathcal{O}^{\text{VLR}}_{\nu d,\alpha\alpha sb}$ may become about ten times as strong as the current bound, whereas the improvement factor for scalar operators is roughly five, and about or less than  two for the other operators. Besides the numerical values of the bounds on the Wilson coefficients, we also provide an approximate lower bound for the associated scale
\begin{equation}
    \Lambda \approx \frac{1}{\sqrt{|C^{\text{XLY}}_{\nu d,\alpha\beta sb}|}}
\end{equation}
at which new physics contributing to the respective operator might reside. Here, tree-level mediation and $\mathcal{O}(1)$ couplings are assumed, and potential (unknown) enhancement or suppression factors are neglected. 

Currently, depending on the operator under consideration, new physics scales between a few TeV and roughly 25
TeV may be seen as (partly) constrained. In the fourth and seventh column of Table~\ref{tab:single_operator_constraints}, we provide the process which gives rise to the indicated bound. If $B\to K^{(*)}\nu\nu$ is indicated, both processes are very similarly competitive. We find that $B\to K\nu\nu$ is most sensitive to scalar operators, whereas tensor operators receive the most stringent constraint from $B\to K^*\nu\nu$. For vector operators, there is no overall trend towards one clearly most competitive observable.

\begin{figure}[bt!]
    \centering
    \includegraphics[width=0.49\linewidth]{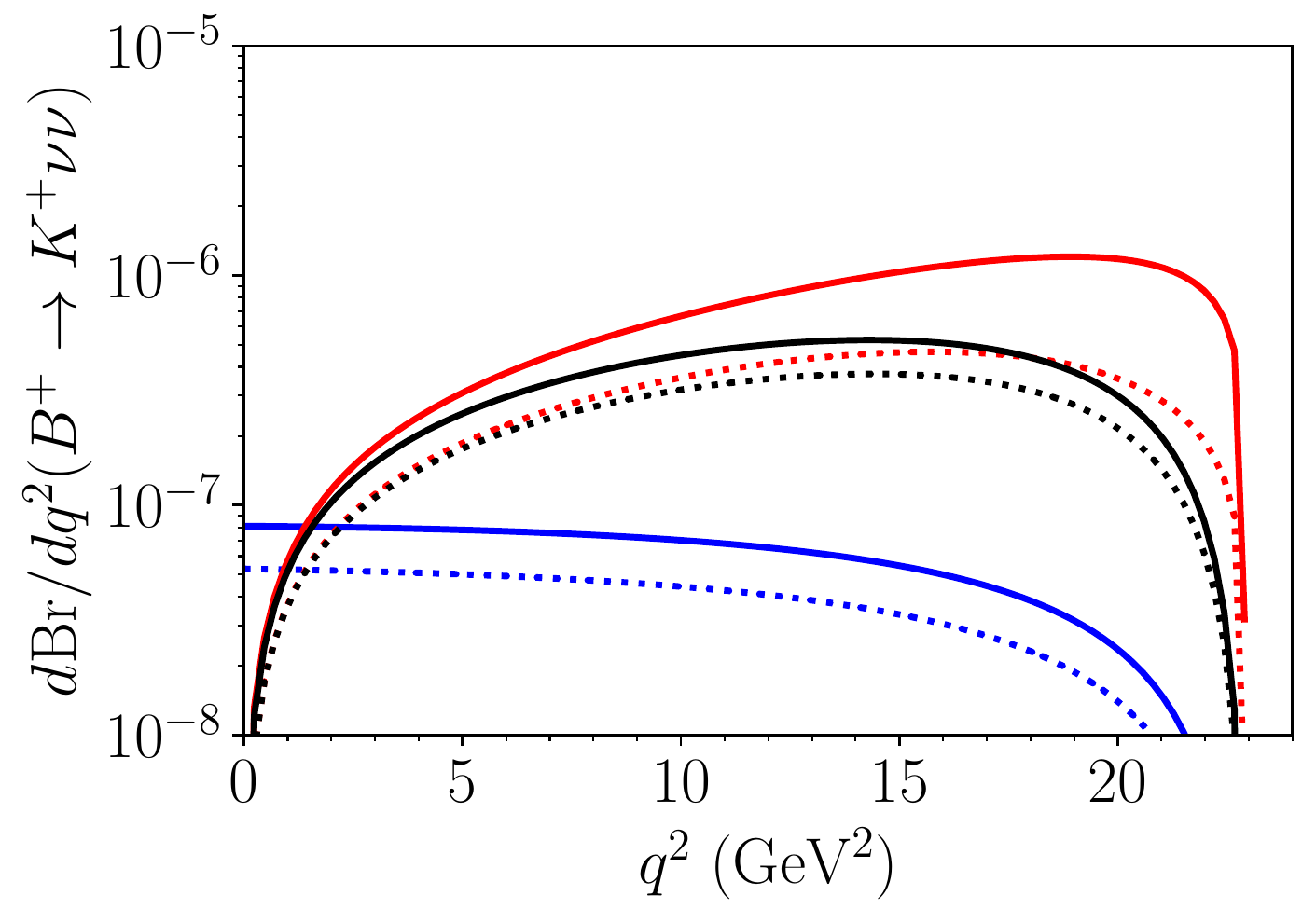}
    \includegraphics[width=0.49\linewidth]{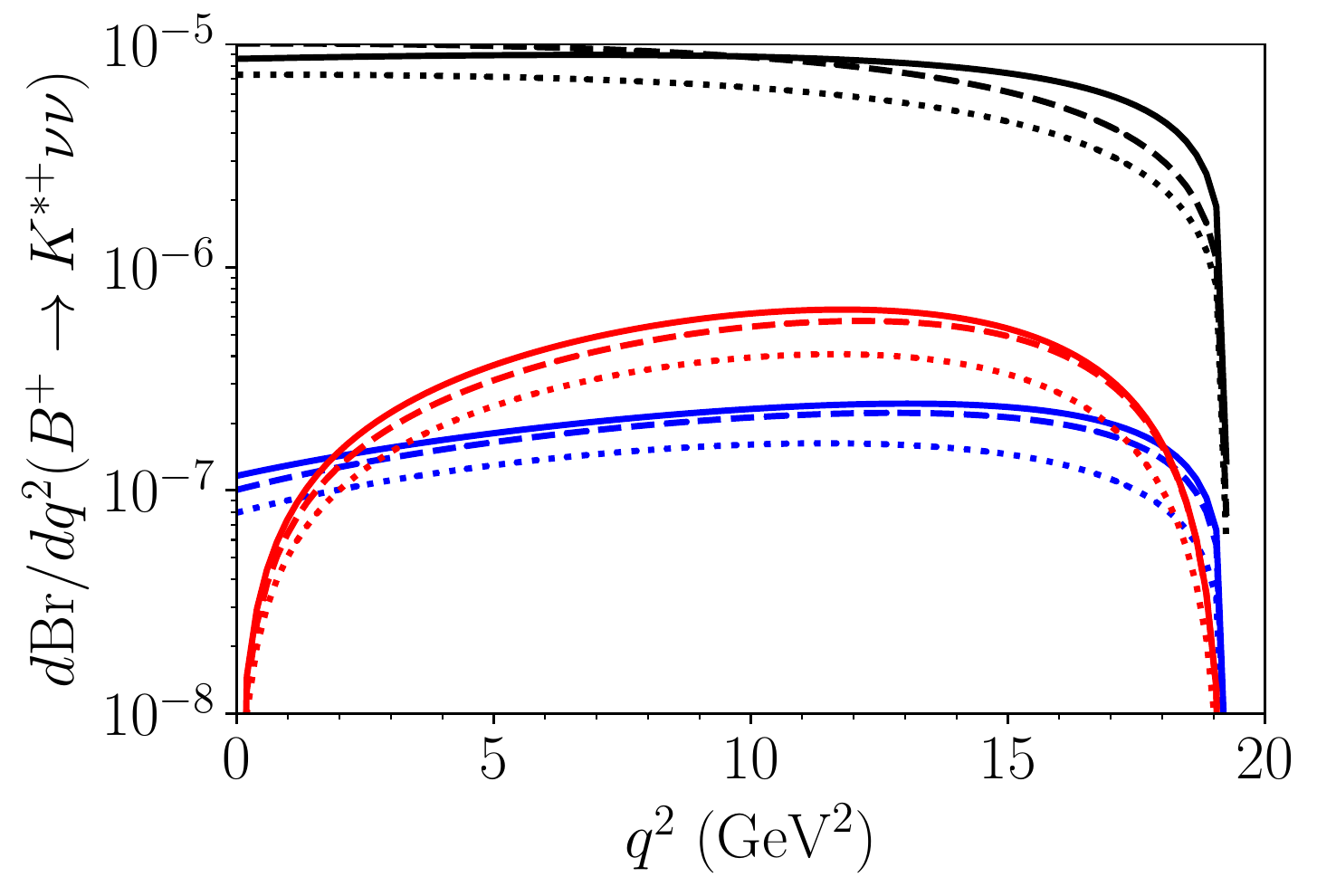}

    \includegraphics[width=0.49\linewidth]{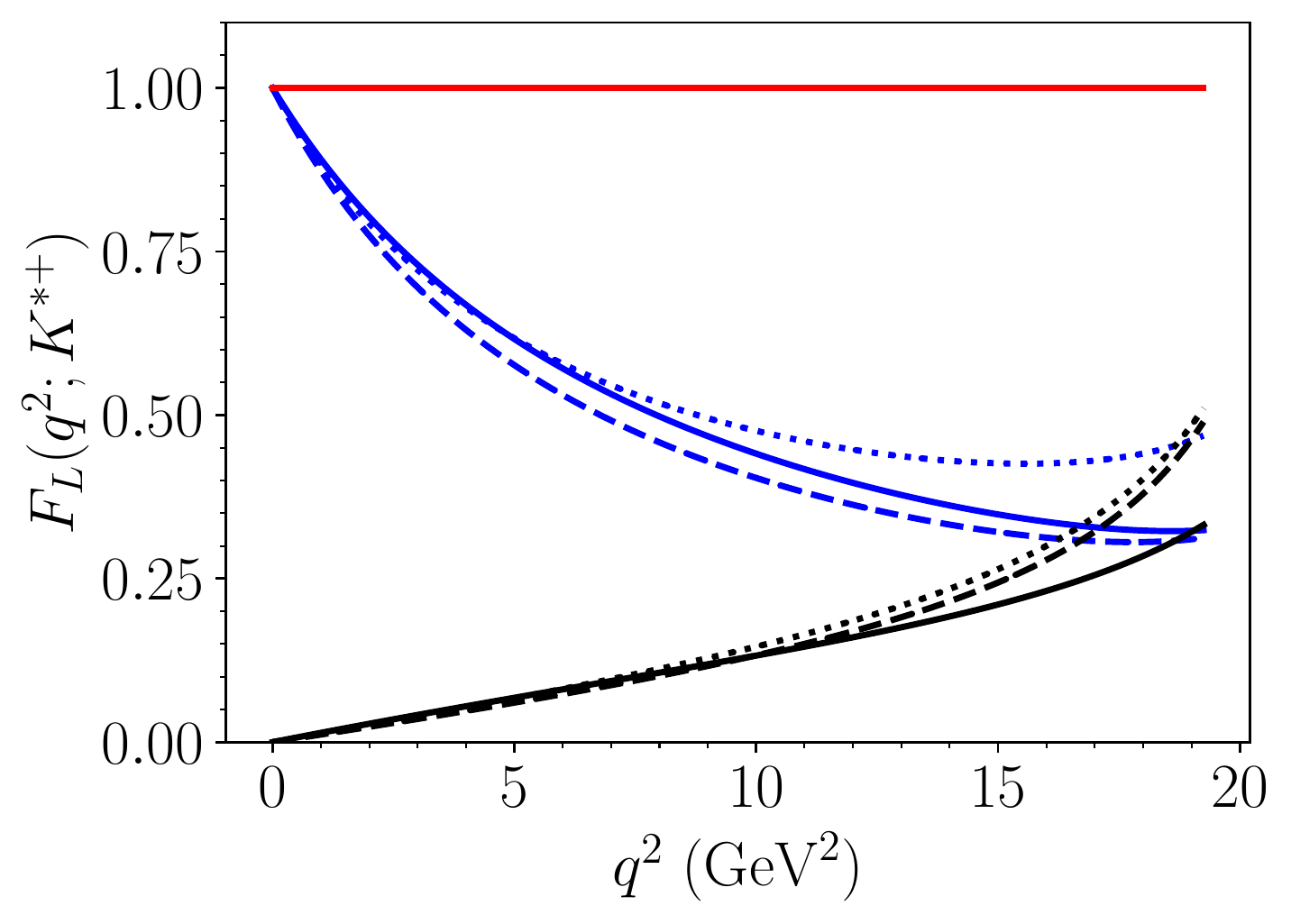}
    \caption{The differential branching ratio distributions (top) and the differential longitudinal polarisation fraction $F_L$ (bottom) generated for different non-zero Wilson coefficients $C^{\rm XLL}_{\nu d,23sb} =0.01 \,\mathrm{TeV}^{-2}$ for X = V,S,T, and choices of form factors. The blue (red) [black] lines stand for the vector (scalar) [tensor] operator, respectively. The solid lines denote the results for the form factors which are used in the analysis, taken from~\cite{Bharucha:2015bzk} for the $B\to K^*$ and from~\cite{Gubernari:2018wyi} for the $B\to K$ form factors. Both sets of form factors are based on a combined fit to LCSR and LQCD data. The dotted lines indicate the form factors based on the LCSR fit in \cite{Gubernari:2018wyi} and the dashed lines show the $B\to K^*$ form factors obtained using a combined fit to LCSR and LQCD in \cite{Gubernari:2018wyi}. Note that no SM contribution is included here. 
    }
    \label{fig:differential_decay}
\end{figure}

Lastly, we discuss the differential branching ratios as functions of the transferred momentum $q^2$. Figure~\ref{fig:differential_decay} shows the contours for vector (blue), scalar (red) and tensor (black) operators. For each of the curves, one representative non-zero Wilson coefficient is introduced.
We have set $C^{\text{XLL}}_{\nu d,23sb}=0.01\, \mathrm{TeV}^{-2}$ in all cases X = V,S,T.
The different linestyles correspond to different choices for the form factors as detailed in the caption of Figure~\ref{fig:differential_decay}.
One finds that vector operators dominantly contribute to the small-$q^2$ region in the case of $B\to K\nu\nu$, whereas the tensor and scalar operators source this decay more efficiently at intermediate and large $q^2$, respectively. For $B\to K^*\nu\nu$, one instead finds that the contributions from tensor operators are quite large for small and intermediate $q^2$ and then decrease. Here, vector and scalar operators become most efficient for larger $q^2$ values. 
As we use a logarithmic scale on the vertical axes in the top plots of Figure~\ref{fig:differential_decay}, we cannot show the behaviour of the respective curves at the kinematic endpoints which can be intuitively understood in terms of helicity conservation, see for instance Ref.~\cite{Kim:1999waa} for a discussion.
Three $q^2$ bins would most likely already help distinguish potential contributions from different operators for either decay channel. As for the form factors, the different sets are generally in good agreement for each operator. The largest discrepancies arise for $B\to K\nu\nu$ in the case of scalar operators.

As can be seen from the definition in Eq.~\eqref{eq:F_L}, the (unbinned) longitudinal polarisation fraction is not sensitive to the value of the contributing Wilson coefficient if only one is switched on at a time. Scalar-operator contributions do generally not enter the numerator of $F_T \equiv 1 - F_L$ and thus imply $F_L(q^2;K^{*+}) = 1$ (without taking into account the SM contribution).
The behaviour of vector and tensor operators is complementary in the sense that the former gradually reduce the value of $F_L(q^2;K^{*+})$ if $q^2$ increases, whereas the effect of the latter is a complete cancellation of $F_L(q^2;K^{*+})$ for small $q^2$ which then becomes less efficient for larger $q^2$. This is related to the normalisation of the relevant helicity amplitudes with respect to $q^2$, i.e., one has $q^2|H^{V(A)}_{\pm\alpha\beta}|^2\to 0$ and $q^2|H^{T(T_t)}_{0\alpha\beta}|^2\to 0$, but $q^2|H^{V(A)}_{0\alpha\beta}|^2\to \text{const.}$ and $q^2|H^{T(T_t)}_{\pm\alpha\beta}|^2\to \text{const.}$ for $q^2\to 0$, see Eqs.~(\ref{eq:hel_ampl_BtoKs}), since $A_{12}$, 
$T_1$ and $T_2$ do not vanish at $q^2 = 0$. In general, the distributions pertaining to scalar and tensor operators (approximately) converge at the kinematic endpoint of the distribution, only for the ones based on the form-factor set in Ref.~\cite{Gubernari:2018wyi} which employs a combined fit to LQCD and LCSR data there is a slight discrepancy.


\subsection{Two Operators with Massless Neutrinos}
\label{sec:massless_neutrinos_two_operators}

In the following, we discuss the parameter space compatible with non-zero contributions from two operators induced by new physics under the assumption of massless neutrinos, both for sterile states and as an approximation for the very small masses of the active SM neutrinos. The case of massive neutrinos is discussed in Sect.~\ref{sec:massive_neutrinos}.

Depending on the observable and whether the two operators shown in a plot interfere with each other, the parameter space compatible with that observable will in most cases have the shape of an ellipse or of straight bands. Straight bands indicate the possibility of exact cancellations among two operators. This occurs if the observable under consideration depends only on the sum or the difference of the two Wilson coefficients shown. If there is no interference between the two operators, the viable parameter space will in general be elliptic. The same shape arises if the observable under consideration receives contributes both from the sum and from the difference. The cases can be distinguished based on the orientation of the ellipses in parameter space. The occurrence of parabola in the case of $F_L(K^{*+})$ is due to its insensitivity to contributions from a single vector operator or due to cancellations between contributions to the numerator and the denominator, see the bottom-right plot in Figure~\ref{fig:belleII_2}.

\begin{figure}[ht!]
    \centering
    \includegraphics[width=0.49\linewidth]{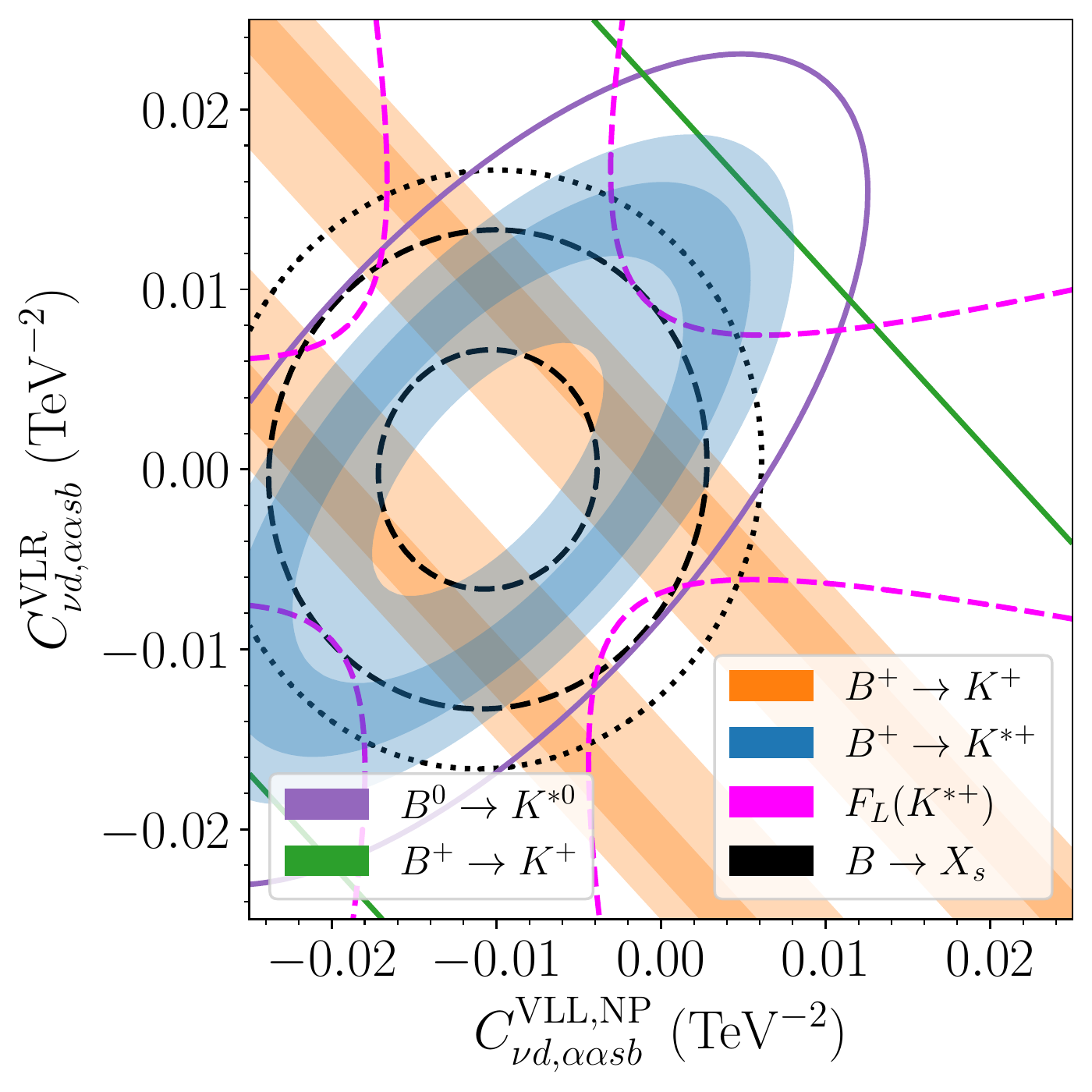}
    \includegraphics[width=0.49\linewidth]{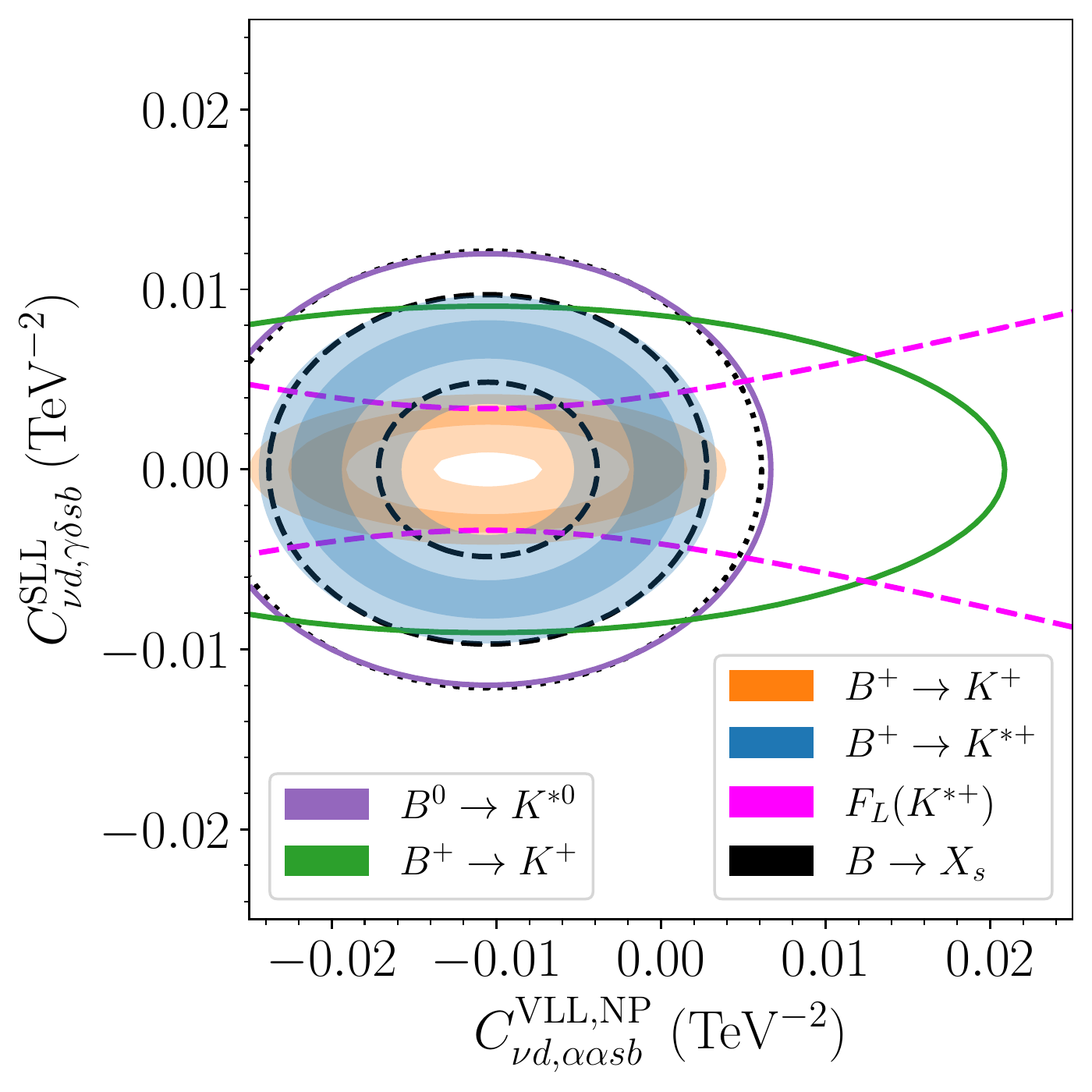}
    
    \includegraphics[width=0.49\linewidth]{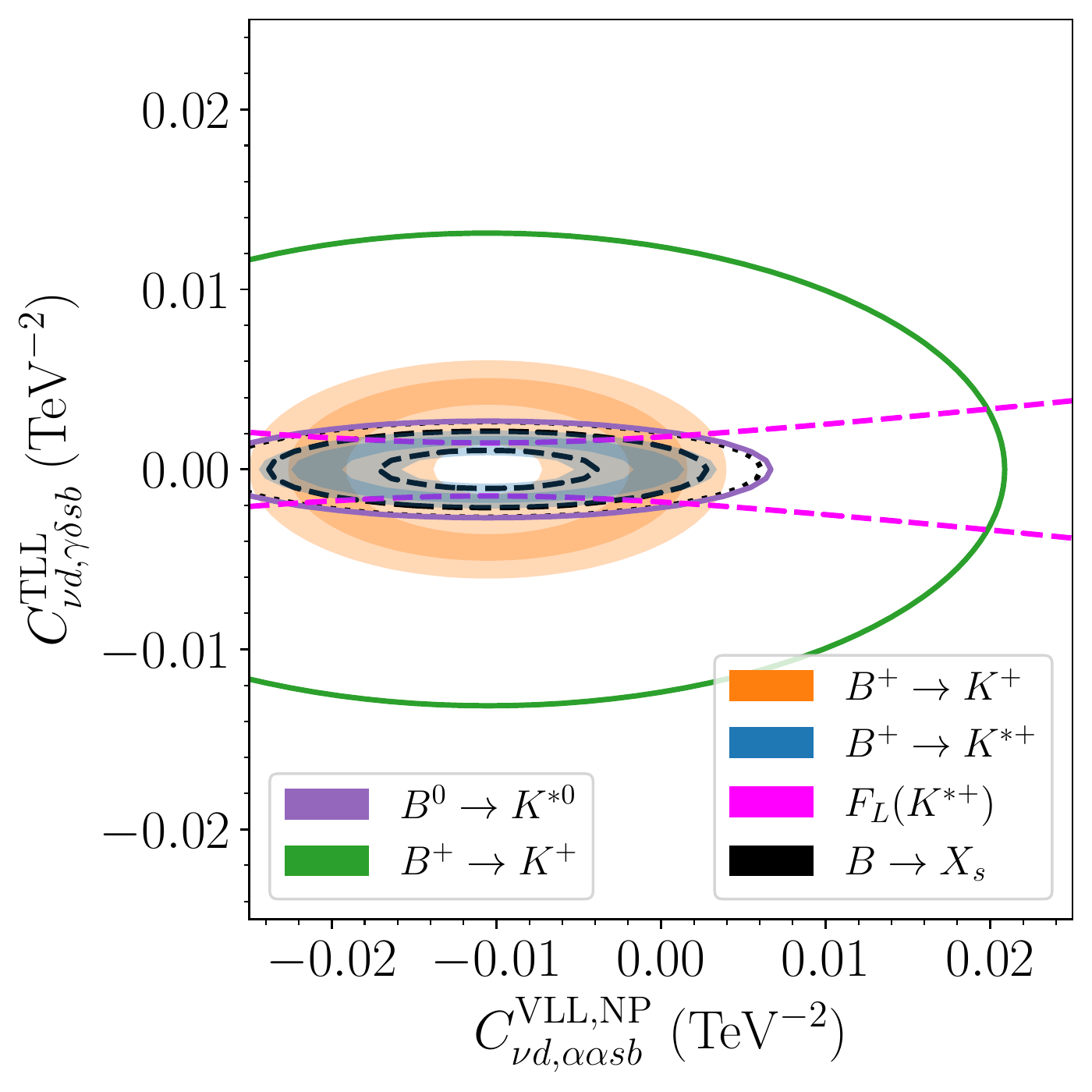}
    
    \caption{The allowed parameter space for the Wilson coefficients under the assumption that the Belle-II results for $5\;\text{ab}^{-1}$ (light shaded regions) and for $50\;\text{ab}^{-1}$ (dark shaded regions, dashed lines) for several $b\to s\nu\nu$ observables will confirm the SM predictions. In the shown cases, interference with the SM occurs. We use the sensitivities referenced in \cite{Kou:2018nap} and assume an experimental uncertainty of 50\% (dotted lines) and 20\% (dashed lines) for the inclusive decay $B\to X_s\nu\nu$, respectively. The solid dark purple and green lines reflect the current experimental bounds, see Table \ref{tab:BelleII}. For the neutrino flavor indices, $\alpha\in (1,2,3)$, while $\gamma$ and $\delta$ are arbitrary.
}
    \label{fig:belleII_1}
\end{figure}

In the plots in Figure~\ref{fig:belleII_2}, no interference with the SM contribution occurs. In the plots in Figure~\ref{fig:belleII_1}, the Wilson coefficient shown on the horizontal axis interferes with the SM contribution. This implies an overall shift of the centre of the resulting viable parameter space, i.e.~the intersection of the regions pertaining to the different observables, from $(0,0)$ to $(-1,0)$ in units of $|C^{\text{VLL},\text{SM}}_{\nu d,\alpha\alpha sb}| \approx 0.01\,\text{TeV}^{-2}$. Moreover, a region containing that point will be excluded as well, since destructive interference would render the respective decays unobservable in there, contrary to our assumption that Belle II will confirm the SM predictions.
If instead the measured branching ratios turned out to be larger than expected, the viable regions would in general get inflated, since there would necessarily have to be non-zero contributions from new physics to induce the measured excess. In the case of no interference with the SM, an excluded region containing $(0,0)$ would appear. The excluded region containing $(-|C^{\text{VLL},\text{SM}}_{\nu d,\alpha\alpha sb}|,0)$ in the plots in Figure~\ref{fig:belleII_1} would also grow, since a cancellation of the SM contribution would be even more strongly disfavoured. On the contrary, if Belle II turned out to measure smaller branching ratios than expected, this would imply that there has to be cancellation of the SM contribution. Thus, the viable region in the plots would generally shrink towards their respective centre points.

If the constraints from all decay channels are combined, there trivially is at least one single connected viable region in parameter space containing $(0,0)$. If neither of the shown operators interferes with the SM contribution, it is the only viable region. In the case of interference, a region enclosing the point $(-2,0)$ in units of $|C^{\text{VLL},\text{SM}}_{\nu d,\alpha\alpha sb}|$ will be viable as well. This is because the new-physics contribution will result only in a sign flip of $C^{\text{VLL}}_{\nu d,\alpha\alpha sb}$ which has no observable effect.\footnote{Throughout this work, $C^{\text{VLL}}_{\nu d,\alpha\alpha sb}$ = $C^{\text{VLL},\text{SM}}_{\nu d,\alpha\alpha sb} + C^{\text{VLL},\text{NP}}_{\nu d,\alpha\alpha sb}$ is understood.}

In the case of vector operators, a region compatible with $B\to K\nu\nu$ has the shape of a straight band as can be seen in the top-left plot of Figure~\ref{fig:belleII_1} and the bottom-left plot in Figure~\ref{fig:belleII_2}, because the observable only depends on the (squared) sum of $\mathcal{O}^{\text{VLL}}_{\nu d,\alpha\beta sb}$ and $\mathcal{O}^{\text{VLR}}_{\nu d,\alpha\beta sb}$ where $\beta$ may be equal to $\alpha$. Thus, there are exact cancellations between opposite-sign contributions from new physics to these two Wilson coefficients. Put differently, $B\to K\nu\nu$ bounds new-physics contributions to left- and right-handed vector operators of equal sign. On the contrary, $B\to K^*\nu\nu$ depends both on the sum and on the difference of $\mathcal{O}^{\text{VLL}}_{\nu d,\alpha\beta sb}$ and $\mathcal{O}^{\text{VLR}}_{\nu d,\alpha\beta sb}$, each being multiplied by different combinations of form factors and constants. Hence, the parameter space compatible with $B\to K^*\nu\nu$ is always elliptic in the case of vector operators, see the top-left plot in Figure~\ref{fig:belleII_1} and the bottom-left plot in Figure~\ref{fig:belleII_2}.

Interference between $\mathcal{O}^{\text{VLL}}_{\nu d,\alpha\alpha sb}$ and $\mathcal{O}^{\text{VLR}}_{\nu d,\alpha\alpha sb}$ as visible in the top-left plot in Figure~\ref{fig:belleII_1} can slightly weaken the current single-operator constraints listed in Table~\ref{tab:single_operator_constraints} to  $-0.033\;\text{TeV}^{-2} \lesssim C^{\text{VLL},\text{NP}}_{\nu d,\alpha\alpha sb} \lesssim 0.012\;\text{TeV}^{-2}$
or  $-0.022\;\text{TeV}^{-2} \lesssim C^{\text{VLR}}_{\nu d,\alpha\alpha sb} \lesssim 0.022\;\text{TeV}^{-2}$
which amounts to an effect of roughly 18\% and 5\%, respectively,
and the implied lower bounds on the new-physics scale become $\Lambda\gtrsim 6\;\text{TeV}$ and $\Lambda \gtrsim 7\;\text{TeV}$. Note, though, that the future sensitivities are not noticeably affected in a similar way in the case of vector operators due to the fact that $B\to K\nu\nu$ and $B\to K^*\nu\nu$ will become almost equally competitive.

For $\mathcal{O}^{\text{VLL}}_{\nu d,\alpha\alpha sb}$ and $\mathcal{O}^{\text{VLR}}_{\nu d,\alpha\alpha sb}$, regions containing the point $(-1,\pm 1)$ in units of $|C^{\text{VLL},\text{SM}}_{\nu d,\alpha\alpha sb}|$ are viable as well, see the top-left plot in Figure~\ref{fig:belleII_1}. 
Thus, for these two operators, an experimental ``{}confirmation" of the SM will restrict any deviation of the new-physics contribution from the points $(0,0)$, $(0,-2)$ and $(-1,\pm 1)$ in units of $|C^{\text{VLL},\text{SM}}_{\nu d,\alpha\alpha sb}|$ to less than roughly $0.002\;\text{TeV}^{-2}$, respectively. If the sign of this deviation is the same (opposite) for $C^{\text{VLL},\text{NP}}_{\nu d,\alpha\alpha sb}$ and $C^{\text{VLR}}_{\nu d,\alpha\alpha sb}$, the relevant bound will be set by $B\to K\nu\nu$ ($B\to K^*\nu\nu$). For the region containing $(0,0)$, this would infer a prospective bound on $C^{\text{VLL},\text{NP}}_{\nu d,\alpha\alpha sb}$ which is numerically very similar to the one on $C^{\text{VLR}}_{\nu d,\alpha\alpha sb}$ in Table~\ref{tab:single_operator_constraints}. A region containing $(-1,\pm 1)|C^{\text{VLL},\text{SM}}_{\nu d,\alpha\alpha sb}|$ means that sizeable $\mathcal{O}(|C^{\text{VLL},\text{SM}}_{\nu d,\alpha\alpha sb}|)$ new-physics contributions to two Wilson coefficients effectively relocate the source of the processes under consideration from $\mathcal{O}^{\text{VLL}}_{\nu d,\alpha\alpha sb}$, as it is the case in the SM, to $\mathcal{O}^{\text{VLR}}_{\nu d,\alpha\alpha sb}$ without altering the experimentally accessible signal. Thus, the possibility of the existence of two further relatively small, disjoint windows for new physics will persist, distinguished by the sign of $C^{\text{VLR}}_{\nu d,\alpha\alpha sb}$, with an associated scale of roughly 10 TeV.

The bottom-left plot in Figure~\ref{fig:belleII_2} shows a situation where two vector operators interfere among themselves, but not with the SM. Here, it is sufficient to discuss the constraints for non-negative $C^{\text{VLR}}_{\nu d,\alpha \beta sb}$ where $\alpha\neq\beta$ for $\alpha < 4$ or $\beta < 4$, as there is no change under swapping $C^{\text{VLL}}_{\nu d,\alpha \beta sb}\leftrightarrow C^{\text{VLR}}_{\nu d,\alpha \beta sb}$ or a sign flip of the contributions. The current constraint may weaken to $-0.018\;\text{TeV}^{-2}\lesssim C^{\text{VLL}}_{\nu d,\alpha \beta sb} \lesssim 0.018\;\text{TeV}^{-2}$ if $C^{\text{VLR}}_{\nu d,\alpha \beta sb} \approx (-)0.011\;\text{TeV}^{-2}$ at the upper (lower) bound, which amounts to a relaxation of the constraint on $\mathcal{O}^{\text{VLL}}_{\nu d,\gamma\delta sb}$ in Table~\ref{tab:single_operator_constraints} by roughly 30\%, and may be interpreted as the possibility of new physics residing at roughly 7 TeV.

\begin{figure}[tb!]
    \includegraphics[width=0.49\linewidth]{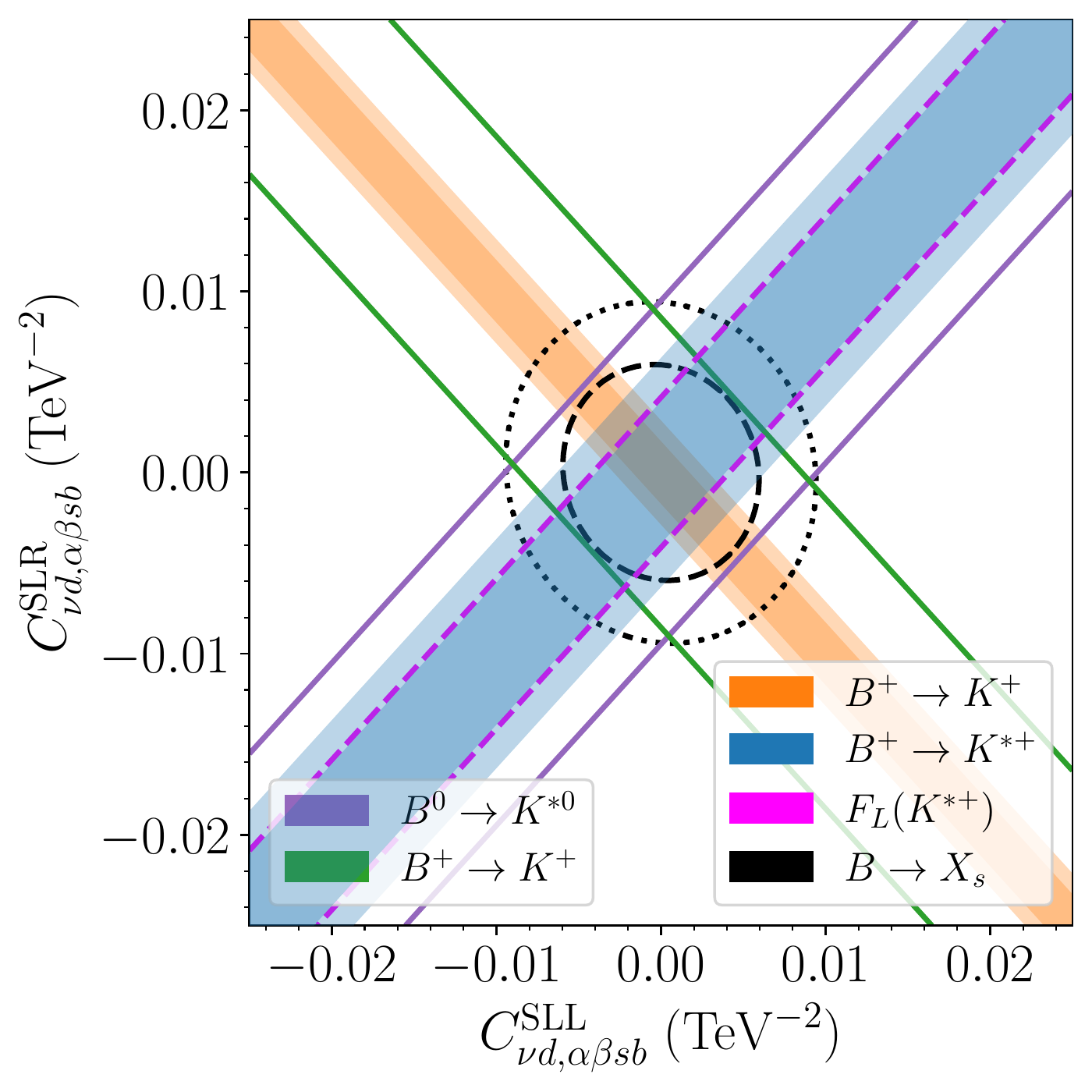}
    \includegraphics[width=0.49\linewidth]{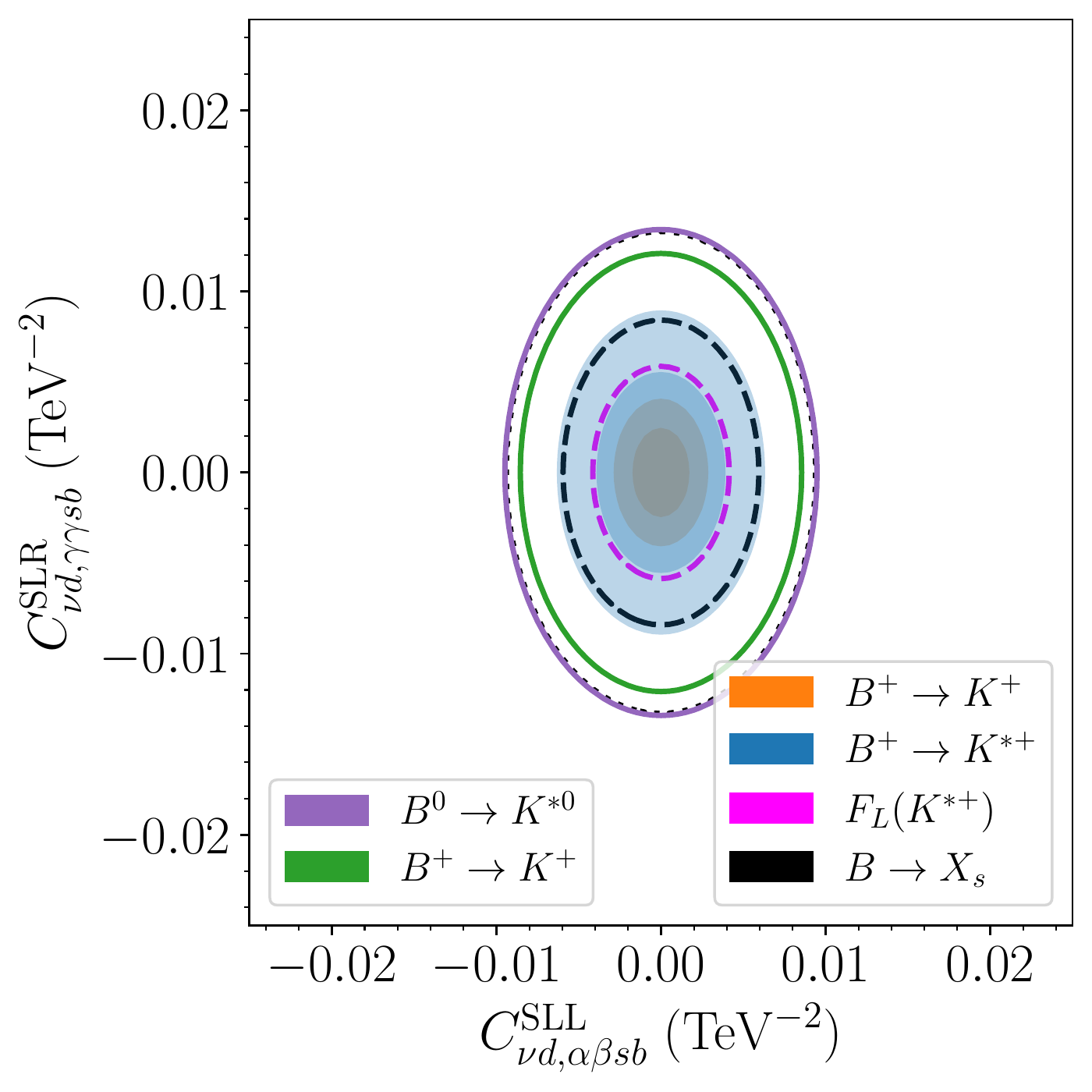}
    
    \includegraphics[width=0.49\linewidth]{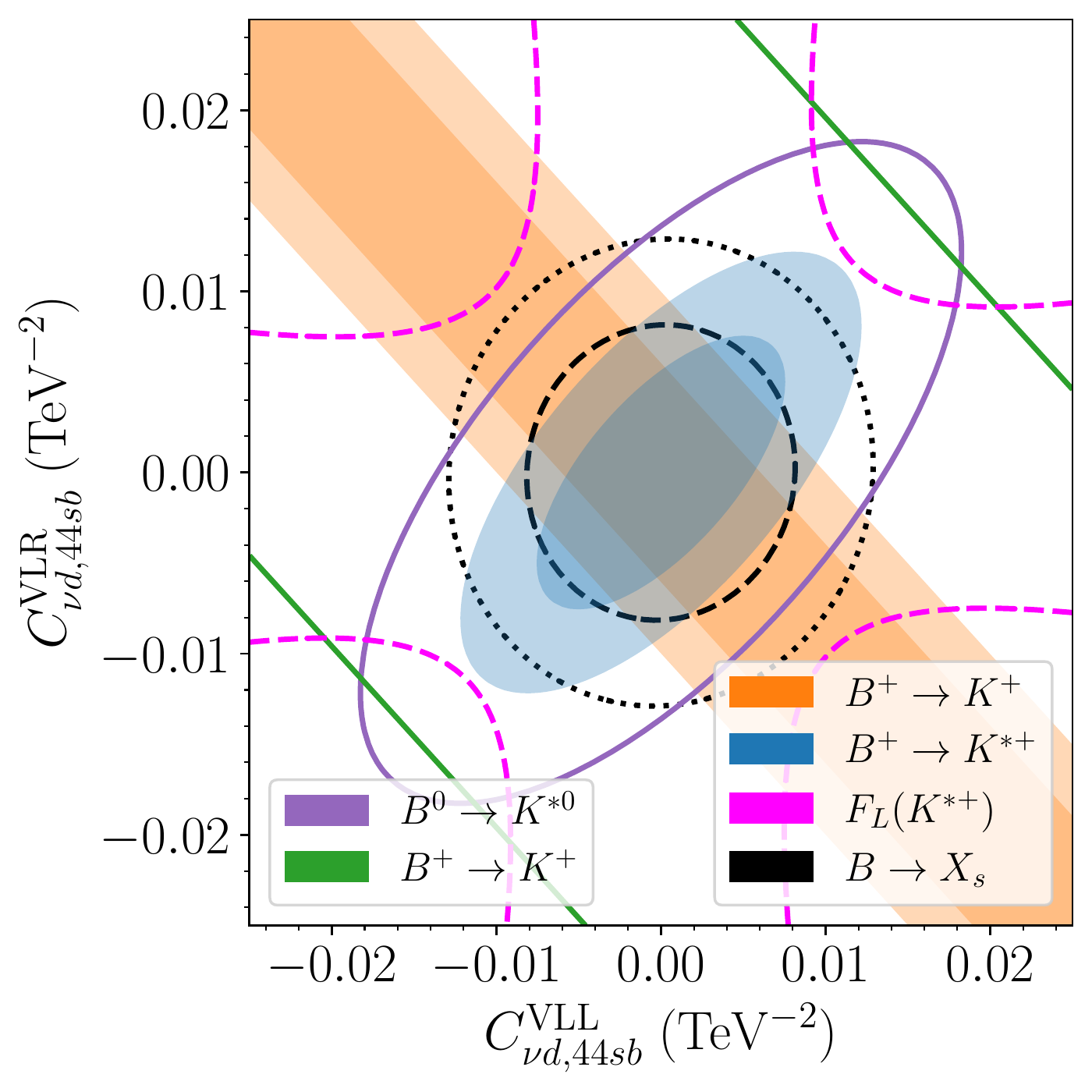}
    \includegraphics[width=0.49\linewidth]{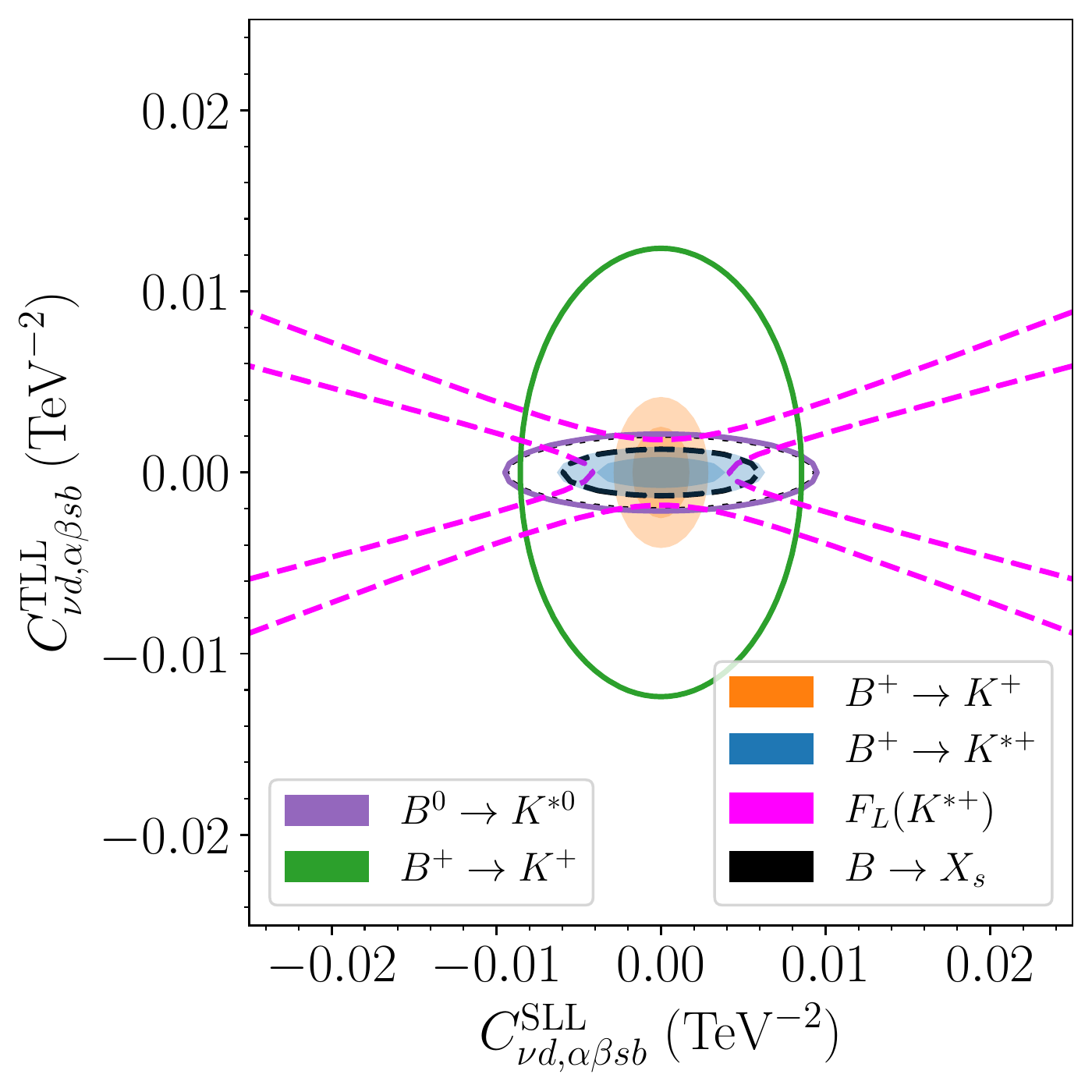}
 
    \caption{Continuation of Figure~\ref{fig:belleII_1}, but here the neutrino flavour indices are arbitrary with $\alpha \neq \beta$.}
   \label{fig:belleII_2}
\end{figure}

Vector operators are a suitable instance to make a case for efforts to experimentally access the inclusive mode $B\to X_s\nu\nu$. While this is very challenging, our results show that with an assumed sensitivity of 50\% one can already expect to (almost) exclusively probe parameter space which has been inaccessible so far. Also, note that the inclusive mode is less prone to cancellations among contributions from left- and right-handed operators than $B\to K^*\nu\nu$ in parts of parameter space.

In the case of scalar operators, $B\to K^*\nu\nu$ only depends on the difference of $\mathcal{O}^{\text{SLL}}_{\nu d,\alpha\beta sb}$ and $\mathcal{O}^{\text{SLR}}_{\nu d,\alpha\beta sb}$. Thus, the region compatible with this observable also has the shape of a straight band, as it can be seen in the top-left plot in Figure~\ref{fig:belleII_2}. $B\to K^*\nu\nu$ ($B\to K\nu\nu$) hence provides a bound on new-physics contributions to left- and right-handed scalar operators of opposite (equal) sign.

Note that interference between contributions to $O^{\text{SLL}}_{\nu d,\alpha\beta sb}$ and $O^{\text{SLR}}_{\nu d,\alpha\beta sb}$ cannot significantly relax the relevant current constraints indicated in Table~\ref{tab:single_operator_constraints}, but $|C^{\text{SLL}}_{\nu d,\alpha\beta sb}| \lesssim 0.003\;\text{TeV}^{-2}$ and a corresponding new-physics scale of roughly
20 TeV, which amounts to a loosening of the single-operator bounds by roughly 50\%, may still be viable in the future. This is due to the fact that the new-physics reach of $B\to K\nu\nu$ will become clearly dominant in the case of scalar operators, whereas currently $B\to K^*\nu\nu$ is only slightly inferior. In that sense, the situation is contrary to the one for vector operators where interference can only noticeably affect the current constraints.

Generally, the observable $F_L(K^{*+})$ is very suitable to test contributions to scalar operators because they only modify the denominator in $F_T \equiv 1 - F_L$, see Eq.~\eqref{eq:F_T}, whereas vector and tensor operators also alter the numerator. Furthermore, note that a single contribution to $\mathcal{O}^{\text{VLL}}_{\nu d,\alpha\alpha sb}$ can be removed from the $q^2$ integral in the numerator and the denominator and thus $F_L(K^{*+})$ is not sensitive to its value. If new physics contributes to $\mathcal{O}^{\text{SLL}}_{\nu d,\alpha\beta sb}$ and $\mathcal{O}^{\text{SLR}}_{\nu d,\alpha\beta sb}$ with opposite signs, $B\to K^*\nu\nu$ (for 50 ab$^{-1}$) and $F_L(K^{*+})$ provide similarly competitive constraints, as can be seen in the top-left plot of Figure~\ref{fig:belleII_2}. The top-right plot in Figure~\ref{fig:belleII_1} demonstrates that combining $F_L(K^{*+})$ with $B\to K^*\nu\nu$ would prove efficient in probing the scenario of new physics contributing to $\mathcal{O}^{\text{VLL}}_{\nu d,\alpha\alpha sb}$ and a scalar operator (or a tensor operator as shown in the bottom diagram). In this case, both observables related to $B\to K^*$ can already considerably tighten the existing bounds, and leveraging $B\to K\nu\nu$ as well would imply only a moderate further improvement especially in the case of tensor operators.  

The current single-operator bound on, say, $\mathcal{O}^{\text{SLR}}_{\nu d,\alpha\beta sb}$ does not significantly loosen if at the same time the SM contribution would be (partly) cancelled by new physics. In fact, an efficient cancellation of the SM contribution and a simultaneous contribution to scalar operators would already come under severe pressure if the 5 ab$^{-1}$ data set confirmed the respective SM predictions for $B\to K\nu\nu$ and $B\to K^*\nu\nu$, as there is only little overlap between the relevant light-shaded regions in the top-right plot in Figure~\ref{fig:belleII_2}. On the contrary, in the top-left plot the intersection of the viable regions pertaining to the 5 ab$^{-1}$ data set is even disconnected in parameter space, but cancellations of the SM contribution could not be excluded at all. Still, a scenario with $|C^{\text{SLR}}_{\nu d,\alpha\beta sb}| \lesssim 0.004\;\text{TeV}^{-2}$ and a less efficient cancellation of the SM contribution could only be ruled out with the 50 ab$^{-1}$ data set. This illustrates that Belle II can be expected to be quite efficient in constraining new physics which sources only one scalar operator with different final-state neutrinos, hence, more contributions would be necessary to ``{}mimic" the SM expectation.

For the $b\to s\nu\nu$ processes, tensor operators only exist with left-handed projectors in the fermion bilinears (together with their hermitean conjugates), thus they can never interfere with one another. As in the case of scalar operators, the 5 ab$^{-1}$ data set will not entirely suffice either to rule out the scenario that $|C^{\text{VLL},\text{SM}}_{\nu d,\alpha\alpha sb}|$ gets (partially) cancelled by new physics and the relevant decays under consideration are instead induced by tensor operators, but 50 ab$^{-1}$ will provide a conclusive answer, see the bottom plot in Figure~\ref{fig:belleII_1}. Note that if $\mathcal{O}^{\text{VLL},\text{NP}}_{\nu d,\alpha\alpha sb}$ and $\mathcal{O}^{\text{TLL}}_{\nu d,\gamma\delta sb}$ contribute together, instead of $B\to K^*\nu\nu$ one could consider $F_L(K^{*+})$ together with $B\to K\nu\nu$ without a significant loss of constraining power.


\subsection{Massive Neutrinos}
\label{sec:massive_neutrinos}

\begin{figure}[hbpt!]
    \centering
    \includegraphics[width=0.495\linewidth]{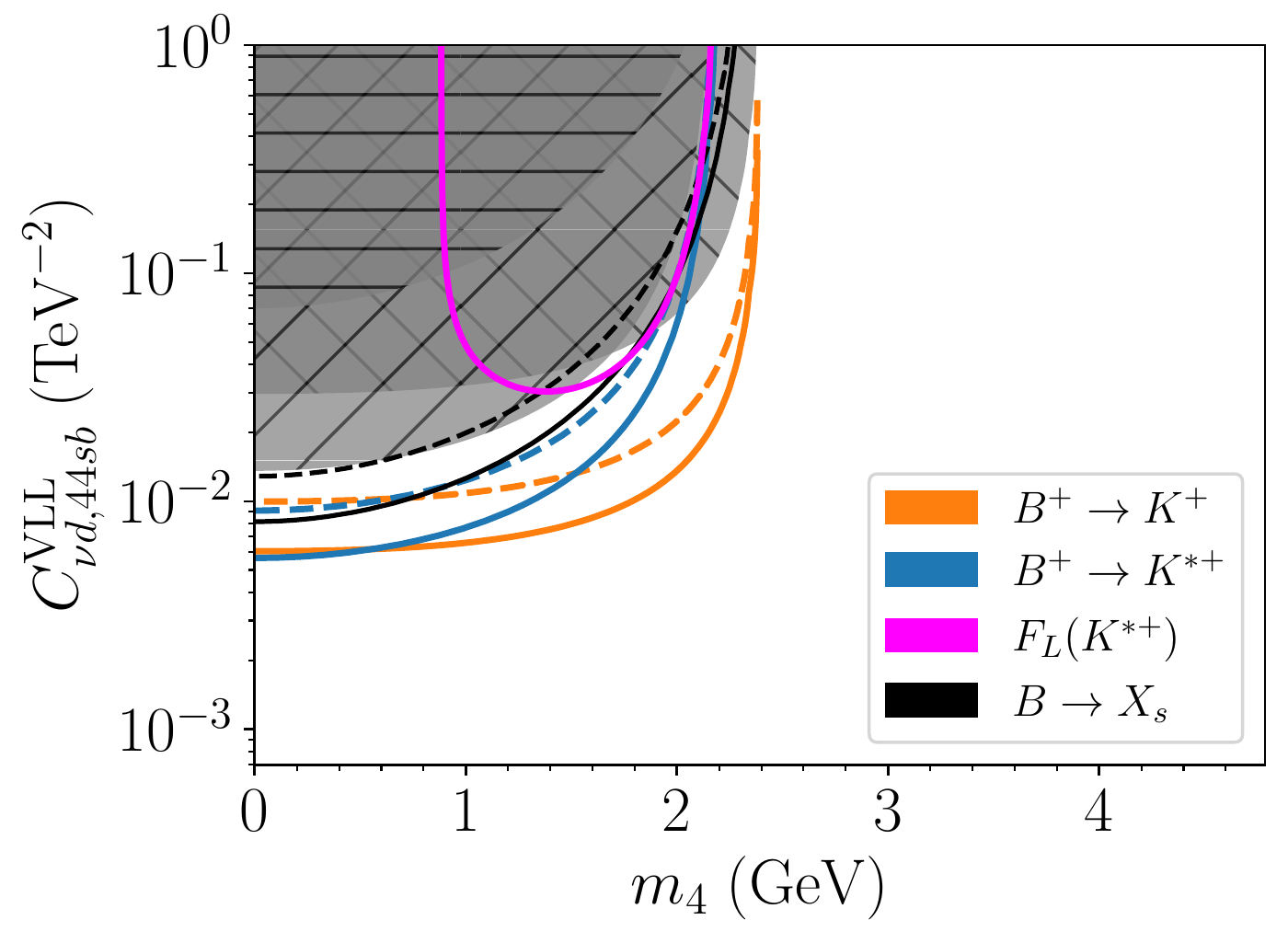}
    \includegraphics[width=0.495\linewidth]{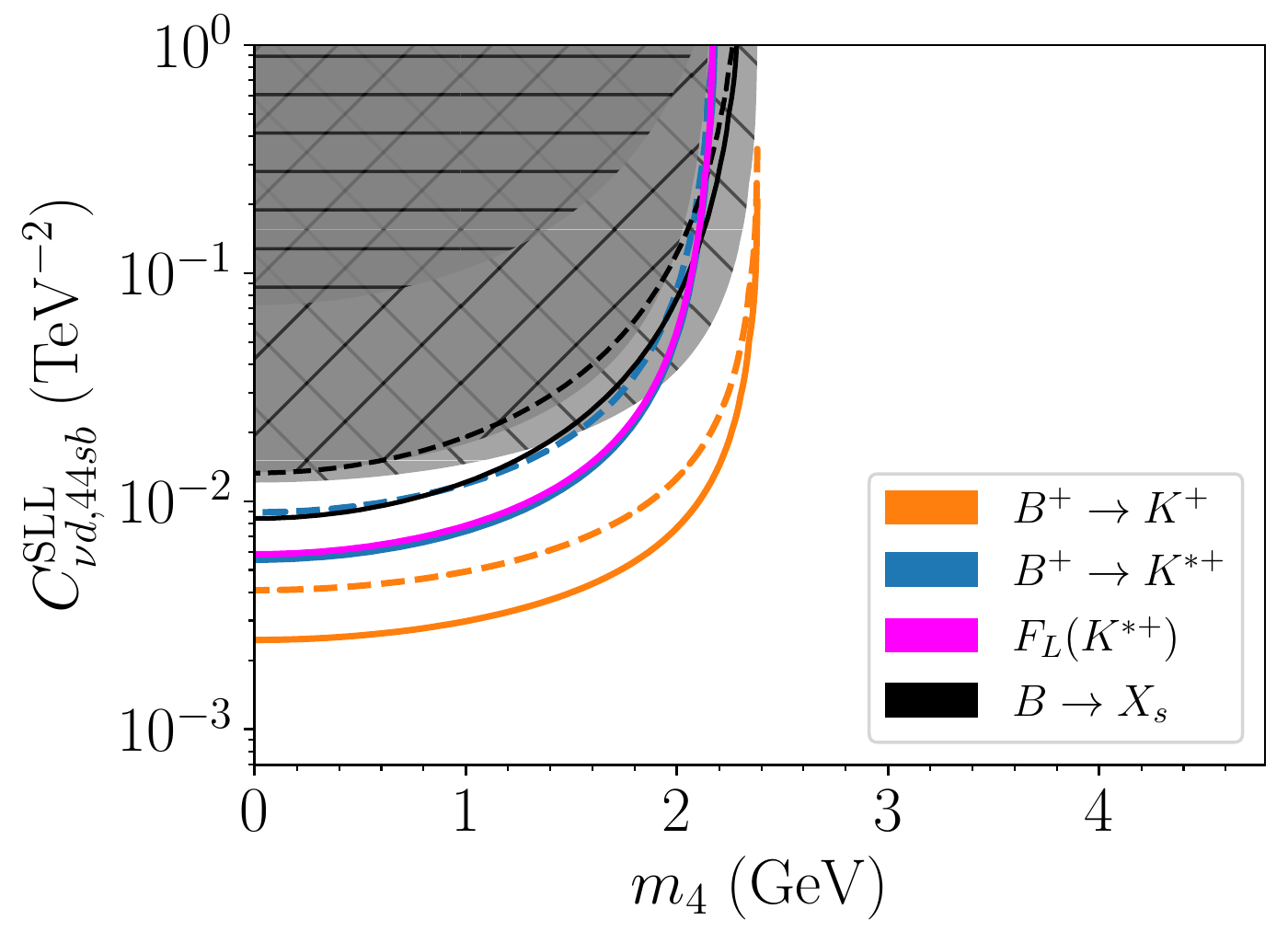}
    
    \includegraphics[width=0.495\linewidth]{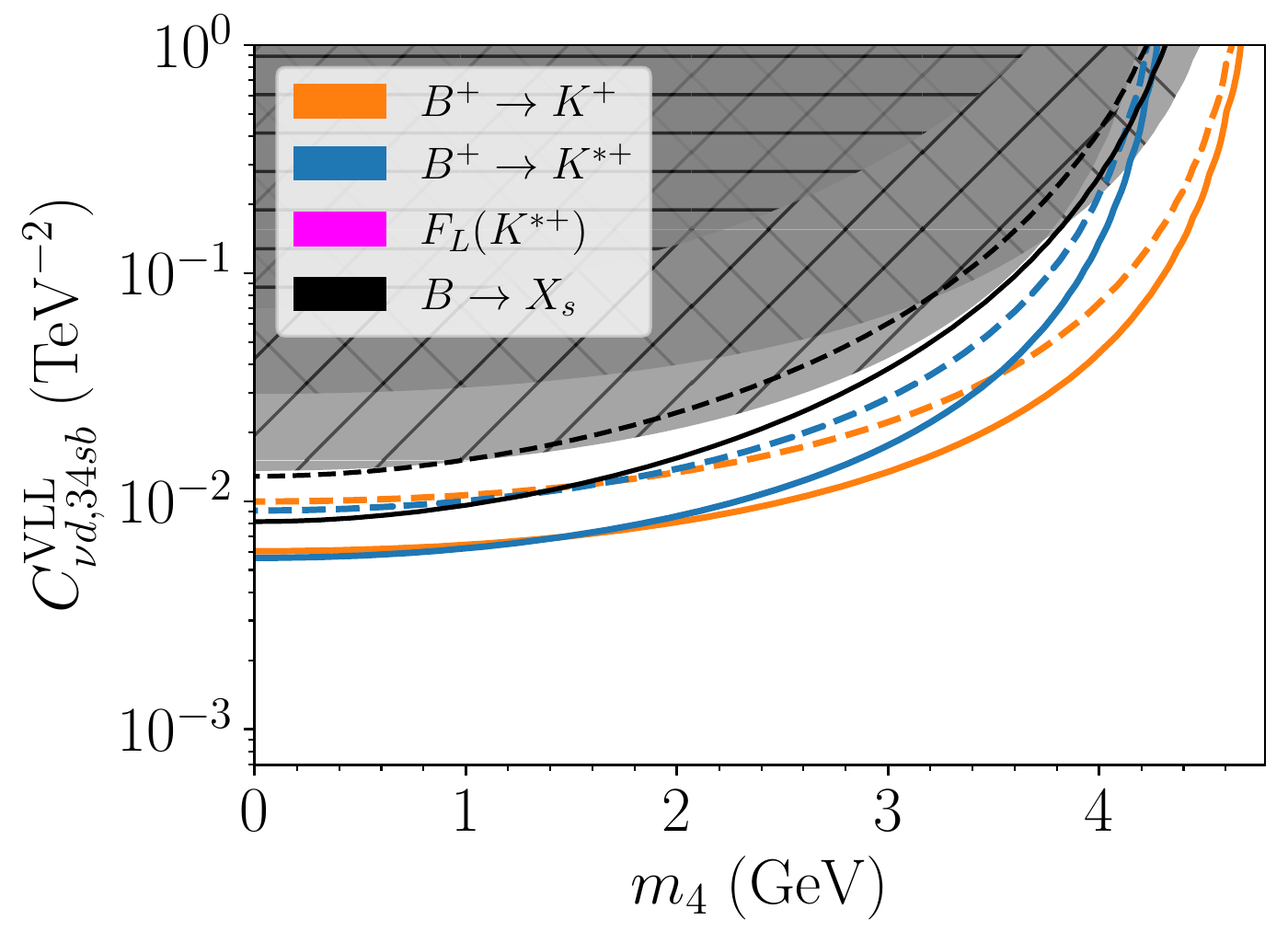}
    \includegraphics[width=0.495\linewidth]{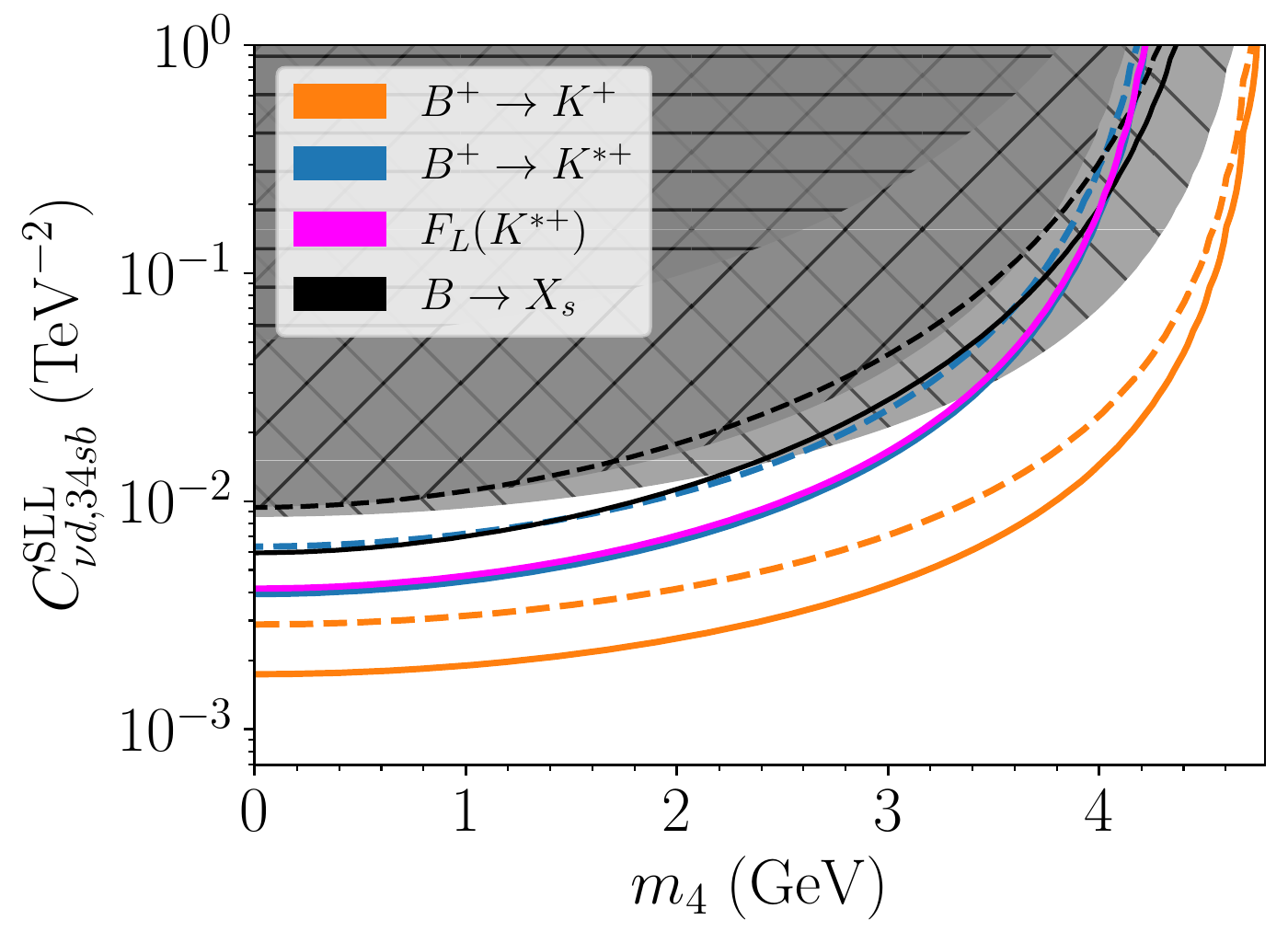}
    \includegraphics[width=0.495\linewidth]{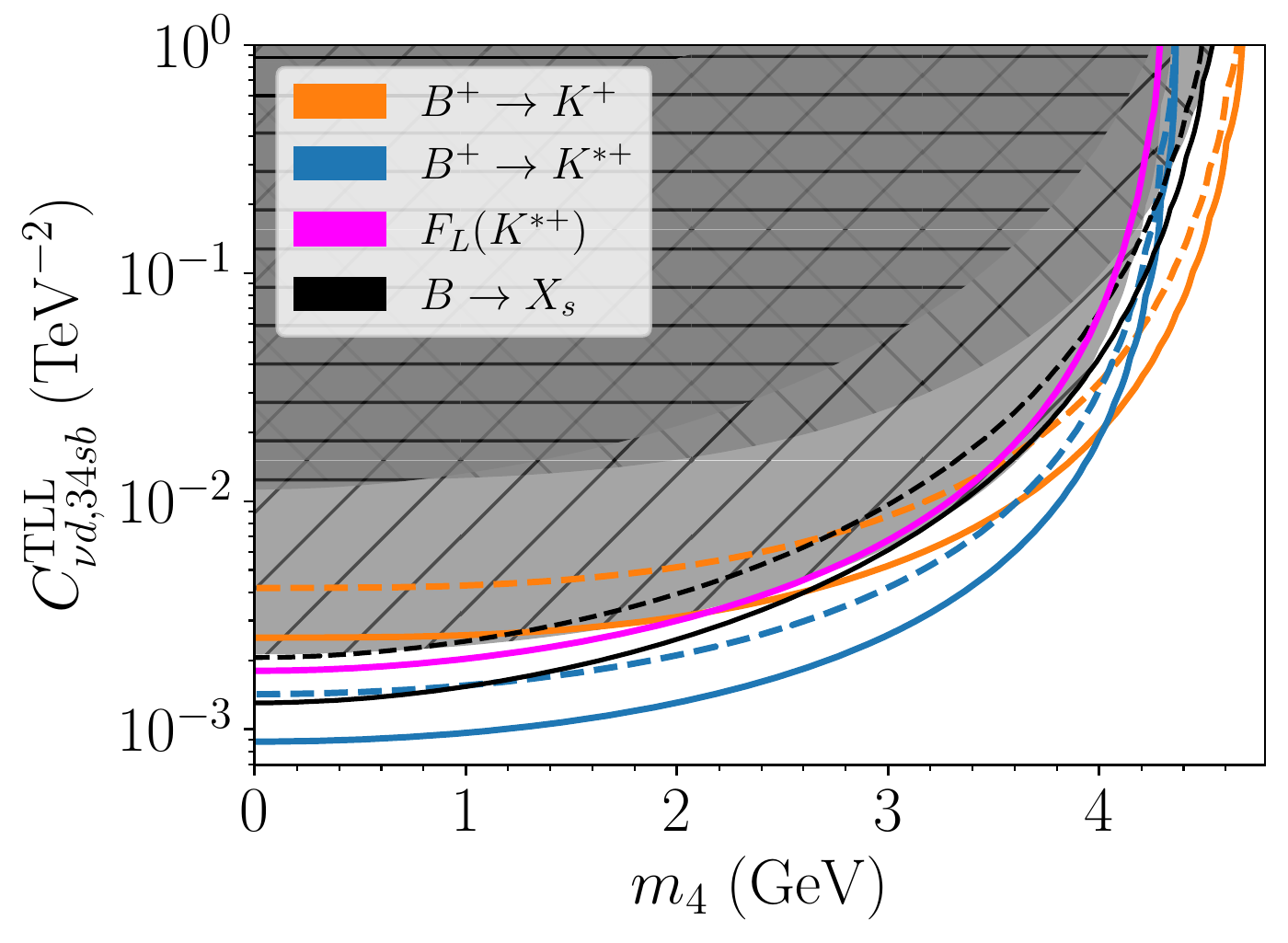}
    \caption{Current (shaded regions) and future sensitivities (lines) on a single Wilson coefficient as a function of the mass of two sterile neutrinos (top panel) and one sterile neutrino (middle panel and bottom plot) in the final state, respectively. We assume that the Belle-II results for $5\;\text{ab}^{-1}$ (dashed lines) and for $50\;\text{ab}^{-1}$ (solid lines) for several $b\to s\nu\nu$ observables will confirm the SM predictions. We use the sensitivities referenced in \cite{Kou:2018nap} and assume an experimental uncertainty of 50\% (solid line) and 20\% (dashed line) for the inclusive decay $B\to X_s\nu\nu$. Regions with \textbackslash\textbackslash\;(//) [--] hatching are excluded via the current bounds on $B^+\to K^+\nu\nu$ ($B^0\to K^{*0}\nu\nu$) [$B\to X_s\nu\nu$]. The constraints are identical if exchanging the third neutrino flavour $\alpha=3$ for $\alpha = 1,2$.}
    \label{fig:mass_vs_WC}
\end{figure}

For the discussion of the impact of non-zero neutrino masses, we start with the current constraints on and future sensitivities to a single Wilson coefficient, respectively, as a function of the mass of a sterile neutrino, as shown in Figure~\ref{fig:mass_vs_WC}. In each case, all other operators are switched off. It is assumed that the SM is extended by only one sterile neutrino, that is, there cannot be two sterile neutrinos in the final state with different masses.  We study the entire range from massless neutrinos up to the respective kinematic threshold for each process. Regarding the final-state neutrino flavours, we consider a representative off-diagonal element as well as the diagonal one with two sterile neutrinos with identical masses in the final state for the vector and scalar operator.

Note that for a sterile-neutrino mass larger than $m_4\gtrsim1.7 \;(3.7)$ GeV for (off-)diagonal elements of vector operators, $B\to K\nu\nu$ is currently more constraining than $B\to K^*\nu\nu$. In terms of future sensitivities, $B\to K\nu\nu$ and $B\to K^*\nu\nu$ are very similarly competitive in the (approximately) massless case, but for heavier sterile neutrinos $B\to K^*\nu\nu$ also grows more and more inferior. Indeed, for $m_4\gtrsim 1.5\;(3.5)$ GeV for (off-)diagonal vector-operator elements, even the results for $B\to K\nu\nu$ based on the 5 ab$^{-1}$ data set are projected to outperform the bounds imposed by all other observables. The plots in Figure~\ref{fig:mass_vs_WC} also reflect the previously made observation that $F_L(K^{*+})$ is not sensitive to a single contribution to $\mathcal{O}^{\text{VLL}}_{\nu d,\alpha\beta sb}$ in the case of massless neutrinos. A sizeable deviation from that only occurs for two massive neutrinos in the final state as can be seen in Eqs.~\eqref{eq:G000} and \eqref{eq:G200}, and thus $F_L(K^{*+})$ cannot impose a constraint on $C^{\text{VLL}}_{\nu d,34 sb}$ where the mass only reduces the available phase space. For $\mathcal{O}^{\text{VLL}}_{\nu d,44sb}$, a prospective constraint arises for $m_4\gtrsim 0.9$ GeV which nonetheless will only imply a (moderate) improvement over the current bounds in the range $1.5\;\text{GeV}\lesssim m_4\lesssim 1.9\;\text{GeV}$, and is generally not competitive.

For the entire accessible neutrino-mass range, $B\to K\nu\nu$ accounts for the highest future sensitivity to as well as the most stringent current bound on scalar operators, although this dominant role is not very pronounced in the latter case for very small or zero neutrino masses. Furthermore, irrespective of their symmetry properties, scalar operators are always more strongly constrained than vector operators also for non-zero neutrino masses. In the case of tensor operators, $B\to K^*\nu\nu$ imposes the most competitive bound for almost the entire neutrino-mass range. Indeed, for $1.1\;\text{GeV} \lesssim m_4\lesssim 3.6\;\text{GeV}$, even the results for $B\to K^*\nu\nu$ based on the 5 ab$^{-1}$ data set will outperform the bounds imposed by all other observables.
Only if the sterile neutrino is heavier than $m_4\gtrsim 4$ GeV, $B\to K\nu\nu$ becomes more competitive, and in this range the tensor operator will also be slightly less stringently constrained than the scalar operator. Hence, $B\to K\nu\nu$ is indeed a very powerful probe of new physics in $b\to s\nu\nu$ processes for all considered operators. Note that up to $m_4 \lesssim 2.6$ GeV, we find that the inclusive mode is more sensitive to tensor-operator contributions than $B\to K\nu\nu$.

\begin{figure}[bht!]
    \centering
    \includegraphics[width=0.49\linewidth]{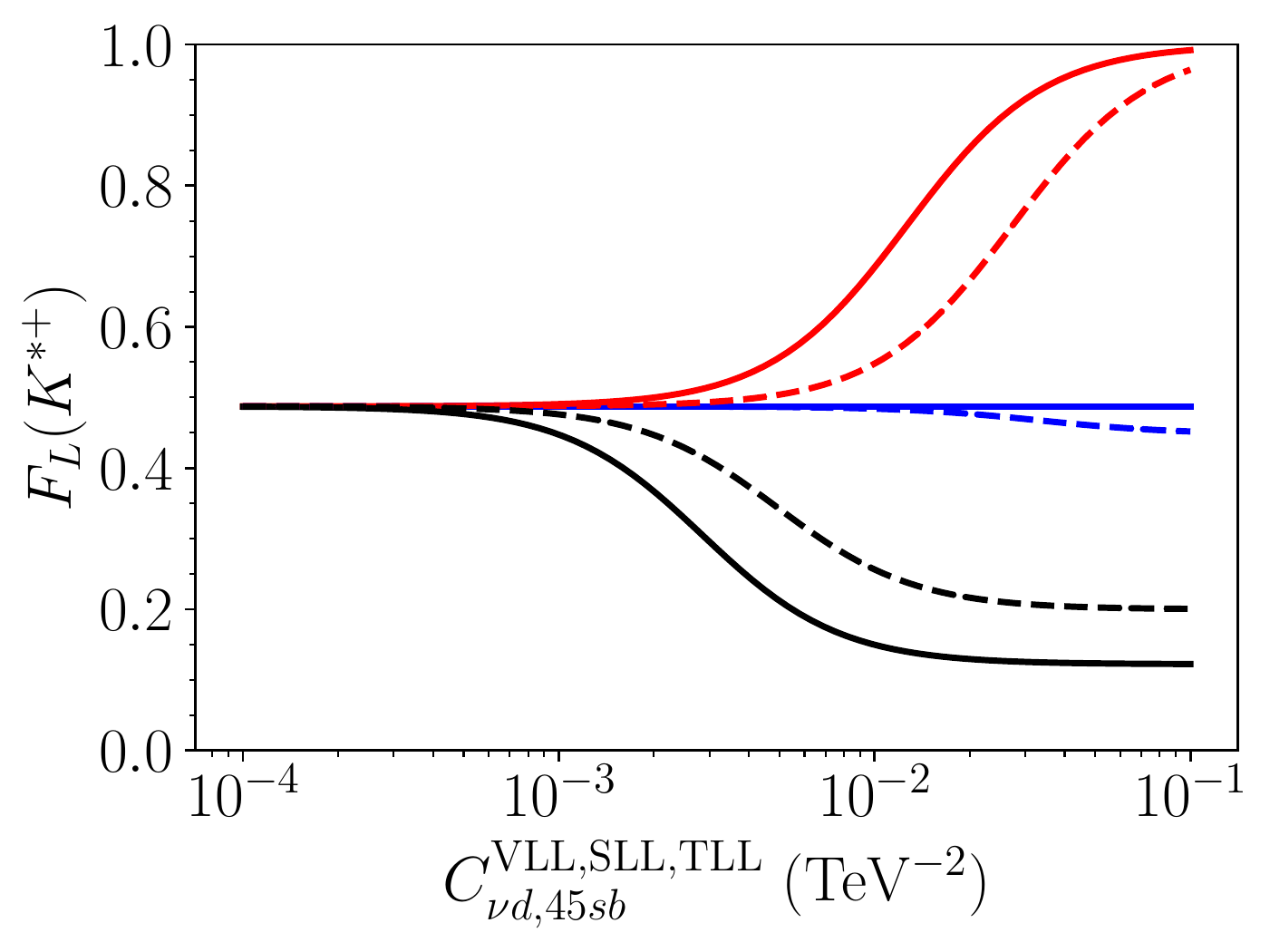}
    \includegraphics[width=0.49\linewidth]{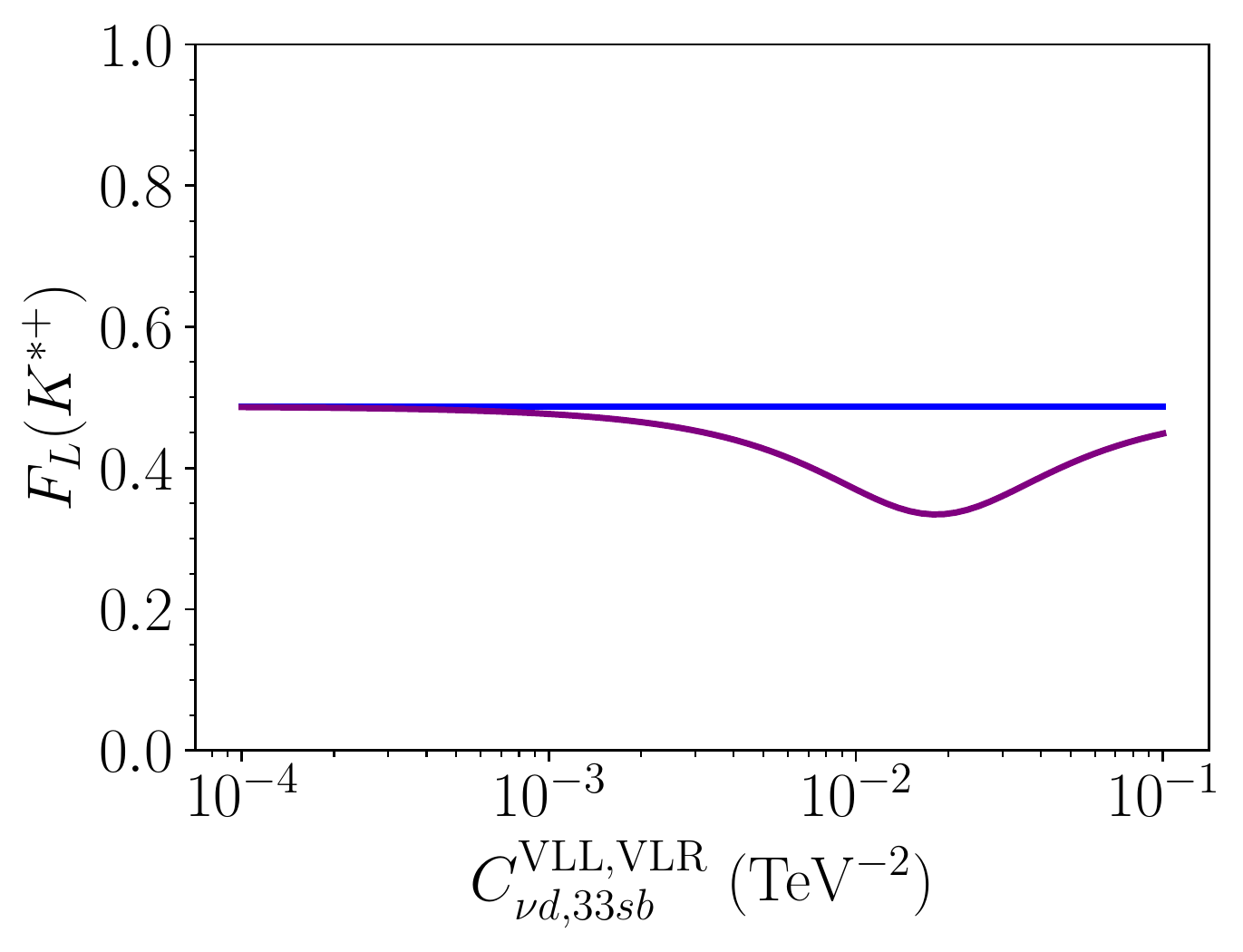}
    
    \caption{Binned longitudinal polarisation fraction $F_L(K^{*+})$ as a function of one new-physics Wilson coefficient at a time (including the SM contribution). On the left-hand side the blue (red) [black] lines stand for the vector (scalar) [tensor] operator, respectively. Solid [dashed] contours signify $m_{4,5} = 0$ GeV [$m_{4,5} = 1.5$ GeV]. On the right-hand side the blue (purple) lines stand for the Wilson coefficients $C^{\text{VLL}}_{\nu d,33sb}$ ($C^{\text{VLR}}_{\nu d, 33sb}$).  Note the binned longitudinal polarisation fraction $F_L(K^{*+})$ is obtained by separately binning the numerator and denominator, see Eq.~\eqref{eq:def_FL}, and not by integrating the distributions $F_L(q^2;K^{*+})$ shown in Figure~\ref{fig:differential_decay}.
    }
    \label{fig:FL}
\end{figure}

Linearly adding up the theoretical and experimental uncertainties for $F_L(K^{*+})$ as given in Table~\ref{tab:BelleII}, one finds that only a result in the range $(0.37,0.61)$ would be compatible with the SM expectation at $1\sigma$. Hence, a measurement of $F_L(K^{*+})$ in principle allows for a sharp distinction between the case of dominant contributions to only the scalar operator, or only the tensor operator, as can be seen in Figure~\ref{fig:FL} on the left-hand side. As scalar operators do only contribute to the denominator, but not to the numerator of $F_T$, increasing the Wilson coefficient only implies a shrinkage of the difference of $F_L(K^{*})$ from 1. Note that with one massive neutrino in the final state, $F_L(K^{*+})$ is affected by new physics contributing to $\mathcal{O}^{\text{VLL}}_{\nu d,\alpha\alpha sb}$ only via phase-space suppression which does not result in a competitive bound, see Eqs.~\eqref{eq:G000} and~\eqref{eq:G200}. While $F_L(K^{*+})$ is sensitive to new-physics contributions to left-handed vector operators with two massive neutrinos, unambiguously discerning a deviation from the SM expectation might be challenging in this case. On the contrary, a contribution to a right-handed vector operator can induce a signal in $F_L(K^{*+})$ also for massless neutrinos, see the plot on the right in Figure~\ref{fig:FL}, which should be distinguishable from the SM case at least for a Wilson coefficient value close to $C^{\text{VLR}}_{\nu d,\alpha\alpha sb}\approx 0.02\;\text{TeV}^{-2}$.

A generic effect of the introduction of sterile-neutrino masses is a larger phase-space suppression and thus a relaxation of the implied bounds on the new-physics Wilson coefficients. Note that the masses have to be quite large to induce a noticeable deviation, for instance, a decrease of the bounds on the respective Wilson coefficients by a factor of 2 occurs only for sterile-neutrino masses of at least roughly $m_4 = 1\;(2)\;\text{GeV}$ or larger in the case of (off-)diagonal operator elements, that is, for about half of the kinematically allowed range there is only a small effect. Indeed, the structure of the respective viable regions for two non-zero operators does not substantially change either if massive neutrinos are considered. In particular, neutrino masses do not spoil the possibility of exact cancellations between contributions from vector and scalar operators of opposite chirality, respectively, to $B\to K\nu\nu$ and $B\to K^*\nu\nu$. 

Non-zero neutrino masses allow for interference between vector operators and scalar or tensor operators. Still, as the contributions are proportional to (the sum or difference of) the final-state neutrino masses, the discussion of potential interference of new physics with the SM contribution in the last section will not be noticeably impacted if the tiny masses of the active SM neutrinos were taken into account. A non-trivial consequence of two massive sterile neutrinos in the final state, though, are non-vanishing contributions from interference among scalar operators with different quark-flavour orderings, $sb$ and $bs$. This can also occur for tensor operators.\footnote{Vector operators with the quark-flavour ordering $bs$ are trivially related to those with $sb$ via Hermitian conjugation.} As the amplitudes for the decays under consideration receive contributions from the Hermitian conjugates of the $bs$ operators which amounts to 
a chirality flip in
the neutrino bilinears, interference with $sb$ operators only occurs if both neutrinos in the final state are massive.

\begin{figure}[hbt!]
    \centering
    \includegraphics[width=0.49\linewidth]{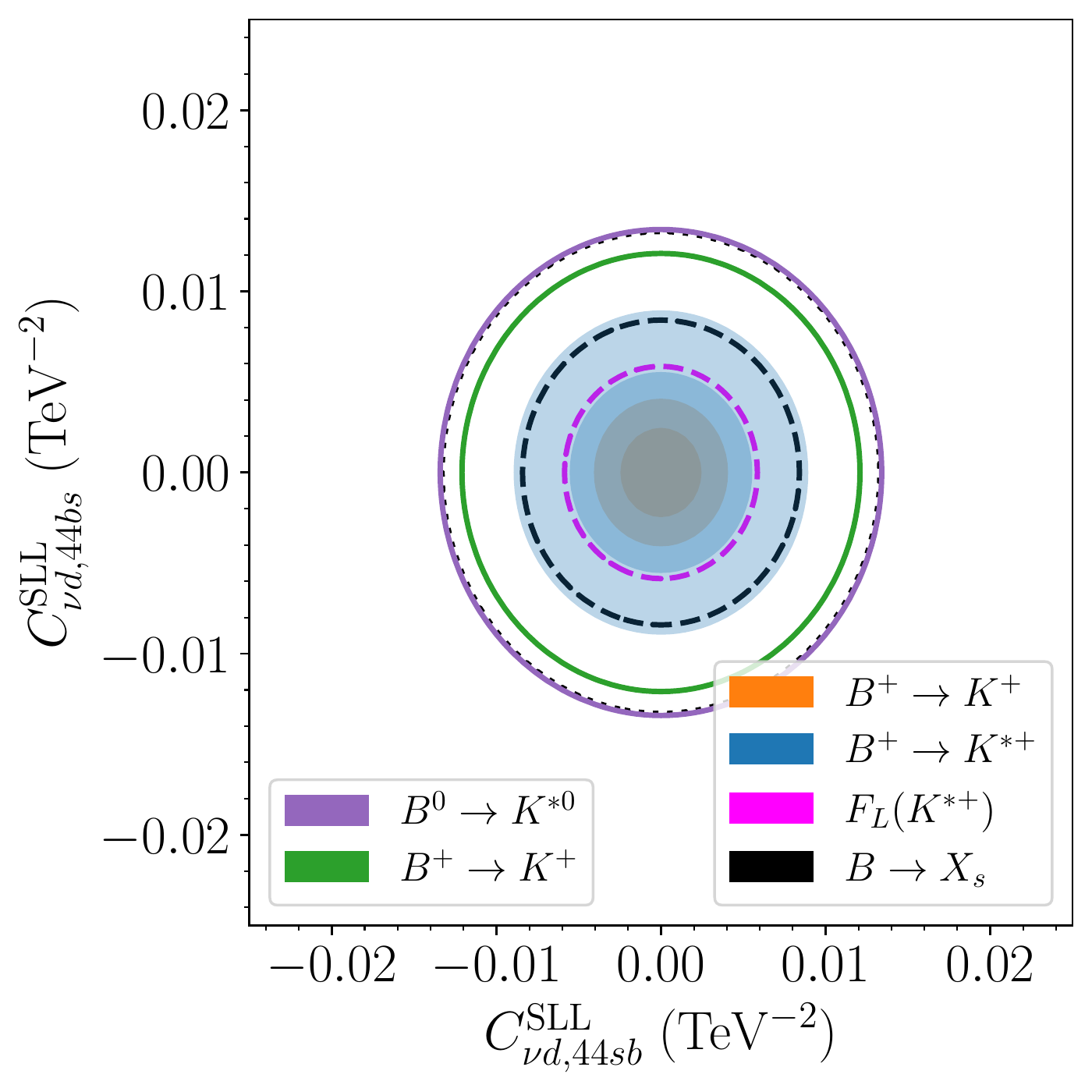}
    \includegraphics[width=0.49\linewidth]{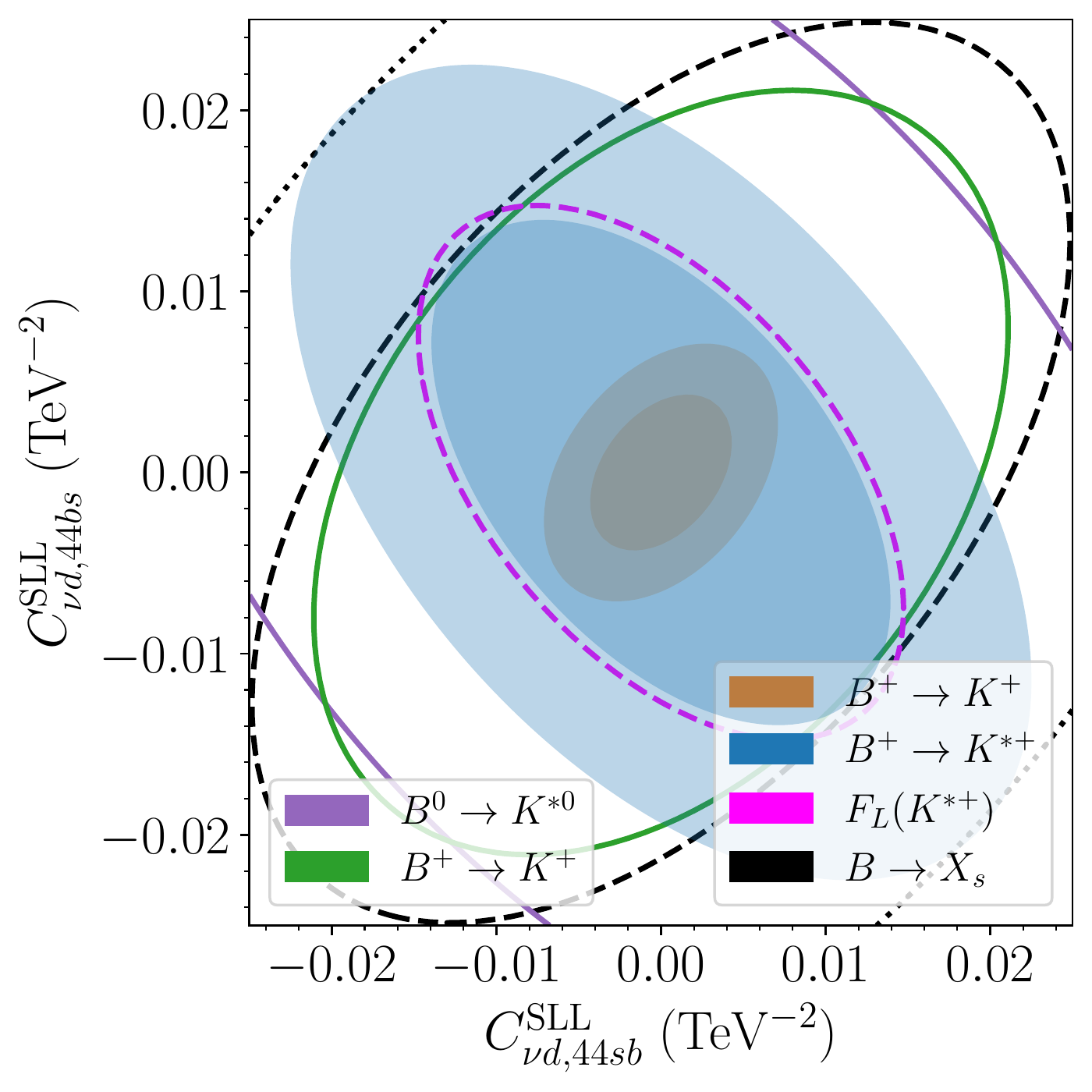}
    
    \includegraphics[width=0.49\linewidth]{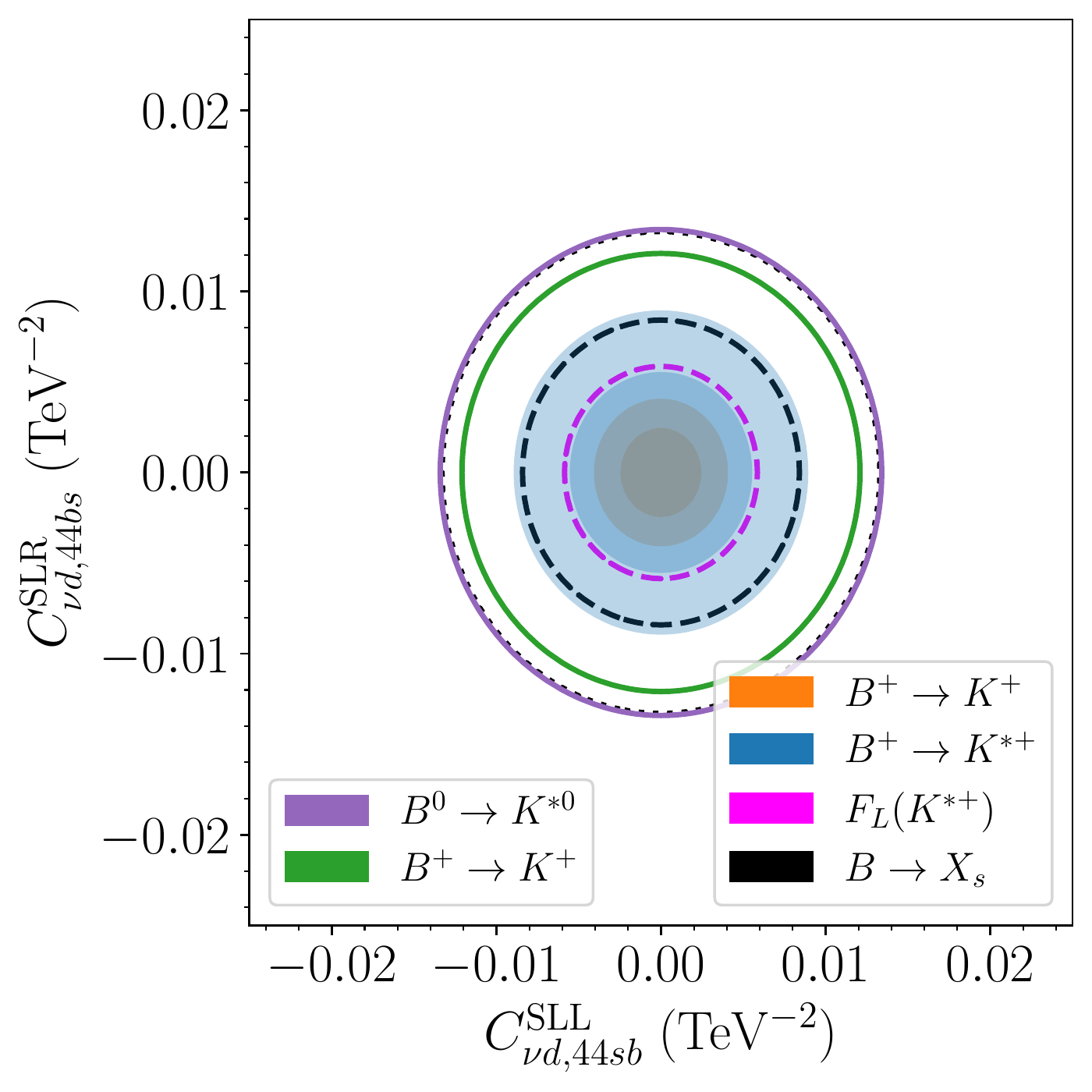}
    \includegraphics[width=0.49\linewidth]{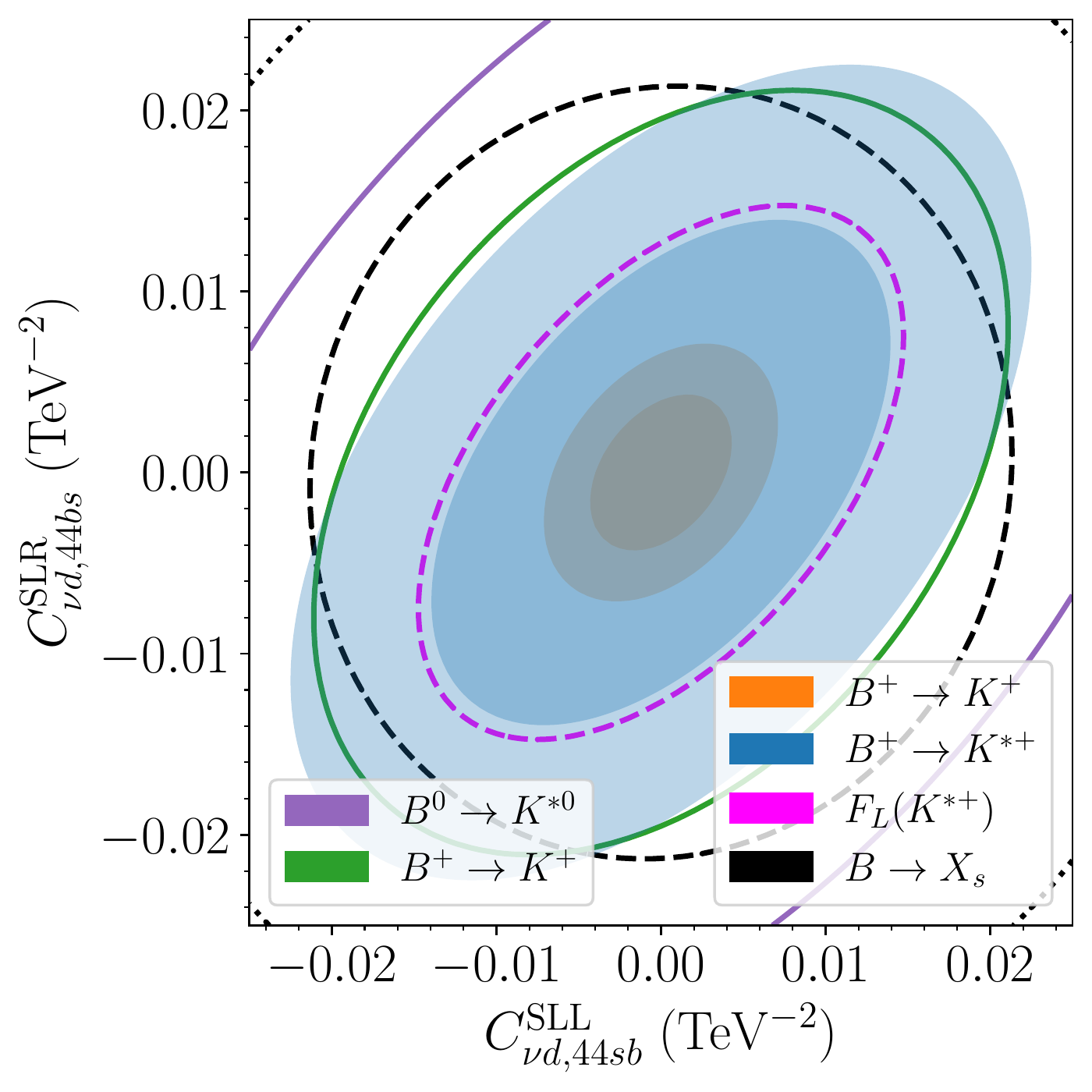}
    
    \caption{
   Future sensitivity of Belle II for $5\;\text{ab}^{-1}$ (light shaded regions) and for $50\;\text{ab}^{-1}$ (dark shaded regions, dashed lines) to scalar Wilson coefficients with $sb$ and $bs$ quark-flavour ordering and massive sterile neutrinos following the same analysis as in Figure~\ref{fig:belleII_1}. 
   The solid dark purple and green lines indicate the current experimental bound, see Table~\ref{tab:BelleII}. 
    Left: $m_{4} =0$ GeV; Right: $m_{4}= 1.5$ GeV.
    }
    \label{fig:sb_vs_bs}
\end{figure}

The plots in Figure~\ref{fig:sb_vs_bs} indicate that the interference effect could in principle be exploited to distinguish between interfering contributions from $sb$ and $bs$ quark-flavour scalar operators with  of the same quark chirality from those of opposite chirality. In particular, the orientation of the ellipses indicating the parameter space compatible with $B\to K^*\nu\nu$ and $F_L(K^{*+})$ changes. Note that especially for two operators of the same chirality, a measurement of either observable can be expected to already imply a substantial improvement over the current bound imposed by $B\to K\nu\nu$.
Still, in the considered scenario with only two contributing operators, the latter will retain the best future sensitivity to new physics. Nonetheless, $B\to K\nu\nu$ lacks the feature of distinguishing between chiralities of scalar operators. Thus it is conceivable that in scenarios with more contributions, for instance also to $C^{\text{SLR}}_{\nu d,44sb}$ with the sign opposite to that of $C^{\text{SLL}}_{\nu d,44sb}$, interference effects render $B\to K^*\nu\nu$ and/or $F_L(K^{*+})$ entirely competitive with $B\to K\nu\nu$
and the shape of the combined viable parameter space carries information about the chiralities.


\subsection{A Hint for New Physics?}
\label{sec:average}

\begin{figure}[hbpt!]
    \centering
    \includegraphics[width=0.49\linewidth]{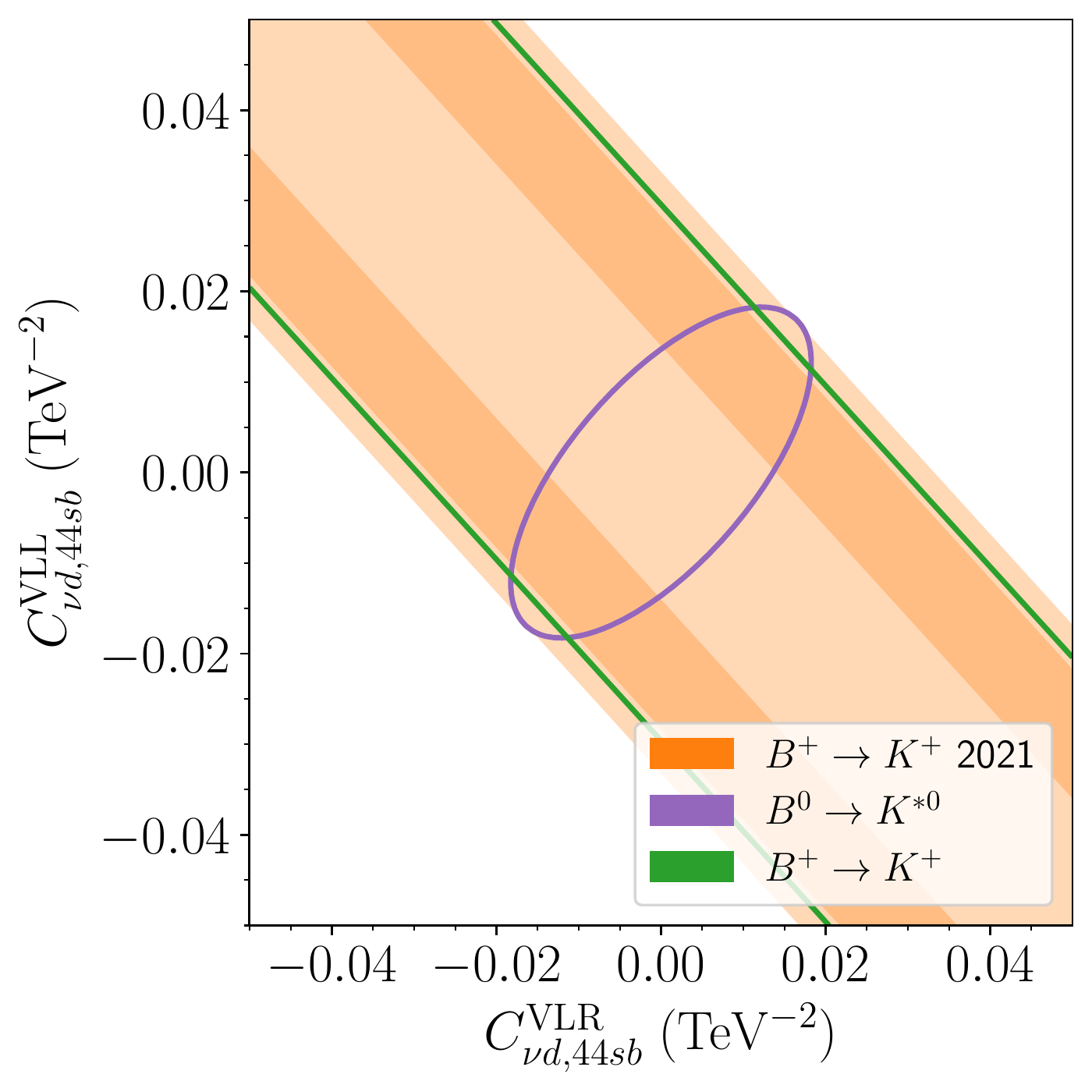}
    \includegraphics[width=0.49\linewidth]{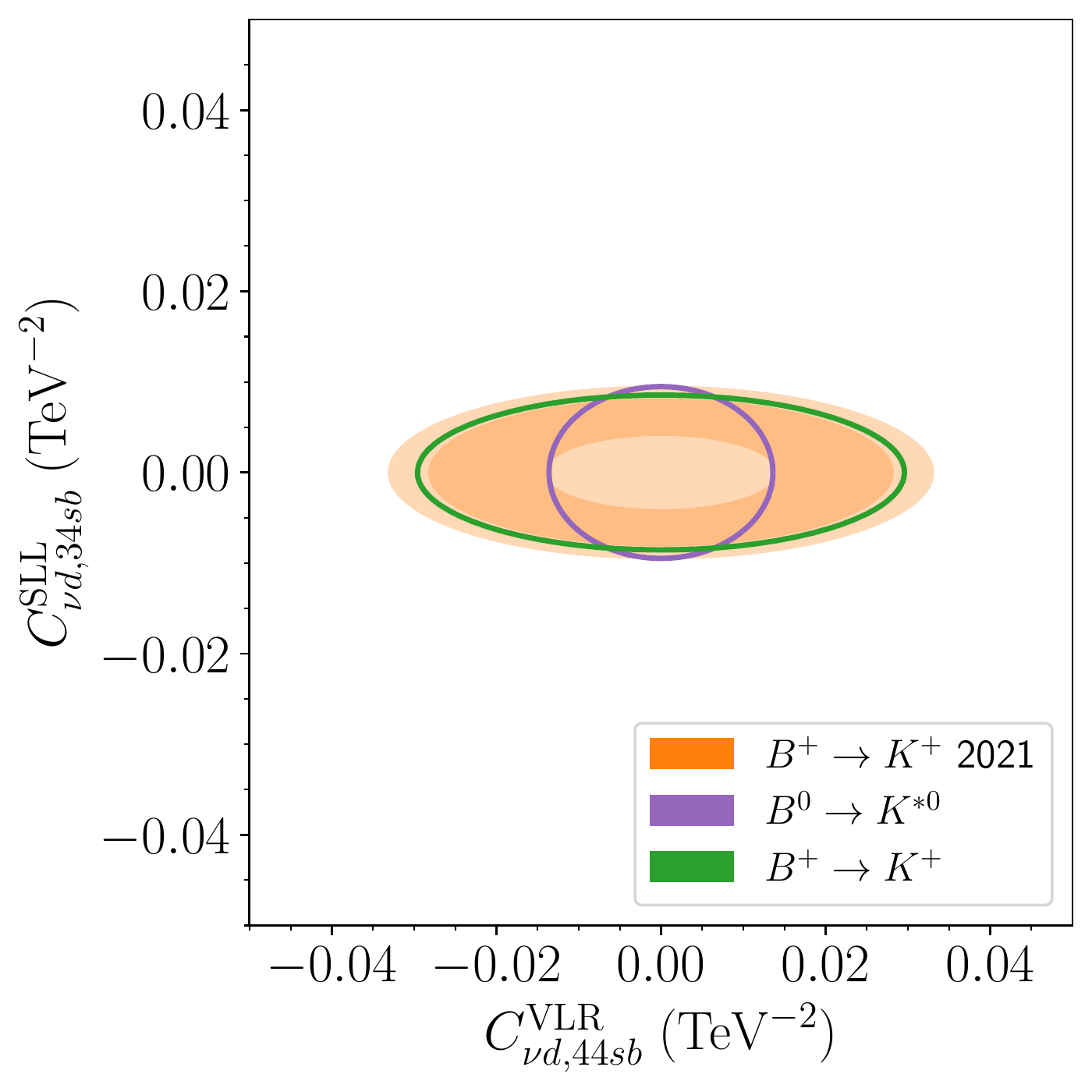}
    
    \includegraphics[width=0.49\linewidth]{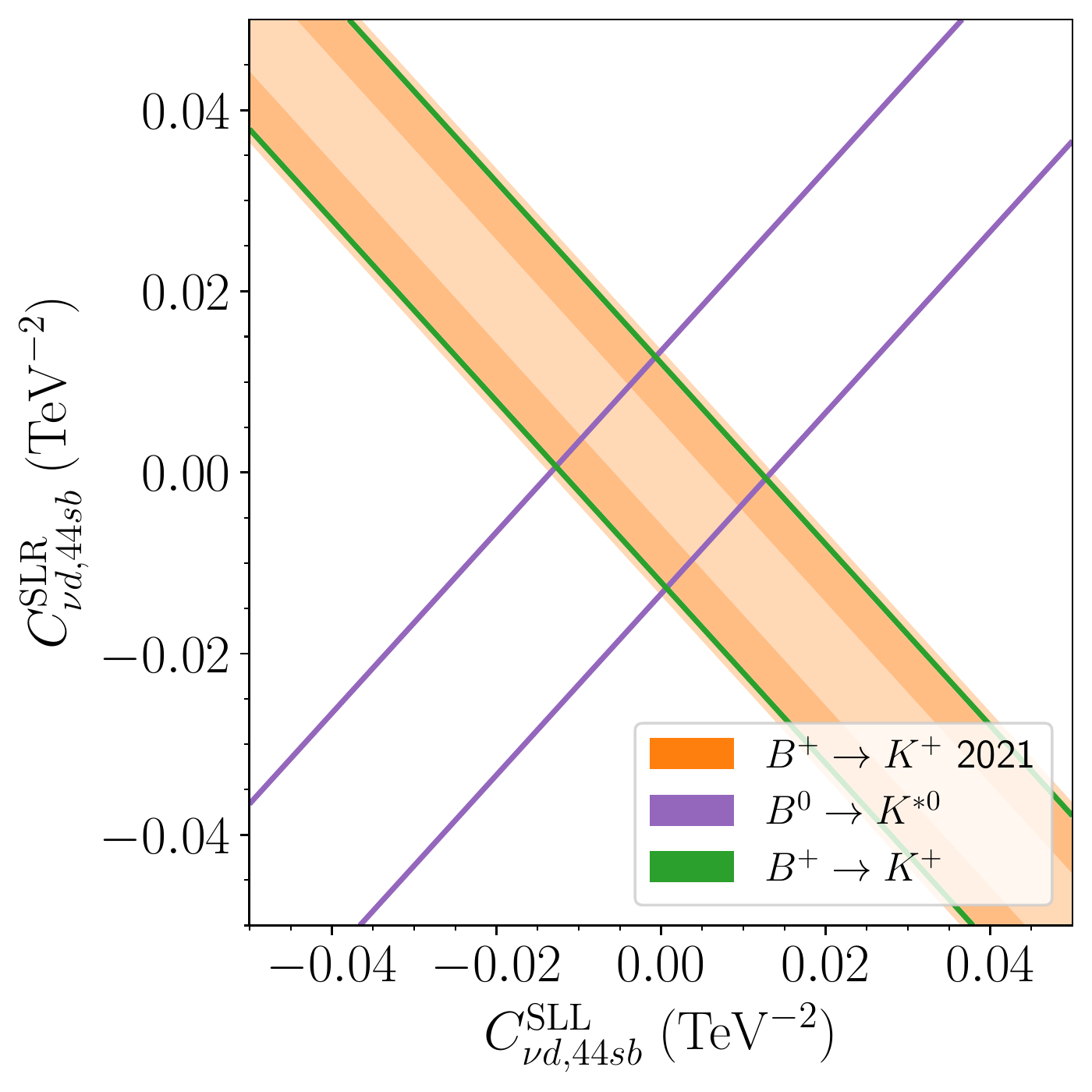}
    \includegraphics[width=0.49\linewidth]{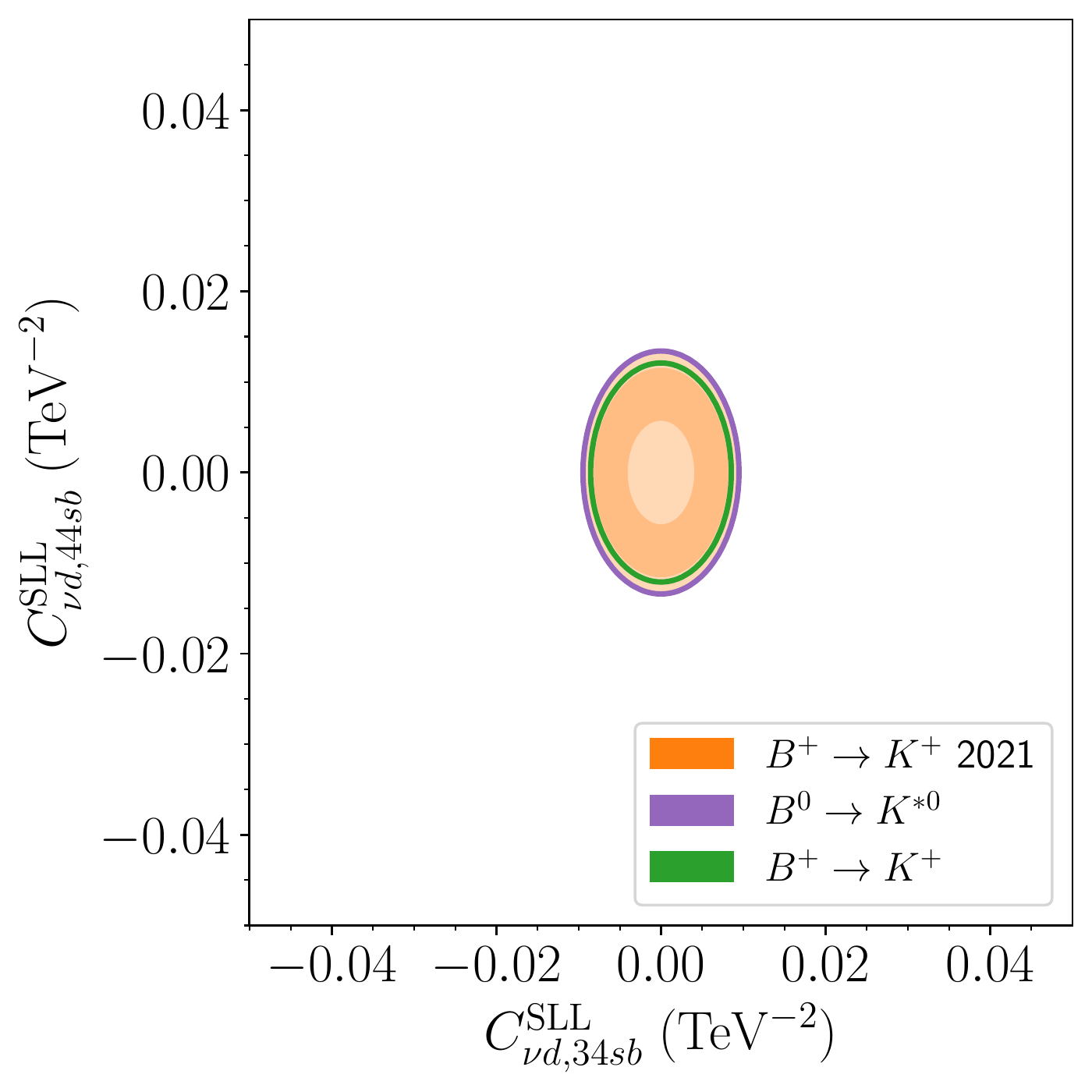}
    
    \includegraphics[width=0.49\linewidth]{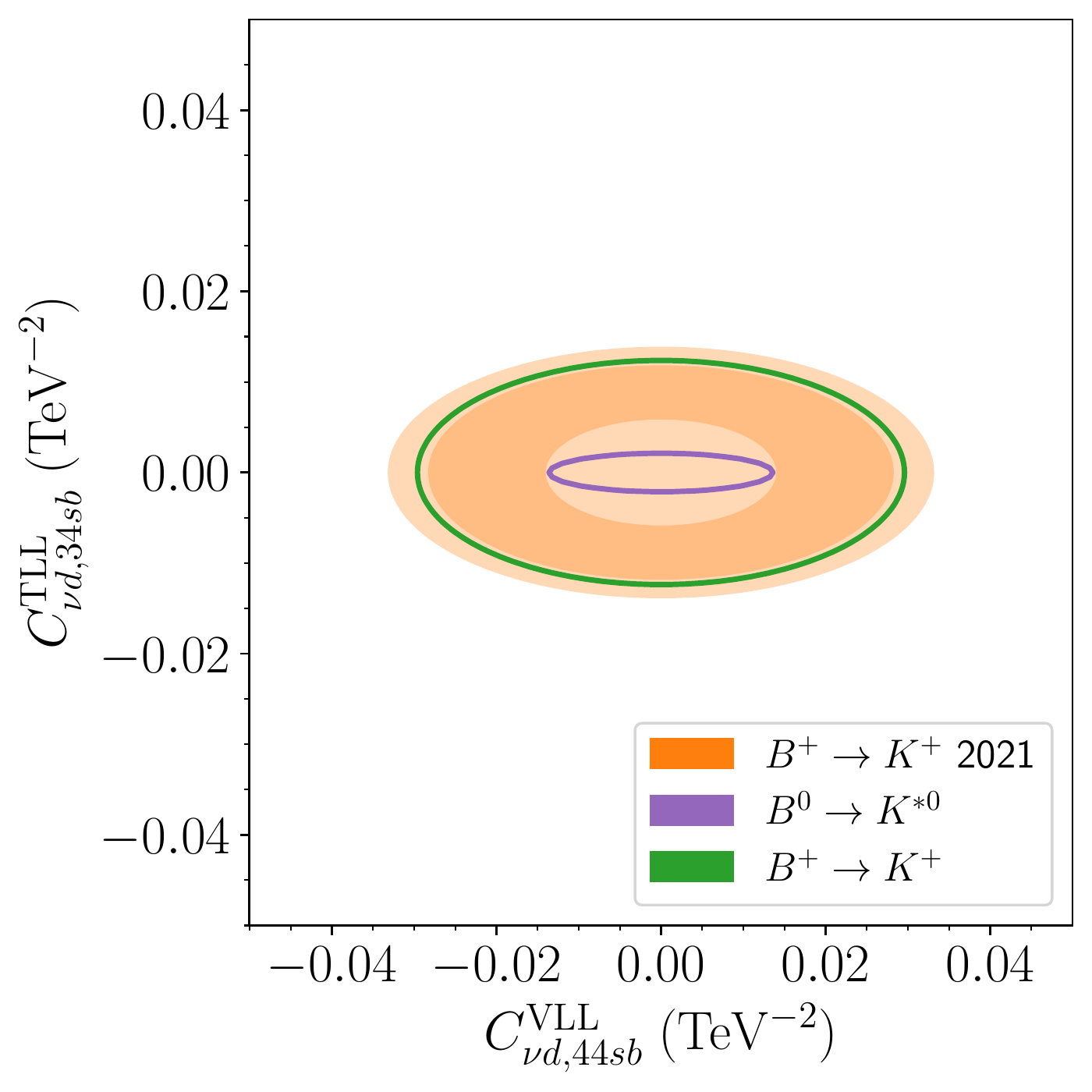}
    \includegraphics[width=0.49\linewidth]{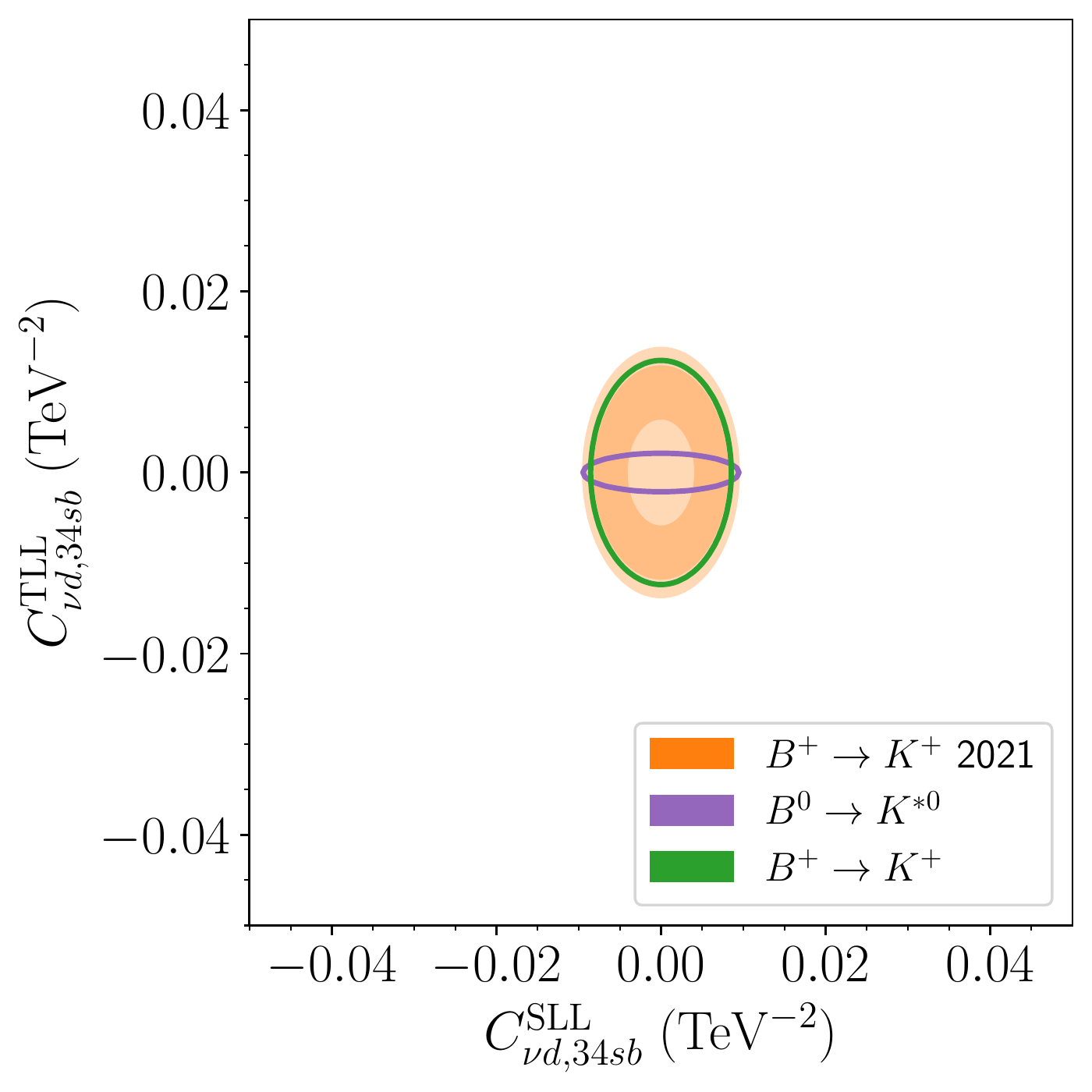}
     \caption{Parameter space which is compatible with the non-zero simple weighted average of Br($B^+ \to K^+\nu\nu$)~\cite{Dattola:2021cmw,Belle-II:2021rof} at $1\sigma$ ($2\sigma$) [darker(lighter)-orange shaded region] and the current bounds on $B^+\to K^+\nu\nu$ and $B^0\to K^{*0}\nu\nu$.
     }
    \label{fig:EXP_1}
\end{figure}

Recently, the Belle-II collaboration reported an upper limit Br($B^+\to K^+\nu\nu)< 4.1\times 10^{-5}$~\cite{Belle-II:2021rof} at the 90\% confidence level. As part of the analysis a simple weighted average of the branching ratio with previous analyses~\cite{Belle:2013tnz,BaBar:2013npw,Belle:2017oht} was presented with Br$(B^+\to K^+ \nu\nu)=(1.1\pm0.4)\times 10^{-5}$~\cite{Dattola:2021cmw,Belle-II:2021rof} which is above the SM expectation Br$(B^+\to K^+\nu\nu)=(4.4\pm0.7)\times 10^{-6}$~\cite{Straub:2018kue}. In this section, we interpret it as a hint for new physics and discuss its implication for and interplay with the existing bounds on the other decay channels. With the sets of form factors employed in this work, the SM prediction is contained in the $2\sigma$ region of the average. At $3\sigma$, the latter is still well compatible with zero.

\begin{table}[tb!]
    \centering
    \begin{tabular}{ccccccc}\toprule
        & $\mathcal{O}^{\text{VLL}}_{\nu d,44sb}$ & $\mathcal{O}^{\text{SLL}}_{\nu d,44sb}$ & $\mathcal{O}^{\text{SLL}}_{\nu d,34sb}$ & $\mathcal{O}^{\text{TLL}}_{\nu d,34sb}$ & Bound & SM\\ \midrule
        WC $(10^{-3}\;\text{TeV}^{-2})$ & $22.3^{+5.97}_{-8.31}$ & $9.12^{+2.44}_{-3.40}$ & $6.45^{+1.72}_{-2.40}$ & $9.33^{+2.50}_{-3.48}$ & & 0 \\
        Br$(B^0\to K^{*0}\nu\nu)/10^{-5}$ & $2.89\pm 1.05$ & \multicolumn{2}{c}{$1.45\pm 0.18$} & $13.5\pm 7.5$ & 1.8 & $1.16\pm 0.11$ \\ Br$(B^+\to K^{*+}\nu\nu)/10^{-5}$ & $3.11\pm 1.13$ & \multicolumn{2}{c}{$1.57\pm 0.20$} & $14.6\pm 8.1$ & 4.0 & $1.24\pm 0.12$ \\
        Br$(B\to X_s\nu\nu)/10^{-4}$ & $1.01\pm 0.37$ & \multicolumn{2}{c}{$0.494\pm 0.055$} & $4.57\pm 2.53$ & 6.4 & $0.27\pm 0.02$ \\
        \bottomrule
    \end{tabular}
    \caption{Implication of the non-zero simple weighted average of Br$(B^+\to K^+\nu\nu)$ for the contributing WCs and the other decay channels. The indicated upper and lower ranges reflect the uncertainty at $1\sigma$. Note that our new-physics predictions for the inclusive mode do not take into account QCD and HQET corrections, as indicated in Section~\ref{sec:observables}, and are thus expected to be overestimated by $\mathcal{O}(10-20\%)$. All bounds and SM predictions are the same as in Table~\ref{tab:BelleII}.}
    \label{tab:AverageBK}
\end{table}

We take the SM to be extended by one massless sterile neutrino which accounts for the non-zero branching ratio Br($B^+\to K^+\nu\nu$). We further assume one non-zero Wilson coefficient at a time, and compute the resulting branching ratios for the other decay channels. The results (at the scale $\mu = m_Z$) are summarised in Table~\ref{tab:AverageBK}. The constraints for the right-handed vector and scalar operators would be the same.

The comparatively large vector Wilson coefficient reflects that $B\to K\nu\nu$ is generally less sensitive to the vector operator than the scalar operator. Besides, one has $\mathcal{O}^{\text{TLL}}_{\nu d,34sb} = -\mathcal{O}^{\text{TLL}}_{\nu d,43sb}$ and thus the combined contribution from the components of the tensor operator is also fairly large. The non-zero branching ratio Br($B^+\to K^+\nu\nu$) also directly implies Br$(B^0\to K^{0}\nu\nu) = (1.02\pm0.37)\times 10^{-5}$ which is perfectly compatible with the current bound Br$(B^0\to K^{0}\nu\nu) < 2.6\times 10^{-5}$.

As a general result, one may assert that the non-zero weighted average can be most compellingly explained in terms of a contribution from scalar operators, since the relative uncertainties of the predicted branching ratios are at most roughly 13\% at $1\sigma$ and thus fairly small. More specifically, the prediction for the neutral (charged) mode of $B\to K^*\nu\nu$ is roughly 20\% (60\%) smaller than (and hence perfectly compatible with) the current bound, but also still agrees with the SM prediction at $2\sigma$. In particular, as the predictions are slightly larger than in the SM, this scenario will definitely be tested at Belle II. The neat agreement with the current bounds is reflected by the fact that if two operators contribute, the viable region in parameter space compatible with the average at $1\sigma$ in the plots in Figure~\ref{fig:EXP_1} is connected only in the case of two non-interfering scalar operators.

An explanation via vector operators is less preferred due to some tension with $B^0\to K^{*0}\nu\nu$
of which the $1\sigma$ region is already excluded. Still, the prediction for this channel is compatible with zero at $3\sigma$. The latter statement also holds for $B^+\to K^{*+}\nu\nu$, but its prediction respects the current bound at large parts of the $1\sigma$ range. Arguably, contributions to tensor operators are the least elegant way to account for the non-zero weighted average of Br($B^+\to K^+\nu\nu$). The implied predictions for both the neutral mode and the charged mode of $B\to K^*\nu\nu$ are already ruled out at much more than $1\sigma$. Indeed, Figure~\ref{fig:EXP_1} indicates that current bounds already constrain a possible contribution from tensor operators 
to be quite small, i.e.~$|C^{\text{TLL}}_{\nu d,34sb}| \lesssim 0.002\;\text{TeV}^{-2}$, whereas $|C^{\text{TLL}}_{\nu d,34sb}| \gtrsim 0.006\;\text{TeV}^{-2}$ would be needed to explain the non-zero average at $1\sigma$. Conversely, the uncertainties of the predictions are so large that they are compatible with zero at less than $2\sigma$.

\section{Conclusions}
\label{sec:conclusions}

In this paper, we have studied how new physics contributing to $b\to s\nu\nu$ transitions is constrained by current bounds on the branching ratios of $B\to K\nu\nu$, $B\to K^*\nu\nu$, and $B\to X_s\nu\nu$, and what improvements can be expected from the projected measurement of these processes at Belle II. We have also taken into account the longitudinal polarisation fraction $F_L(B\to K^*\nu\nu)$. Throughout the analyses, we have assumed that the Belle-II results will confirm the SM expectations. Our investigation is based on the most general set of dimension-6 operators in low-energy effective theory (LEFT)
which contribute to $b\to s\nu\nu$~\cite{Aebischer:2017gaw,Jenkins:2017jig} including massive sterile neutrinos, except for the dimension-5 dipole operator the contribution of which can be expected to be very suppressed. We employ the form factors provided in~\cite{Gubernari:2018wyi} for $B\to K\nu\nu$ and the ones from~\cite{Bharucha:2015bzk} for the observables related to $B\to K^*\nu\nu$, both of which are based on a combined fit to LQCD and LCSR data. Finite-width effects are taken into account for the $B\to K^*$ form factors via increasing them by 10\% following \cite{Descotes-Genon:2019bud}.
The implementation of the exclusive decays makes use of the general formalism developed in Ref.~\cite{Gratrex:2015hna}. We also provide the leading-order expression for the inclusive decay mode which we computed with \texttt{FeynCalc} \cite{Shtabovenko:2016sxi,Shtabovenko:2020gxv}.

We started our discussion with a consideration of the bounds in the case of new physics (dominantly) contributing to only one operator. We found that currently the vector operator is the least constrained one, whereas the most stringent bound holds for the tensor operator. The associated scale of new physics might reside at roughly 25 TeV in the latter case, which Belle II can be expected to refine to approximately 35 TeV. One should stress that the scalar and tensor operators exhibit symmetries under the exchange of the final-state neutrino flavours, and thus a contribution from a $\alpha\beta$ operator element with $\alpha\neq\beta$ always implies that the $\beta\alpha$ element also contributes with equal strength, which we do not compensate for in our basis.

The bulk of our paper is dedicated to the case of non-zero new-physics contributions to two different operators, as this allows the discussion of effects of interference between the operators, and complementarities between different observables to probe these contributions. We have also considered new-physics contributions to $\mathcal{O}^{\text{VLL}}_{\nu d,\alpha\alpha sb}$ which is the only non-vanishing operator in the SM at leading-order.
Since we assume that Belle II will not find significant deviations from the SM expectation, we exclude the possibility of 
efficient cancellations, and thus there is generally less parameter space available in this scenario. Only a simultaneous compensating contribution to $\mathcal{O}^{\text{VLR}}_{\nu d,\alpha\alpha sb}$ could potentially make the experimental results appear consistent with the SM predictions.

Our results show that the combination of the processes $B\to K\nu\nu$ and $B\to K^*\nu\nu$ is generally the most powerful probe of new physics. Partly, this is due to the fact that $B\to K\nu\nu$ depends on the sum of left- and right-handed vector operators and scalar operators, respectively, while $B\to K\nu\nu$ is dominantly sensitive to the respective difference of these operators. Thus, these processes probe largely different regions in parameter space. Moreover, the experimental uncertainties for these processes are projected to be as small as ca.~10\% with the 50 ab$^{-1}$ data set.
In the case of massive neutrinos, the bound imposed by $B\to K\nu\nu$ becomes completely superior in the case of large neutrino masses for all considered operators. Still, as indicated above, these observables individually are not safe from the possibility of cancellations among interfering contributions from  different Wilson coefficients, in which case independent information from other processes is needed. 

In particular, throughout our study a bound on the inclusive mode always translates into an unambiguous bound on each contributing operator.
$B\to X_s\nu\nu$ is a suitable probe especially in the case of interfering vector operators, but it is also useful to constrain tensor operators for which it outperforms $B\to K\nu\nu$ for sterile-neutrino masses below $\lesssim 2.6\;\text{GeV}$.
Our conservative assumptions about the uncertainties associated with $B\to X_s\nu\nu$ could be nullified with a dedicated study of next-to-leading order contributions to the decay rate. Therefore, we wish to make a case for efforts to experimentally access the inclusive decay and to reduce its theory uncertainty.

Conversely, the longitudinal polarisation fraction $F_L$ is very suitable to test the scenario of new physics yielding contributions to scalar operators. In the case of two massive neutrinos in the final state, it can even help distinguish whether the operators are of the same or opposite chirality. Here, it is perfectly competitive with the branching ratio of $B\to K^*\nu\nu$. The latter observable is the most sensitive probe for tensor operators up to a sterile-neutrino mass of $\lesssim 4\;\text{GeV}$ and also competitive with $B\to K\nu\nu$ in the case of vector operators and small neutrino masses.

In summary, we have demonstrated that the search for rare process based on $b\to s$ quark transitions with missing energy in the final state at Belle II will considerably strengthen the current bounds on new physics contributing to these processes, and that the processes under consideration exhibit different and therefore complementary sensitivity to the different operators taken into account. 
Studies of non-leading contributions to the observables as well as the interpretation of the results in terms of SMEFT and their connection to other rare processes are left for future work.


\section*{Acknowledgements}

MS would like to thank Kevin Varvell and Bruce Yabsley for useful discussions on the Belle II experiment during an early stage of the project and Roman Zwicky for useful correspondence on the helicity formalism.
TF would like to thank Diego Tonelli for a helpful clarification regarding the recent Belle-II result. 
We would like to thank Alexander Glazov for pointing out an error in an earlier draft. We acknowledge the use of \texttt{matplotlib}~\cite{Hunter:2007,thomas_a_caswell_2021_5194481}. This research includes computations using the computational cluster Katana supported by Research Technology Services at UNSW Sydney. This work was funded partially by the Australian Government through the Australian Research Council. 

\appendix

\section{Form factors}
\label{sec:FF}
We follow the parametrisation in \cite{Gubernari:2018wyi}.
For the $B \to P$ transition with $P = \pi, \ K, \ \bar{D}$, the form factors  $f_0, \ f_+$ and $f_T$ are defined as in
\begin{equation}
\begin{aligned}
\langle P(k) | \overline{d} \gamma^\mu b | B(p) \rangle & =
\left[ \left( p+k \right)^\mu - \frac{m_B^2-m_P^2}{q^2} q^\mu \right] f_{+}(q^2) + 
\frac{m_B^2-m_P^2}{q^2} q^\mu f_{0}(q^2),\\
\langle P(k) | \overline{d} \sigma^{\mu \nu} q_\nu b | B(p) \rangle & = \frac{if_T(q^2)}{m_B + m_P} \left(q^2 (p+k)^\mu - (m_B^2-m_P^2) q^\mu \right),
\end{aligned}
\end{equation}
where $q^\mu=p^\mu-k^\mu$, k and p are the 4-momenta of the $P$ pseudoscalar meson and the $B$ meson, respectively.

For the $B \to V$ decay with $V= \rho, \ K^*, \ \bar{D}^* $, the non-vanishing form factors $V, \ A_{0,1,2,3}, \ T_{1,2,3}$ are
\begin{equation}
\begin{aligned}
\langle V(k,\eta) | \overline{d} \gamma^\mu b | B(p) \rangle &= \epsilon^{\mu \nu \rho \sigma} \eta^*_\nu p_\rho k_\sigma \frac{2V}{m_B+m_V},\\
\langle V(k,\eta) | \overline{d} \gamma^\mu \gamma_5 b | B(p) \rangle &= i \eta^*_\nu \left[ g^{\mu \nu}(m_B+m_V)A_1 - \frac{(p+k)^\mu q^\nu}{m_B+m_V} A_2 - q^\mu q^\nu \frac{2m_V}{q^2} (A_3 - A_0) \right],\\
\langle V(k,\eta) | \overline{d} i\sigma^{\mu \nu} q_\nu b | B(p) \rangle &= \epsilon^{\mu \nu \rho \sigma} \eta^*_\nu p_\rho k_\sigma 2 T_1,\\
\langle V(k,\eta) | \overline{d} i \sigma^{\mu\nu} \gamma_5 b | B(p) \rangle &= i \eta^*_\nu \Bigg[ (g^{\mu \nu}(m_B^2-m_V^2)-(p+k)^\mu q^\nu) T_2\\
& \ \ \ \ \ \ \ \ + q^\nu \left( q^\mu - \frac{q^2}{m_B^2-m_V^2}(p+k)^\mu \right) T_3 \Bigg],
\end{aligned}
\end{equation}
where $\eta$ is the polarisation vector of the vector meson.
$A_3$ is a redundant quantity and can be expressed in terms of $A_1$ and $A_2$
\begin{equation}
A_3 \equiv \frac{m_B+m_V}{2 m_V}A_1 - \frac{m_B - m_V}{2 m_V} A_2.
\label{eq:A3}
\end{equation}
In practice, it is common to replace $A_2$ and $T_3$ by
\begin{equation}
\begin{aligned}
A_{12} \equiv& \frac{(m_B + m_V)^2 (m_B^2 - m_V^2 - q^2)A_1 - \lambda(q^2, m_B^2, m_V^2) A_2}{16 m_B m_V^2 (m_B + m_V)},\\
T_{23} \equiv& \frac{(m_B^2 - m_V^2) (m_B^2 + 3 m_V^2 - q^2)T_2 - \lambda(q^2, m_B^2, m_V^2) T_3}{8 m_B m_V^2 (m_B - m_V)}.
\label{eq:A12T23}
\end{aligned}
\end{equation}
Furthermore, there are also three identities for the form factors at $q^2=0$:
\begin{equation}
\begin{aligned}
f_+(q^2 = 0) =& f_0(q^2=0),\\
A_0(q^2=0) =& A_3(q^2=0),\\
T_1(q^2=0) =& T_2(q^2=0).
\label{eq:q20}
\end{aligned}
\end{equation}
Combining Eqs.~\eqref{eq:A3}, \eqref{eq:A12T23}, and \eqref{eq:q20}, 
one obtains
\begin{equation}
A_{12}(q^2=0) = \frac{m_B^2 - m_V^2}{8 m_B m_V} A_0(q^2=0).
\end{equation}



\mathversion{bold}
\section{$S,P,V,A,\mathcal{T}$ basis}
\mathversion{normal}
\label{sec:SPVAT}
The LEFT operators can be related to the basis used in \cite{Gratrex:2015hna}. In order to take into account the Majorana nature of neutrinos we include an additional factor of $1/2$ in the effective Lagrangian. This ensures that the leptonic helicity amplitudes have the same form as in the case of Dirac fermions. The effective Lagrangian is thus given by 
\begin{align}\label{eq:Lagrangian_Gratrex}
    \mathcal{L} &= \frac12 c_H \sum_i\sum_{\alpha,\beta} (C_{i,\alpha\beta} O_{i,\alpha\beta} 
    +C_{i,\alpha\beta}^\prime O_{i,\alpha\beta}^\prime)
    \;.
\end{align}
where $i$ runs over $S,P,V,A,\mathcal{T}$ and $c_H$ determines the normalisation of the operators. In this work we choose $c_H=1$.\footnote{Reference \cite{Gratrex:2015hna} uses $c_H=\frac{4G_F}{\sqrt{2}} \frac{\alpha}{4\pi} V_{ts}^* V_{tb}$. Note that we rewrote the effective Hamiltonian in terms of an effective Lagrangian and replaced $\ell$ in \cite{Gratrex:2015hna} by the neutrino fields $\nu$.} The operators are given by
\begin{equation}
    \begin{aligned}
    O_{S(P)\alpha\beta} & = (\overline{s_L} b)  (\overline{\nu_\alpha}(\gamma_5)\nu_\beta)\;,
    &
    O_{V(A)\alpha\beta} & = (\overline{s_L} \gamma^\mu b) (\overline{\nu_\alpha}\gamma_\mu (\gamma_5)\nu_\beta)\;,
    &
    O_{\mathcal{T}\alpha\beta} & = (\overline{s_L} \sigma^{\mu\nu} b) (\overline{\nu_\alpha} \sigma_{\mu\nu}\nu_\beta)
    \;.
    \end{aligned}
\end{equation}
The primed operators are obtained by replacing $s_L\to s_R$, i.e. $O^\prime=O|_{s_L\to s_R}$ where $q_{L,R}\equiv P_{L,R}q$. The notation $O_{9(10)} \equiv O_{V(A)}$ is also commonly found in the literature.
The operators have well-defined symmetry properties: the pseudo(scalar) operators are symmetric in the neutrino flavour indices and the (axial)vector and tensor operators are antisymmetric.
We find for the Wilson coefficients using the $S,P,V,A,\mathcal{T}$ basis 
\begin{equation}
\begin{aligned}
    C_{V\alpha\beta} & = C_{\nu d,[\alpha\beta] sb}^{\text{VLL}} \;,
    &
    C_{A\alpha\beta} & = - C_{\nu d,(\alpha\beta) sb}^{\text{VLL}} \;,
    \\
    C_{V\alpha\beta}^\prime & = C_{\nu d,[\alpha\beta] sb}^{VLR} \;,
    &
    C_{A\alpha\beta}^\prime & = - C_{\nu d,(\alpha\beta) sb}^{VLR}  \;,
    \\
 C_{S\alpha\beta} & =  C_{\nu d,(\alpha\beta) sb}^{\text{SLR}} +  C_{\nu d,(\beta\alpha) bs}^{\text{SLL}*} \;,
 &
 C_{P\alpha\beta} & = - C_{\nu d,(\alpha\beta) sb}^{\text{SLR}} +  C_{\nu d,(\beta\alpha) bs}^{\text{SLL}*}\;,
 \\
 C_{S\alpha\beta}^\prime & =  C_{\nu d,(\alpha\beta) sb}^{\text{SLL}} + C_{\nu d,(\beta\alpha) bs}^{\text{SLR}*}\;,
 &
 C_{P\alpha\beta}^\prime & = - C_{\nu d,(\alpha\beta) sb}^{\text{SLL}} + C_{\nu d,(\beta\alpha) bs}^{\text{SLR}*}\;,
 \\
 C_{\mathcal{T}\alpha\beta} & =  2C_{\nu d,[\beta\alpha] bs}^{\text{TLL}*} \;,
 &
 C_{\mathcal{T}\alpha\beta}^\prime & = 2 C_{\nu d, [\alpha\beta] sb}^{\text{TLL}} \;,
\end{aligned}
\end{equation}
where $\alpha,\beta$ denote the neutrino flavours.
Parentheses (\dots) indicate symmetrisation and square brackets [\dots] indicate anti-symmetrisation of the neutrino flavour indices as in
\begin{align}
    M_{(ab)} &\equiv \frac12 \left(M_{ab}+M_{ba}\right) \;,
    &
    M_{[ab]} &\equiv \frac12 \left(M_{ab}-M_{ba}\right)  
    \;.
\end{align}

\mathversion{bold}
\section{$B\to K\nu_\alpha\nu_\beta$}
\mathversion{normal}
\label{sec:BKnunu}
For the convenience of the reader, we provide the expression for the coefficient $G^{(0)}$ of the Wigner-$D$ function $D^0_{0,0}(\Omega)=1$ for $\bar B\to \bar K \nu_\alpha\nu_\beta$ following \cite{Gratrex:2015hna}.
 Although the vector and tensor operator are antisymmetric in the neutrino flavour indices $\alpha$, $\beta$, 
all combinations which enter the helicity amplitudes are symmetric under exchanging them.
The CP-conjugate process $B\to K\nu_\alpha\nu_\beta$ is obtained via replacing $G^{(0)}$ by $\bar G^{(0)}$ where
the Wilson coefficients in the helicity amplitudes are replaced by their complex conjugates. Note that this complex conjugation does not introduce additional minus signs into the coefficients of the Wigner-$D$ functions for terms with antisymmetric Wilson coefficients.
The CP conjugation also implies a redefinition of the angles, in particular $\theta_K\to \pi-\theta_K$, under which the relevant Wigner-$D$ functions $D^0_{0,0}$ and $D^2_{0,0}$ are invariant though. According to \cite{Gratrex:2015hna},
\begin{equation}
\begin{aligned}
N^{-1} G^{(0)}(q^2) &= \left( 4 \left( E_\alpha E_\beta + m_\alpha m_\beta  \right) + \frac{\lambda_{\gamma^*}}{3 q^2} \right) |h^V_{\alpha\beta}|^2 + \left( 4 \left( E_\alpha E_\beta - m_\alpha m_\beta  \right) + \frac{\lambda_{\gamma^*}}{3 q^2} \right) |h^A_{\alpha\beta}|^2 \\
&+ \left( 4 \left( E_\alpha E_\beta - m_\alpha m_\beta  \right) + \frac{\lambda_{\gamma^*}}{q^2} \right) | h^S_{\alpha\beta} |^2 + \left( 4 \left( E_\alpha E_\beta + m_\alpha m_\beta  \right) + \frac{\lambda_{\gamma^*}}{q^2} \right) | h^P_{\alpha\beta} |^2 \\
&+ 16 \left( E_\alpha E_\beta + m_\alpha m_\beta  - \frac{\lambda_{\gamma^*}}{12 q^2} \right) | h^{T_t}_{\alpha\beta} |^2 + 
8 \left( E_\alpha E_\beta - m_\alpha m_\beta  - \frac{\lambda_{\gamma^*}}{12 q^2} \right) | h_{\alpha\beta}^{T} |^2 \\
&+ 16 \left( m_\alpha E_\beta + m_\beta E_\alpha \right) \,\mathrm{Im}\left[ h^V_{\alpha\beta} h^{T_t*}_{\alpha\beta}\right] + 8 \sqrt{2} \left( m_\alpha E_\beta - m_\beta E_\alpha \right) \,\mathrm{Im} \left[h^A_{\alpha\beta} h^{T*}_{\alpha\beta} \right],
\end{aligned}
\end{equation}
where the normalisation factor $N$, the energies $E_{\alpha,\beta}$ and the kinematic functions $\lambda_{BK,\gamma^*}$ are defined as in
\begin{equation}
\begin{aligned}\label{eq:HelAmplFunctions}
N &=  \frac{\sqrt{\lambda_{BK} \lambda_{\gamma^*}}}{(4\pi)^3 m_B^3 q^2 (1+\delta_{\alpha\beta})}
\;,
&
E_{\alpha,\beta}&=\sqrt{m_{\alpha,\beta}^2+ \frac{\lambda_{\gamma^*}}{4q^2}}
\;,
&
\lambda_{BK}&\equiv \lambda(m_B^2,m_{K}^2,q^2)\;,
&
\lambda_{\gamma^*} &\equiv \lambda(q^2,m_{\alpha}^2,m_{\beta}^2)
\end{aligned}
\end{equation}
and $\lambda(x,y,z) \equiv x^2+y^2+x^2-2xy-2xz-2yz$ denotes the K\"all\'en function.
The symmetry factor for identical neutrinos in the final state is contained in $N$. 

The helicity amplitudes are given by
\begin{align}
    h^V_{\alpha\beta} & 
    = \frac{\sqrt{\lambda_{BK}}}{2\sqrt{q^2}} \left(C_{V\alpha\beta} + C_{V\alpha\beta}^\prime\right) f_+ \;,
     \\
    h^A _{\alpha\beta} & 
    = \frac{\sqrt{\lambda_{BK}}}{2\sqrt{q^2}} \left(C_{A\alpha\beta} + C_{A\alpha\beta}^\prime\right) f_+ \;,
    \\
    h^S_{\alpha\beta}  & 
    = \frac{m_B^2-m_K^2}{2} \left(\frac{C_{S\alpha\beta}+C_{S\alpha\beta}^\prime}{m_b-m_s} + \frac{m_{\alpha}-m_{\beta}}{q^2} (C_{V\alpha\beta}+C_{V\alpha\beta}^\prime)\right)f_0 \;,
    \\
    h^P_{\alpha\beta}  & = \frac{m_B^2-m_K^2}{2} \left(\frac{C_{P\alpha\beta}+C_{P\alpha\beta}^\prime}{m_b-m_s} + \frac{m_{\alpha}+m_{\beta}}{q^2} (C_{A\alpha\beta}+C_{A\alpha\beta}^\prime)\right)f_0 \;,
    \\
    h^T_{\alpha\beta}  & = -i \frac{\sqrt{\lambda_{BK}}}{\sqrt{2}(m_B+m_K)} \left(C_{\mathcal{T}\alpha\beta}-C_{\mathcal{T}\alpha\beta}^\prime\right) f_T \;,
    \\
    h^{T_t}_{\alpha\beta}  & = -i \frac{\sqrt{\lambda_{BK}}}{2(m_B+m_K)} \left(C_{\mathcal{T}\alpha\beta}+C_{\mathcal{T}\alpha\beta}^\prime\right) f_T
\end{align}
in terms of the $S,P,V,A,\mathcal{T}$ basis. We provide the matching to the chiral LEFT basis in App.~\ref{sec:SPVAT}.
For massless neutrinos, the expression reduces to
\begin{equation}
  G^{(0)}(q^2) =  \frac{\sqrt{\lambda_{BK}}q^2}{(4\pi)^3 m_B^3 (1+\delta_{\alpha\beta})} \left(\frac43 |h_{\alpha\beta}^V|^2 + \frac43 |h_{\alpha\beta}^A|^2 + 2 |h_{\alpha\beta}^S|^2 + 2 |h_{\alpha\beta}^P|^2 + \frac83 |h_{\alpha\beta}^{T_t}|^2 + \frac43 |h_{\alpha\beta}^T|^2 \right)\;.
\end{equation}

\mathversion{bold}
\section{$B\to  K^*\nu_\alpha\nu_\beta$}
\mathversion{normal}
\label{sec:BKstarnunu}
As the final-state neutrinos are not observed, we integrate over the neutrino solid angle. Thus there are only two relevant contributions which are described in terms of the coefficients of the Wigner-$D$ functions $D^{0}_{0,0}(\Omega_K)$ and $D^2_{0,0}(\Omega_K)$ which depend on the solid angle $\Omega_K$ of the final-state $K$ meson in the $K^*$ rest frame. They are denoted by $G^{0,0}_0$ and $G^{2,0}_0$ for $\bar B\to \bar K^* \nu_\alpha\nu_\beta$ following \cite{Gratrex:2015hna}. The corresponding coefficients for the CP conjugate process $B\to K^* \nu_\alpha\nu_\beta$ are denoted by $\bar G^{0,0}_0$ and $\bar G^{2,0}_0$ and obtained from $G^{0,0}_0$ and $G^{2,0}_0$ by replacing all Wilson coefficients with their complex conjugates. The coefficient for the Wigner-$D$ function $D^0_{0,0}$ for $\bar B\to \bar K^*\nu_\alpha\nu_\beta$ is~\cite{Gratrex:2015hna}.
\begin{equation}
\begin{aligned}\label{eq:G000}
N^{-1} G^{0,0}_0 &=\frac49\left(3E_\alpha E_\beta+\frac{\lambda_{\gamma^*}}{4q^2}\right) \sum_{a=0,\pm} ( |H_{a\alpha\beta}^V|^2+ |H_{a\alpha\beta}^A|^2)
+\frac{4m_\alpha m_\beta}{3} \sum_{a=0,\pm} (|H_{a\alpha\beta}^V|^2-|H_{a\alpha\beta}^A|^2)
\\
&+\frac43\left(E_\alpha E_\beta-m_\alpha m_\beta +\frac{\lambda_{\gamma^*}}{4q^2}\right) |H^S_{\alpha\beta}|^2
+\frac43\left(E_\alpha E_\beta+m_\alpha m_\beta +\frac{\lambda_{\gamma^*}}{4q^2}\right) |H^P_{\alpha\beta}|^2
\\
&+\frac{16}{9} \left(3(E_\alpha E_\beta +m_\alpha m_\beta ) -\frac{\lambda_{\gamma^*}}{4q^2}\right) \sum_{a=0,\pm}|H_{a\alpha\beta}^{T_t}|^2
\\
&+\frac{8}{9} \left(3(E_\alpha E_\beta -m_\alpha m_\beta ) -\frac{\lambda_{\gamma^*}}{4q^2}\right) \sum_{a=0,\pm}|H_{a\alpha\beta}^{T}|^2
\\
&+\frac{16}{3}(m_\alpha E_\beta +m_\beta E_\alpha)\, \mathrm{Im}\left[ \sum_{a=0,\pm} H_{a\alpha\beta}^V H_{a\alpha\beta}^{T_t*} \right]
\\
&+\frac{8\sqrt{2}}{3}(m_\alpha E_\beta -m_\beta E_\alpha) \,\mathrm{Im}\left[ \sum_{a=0,\pm} H_{a\alpha\beta}^V  H_{a\alpha\beta}^{T*} \right]
\end{aligned}
\end{equation}
and the coefficient for the Wigner-$D$ function $D^{2}_{0,0}$ is
\begin{equation}
\begin{aligned}\label{eq:G200}
N^{-1} G^{2,0}_0 & = 
-\frac49\left(3E_\alpha E_\beta + \frac{\lambda_{\gamma^*}}{4q^2}\right) \sum_{b=V,A}\left(|H_{+\alpha\beta}^b|^2+|H_{-\alpha\beta}^b|^2-2|H_{0\alpha\beta}^b|^2\right)
\\
&-\frac{4m_\alpha m_\beta}{3} \left(|H_{+\alpha\beta}^V|^2+|H_{-\alpha\beta}^V|^2-2|H_{0\alpha\beta}^V|^2 - (V\to A)\right)
\\
&+\frac83\left(E_\alpha E_\beta-m_\alpha m_\beta+\frac{\lambda_{\gamma^*}}{4q^2}\right)|H^S_{\alpha\beta}|^2
+\frac83\left(E_\alpha E_\beta+m_\alpha m_\beta+\frac{\lambda_{\gamma^*}}{4q^2}\right)|H^P_{\alpha\beta}|^2
\\
&-\frac{16}{9}\left(3(E_\alpha E_\beta+m_\alpha m_\beta) -\frac{\lambda_{\gamma^*}}{4q^2}\right) \left(|H_{+\alpha\beta}^{T_t}|^2+|H_{-\alpha\beta}^{T_t}|^2-2|H_{0\alpha\beta}^{T_t}|^2\right)
\\
&-\frac{8}{9}\left(3(E_\alpha E_\beta -m_\alpha m_\beta) -\frac{\lambda_{\gamma^*}}{4q^2}\right) \left(|H_{+\alpha\beta}^{T}|^2+|H_{-\alpha\beta}^{T}|^2-2|H_{0\alpha\beta}^{T}|^2\right)
\\
&-\frac{16}{3}\left(m_\alpha E_\beta + m_\beta E_\alpha\right)\, \mathrm{Im}\left[ H_{+\alpha\beta}^V H_{+\alpha\beta}^{T_t*} + H_{-\alpha\beta}^V H_{-\alpha\beta}^{T_t*} -2 H_{0\alpha\beta}^V H_{0\alpha\beta}^{T_t*}\right]
\\
&-\frac{8\sqrt{2}}{3}\left(m_\alpha E_\beta - m_\beta E_\alpha\right)\, \mathrm{Im}\left[ H_{+\alpha\beta}^V H_{+\alpha\beta}^{T*} + H_{-\alpha\beta}^V H_{-\alpha\beta}^{T*} -2 H_{0\alpha\beta}^V H_{0\alpha\beta}^{T*}\right] \;,
\end{aligned}
\end{equation}
where the normalisation factor $N$, the energies $E_{\alpha,\beta}$ and the kinematic functions $\lambda_{BK^*,\gamma^*}$ are the same as in Eq.~(\ref{eq:HelAmplFunctions}) with the kinematic function $\lambda_{BK}$ replaced by $\lambda_{BK^*} \equiv \lambda(m_B^2,m_{K^*}^2,q^2)$.
The helicity amplitudes for $\bar B \to \bar K^* \nu_\alpha\nu_\beta$ are given by
\begin{eqnarray}
\begin{aligned}\label{eq:hel_ampl_BtoKs}
H^{V}_{0\alpha\beta} & = \frac{4i m_B m_{K^*}}{\sqrt{q^2}} \left(C_{V\alpha\beta}-C_{V\alpha\beta}^\prime\right) A_{12} \;,
\\
H^A_{0\alpha\beta} & = \frac{4i m_B m_{K^*}}{\sqrt{q^2}} \left(C_{A\alpha\beta}-C_{A\alpha\beta}^\prime\right) A_{12} \;,
\\
H^V_{\pm\alpha\beta} & = \frac{i}{2(m_B +m_{K^*})} \left[ \pm\left(C_{V\alpha\beta} + C_{V\alpha\beta}^\prime\right) \sqrt{\lambda_{BK^*}} V -(m_B+m_{K^*})^2 \left(C_{V\alpha\beta} - C_{V\alpha\beta}^\prime\right) A_1 \right] \;,
\\
H^A_{\pm\alpha\beta} & = \frac{i}{2(m_B +m_{K^*})} \left[ \pm\left(C_{A\alpha\beta} + C_{A\alpha\beta}^\prime\right) \sqrt{\lambda_{BK^*}} V -(m_B+m_{K^*})^2 \left(C_{A\alpha\beta} - C_{A\alpha\beta}^\prime\right) A_1 \right] \;,
\\
H^P_{\alpha\beta} & = \frac{i\sqrt{\lambda_{BK^*}}}{2} \left[ \frac{C_{P\alpha\beta} - C_{P\alpha\beta}^\prime}{m_b+m_s} + \frac{m_\alpha+m_\beta}{q^2} \left(C_{A\alpha\beta}-C_{A\alpha\beta}^\prime\right)\right]A_0 \;,
\\
H^S_{\alpha\beta} & = \frac{i\sqrt{\lambda_{BK^*}}}{2} \left[ \frac{C_{S\alpha\beta} - C_{S\alpha\beta}^\prime}{m_b+m_s} + \frac{m_\alpha-m_\beta}{q^2} \left(C_{V\alpha\beta}-C_{V\alpha\beta}^\prime\right)\right]A_0 \;,
\\
H^T_{0\alpha\beta} & = \frac{2\sqrt{2} m_B m_{K^*}}{m_B+m_{K^*}}\left(C_{\mathcal{T}\alpha\beta} + C_{\mathcal{T}\alpha\beta}^\prime\right) T_{23} \;,
\\
H^{T_t}_{0\alpha\beta} & = \frac{2 m_B m_{K^*}}{m_B+m_{K^*}}\left(C_{\mathcal{T}\alpha\beta} -C_{\mathcal{T}\alpha\beta}^\prime\right) T_{23} \;,
\\
H^{T}_{\pm\alpha\beta} & = \frac{1}{\sqrt{2q^2}}\left[\pm\left(C_{\mathcal{T}\alpha\beta} -C_{\mathcal{T}\alpha\beta}^\prime\right) \sqrt{\lambda_{BK^*}} T_1 - \left(C_{\mathcal{T}\alpha\beta} + C_{\mathcal{T}\alpha\beta}^\prime\right)(m_B^2-m_{K^*}^2) T_2 \right] \;,
\\
H^{T_t}_{\pm\alpha\beta} & = \frac{1}{2\sqrt{q^2}}\left[\pm\left(C_{\mathcal{T}\alpha\beta}+ C_{\mathcal{T}\alpha\beta}^\prime\right) \sqrt{\lambda_{BK^*}} T_1 - \left(C_{\mathcal{T}\alpha\beta} - C_{\mathcal{T}\alpha\beta}^\prime\right)(m_B^2-m_{K^*}^2) T_2 \right]
\end{aligned}
\end{eqnarray}
in terms of the $S,P,V,A,\mathcal{T}$ basis. The matching to the chiral LEFT basis is given in App.~\ref{sec:SPVAT}. For massless neutrinos, the normalisation factor reduces to 
\begin{equation}
    N=\frac{\sqrt{\lambda_{BK^*}}}{(4\pi)^3 m_B^3 (1+\delta_{\alpha\beta})}
\end{equation}
and the coefficients of the Wigner-$D$ functions become
\begin{align}
 G^{0,0}_0 = Nq^2 \Bigg[&
\frac49 \sum_{a=0,\pm} ( |H_{a\alpha\beta}^V|^2+ |H_{a\alpha\beta}^A|^2)
+\frac23 |H^S_{\alpha\beta}|^2
+\frac23 |H^P_{\alpha\beta}|^2
+\frac{8}{9} \sum_{a=0,\pm}|H_{a\alpha\beta}^{T_t}|^2
+\frac{4}{9} \sum_{a=0,\pm}|H_{a\alpha\beta}^{T}|^2
\Bigg]\;,
\nonumber\\
 G^{2,0}_0  = Nq^2 \Bigg[&   
-\frac49 \sum_{b=V,A}\left(|H_{+\alpha\beta}^b|^2+|H_{-\alpha\beta}^b|^2-2|H_{0\alpha\beta}^b|^2\right)
+\frac43|H^S_{\alpha\beta}|^2
+\frac43|H^P_{\alpha\beta}|^2
\\\nonumber
&-\frac{8}{9}
\left(|H_{+\alpha\beta}^{T_t}|^2+|H_{-\alpha\beta}^{T_t}|^2-2|H_{0\alpha\beta}^{T_t}|^2\right)
-\frac{4}{9}\left(|H_{+\alpha\beta}^{T}|^2+|H_{-\alpha\beta}^{T}|^2-2|H_{0\alpha\beta}^{T}|^2\right)
\Bigg]\;.
\end{align}
For massless neutrinos, the binned longitudinal polarisation fraction $F_L$ 
for the decay $B\to K^*\nu_\alpha\nu_\beta$ 
can be compactly written in terms of helicity amplitudes $\bar H$ as
\begin{equation}\label{eq:F_L}
    \begin{aligned}
    F_L & 
=\frac{\braket{\bar G^{0,0}_0+\bar G^{2,0}_0}}{3 \braket{\bar G^{0,0}_0}}\\
    &= \frac49 \frac{\braket{Nq^2 \left(|\bar H^V_{0\alpha\beta}|^2 + |\bar H^A_{0\alpha\beta}|^2 + \frac32 |\bar H^S_{\alpha\beta}|^2 + \frac32 |\bar H^P_{\alpha\beta}|^2  +2 |\bar H^{T_t}_{0\alpha\beta}|^2 + |\bar H^T_{0\alpha\beta}|^2\right)}}{\braket{\bar G^{0,0}_0}}\;.
    \end{aligned}
\end{equation}
The helicity amplitudes $\bar H$ are obtained from the corresponding helicity amplitude $H$ by replacing all Wilson coefficients by the complex conjugates.
The corresponding binned transverse polarisation fraction $F_T$ is given by
\begin{equation}\label{eq:F_T}
    \begin{aligned}
 F_T=1-F_L=\frac49\frac{\braket{ Nq^2  \sum_{a=\pm} \left(|\bar H_{a\alpha\beta}^V|^2+  |\bar H_{a\alpha\beta}^A|^2
+2 |\bar H_{a\alpha\beta}^{T_t}|^2
+ |\bar H_{a\alpha\beta}^{T}|^2
\right)}}{  \braket{\bar G^{0,0}_0}}
\;.
    \end{aligned}
\end{equation}


\mathversion{bold}
\section{$B\to X_s\nu_\alpha\nu_\beta$}
\mathversion{normal}
\label{sec:inclusive}

In the following, the different terms contributing to the inclusive differential decay rate which was computed via \texttt{FeynCalc} \cite{Shtabovenko:2016sxi,Shtabovenko:2020gxv} are given. The individual contributions from vector, scalar and tensor operators are given as in
\begin{eqnarray}
\begin{aligned}
    \frac{d\Gamma_{\text{incl,V}}^{\nu_\alpha\nu_\beta}}{dq^2} & = \frac{q^2}{m_b}
    \Bigg(12\frac{m_s}{m_b}\Big[\big(m_\alpha^2 - 4m_\alpha m_\beta + m_\beta^2 - q^2\big)\Re\big(C^{\text{VLL}}_{\nu d,[\alpha\beta]sb}C^{\text{VLR}*}_{\nu d,[\alpha\beta]sb}\big) \\
    & \quad + \big(m_\alpha^2 + 4m_\alpha m_\beta + m_\beta^2 - q^2\big)\Re\big(C^{\text{VLL}}_{\nu d,(\alpha\beta)sb}C^{\text{VLR}*}_{\nu d,(\alpha\beta)sb}\big)\Big] \\
    & \quad - \frac{1}{m_b^2}\Big[g\big(m_\alpha,m_\beta,m_s,\sqrt{q^2},m_b\big)\frac{1}{q^4} - 6m_\alpha m_\beta\big(m_b^2 + m_s^2 - q^2\big)\Big] \\
    & \quad \times \Big[\big|C^{\text{VLL}}_{\nu d,[\alpha\beta]sb}\big|^2 + \big|C^{\text{VLR}}_{\nu d,[\alpha\beta]sb}\big|^2\Big] \\
    & \quad - \frac{1}{m_b^2}\Big[g\big(m_\alpha ,m_\beta,m_s,\sqrt{q^2},m_b\big)\frac{1}{q^4} + 6m_\alpha m_\beta\big(m_b^2 + m_s^2 - q^2\big)\Big] \\
    & \quad \times \Big[\big|C^{\text{VLL}}_{\nu d,(\alpha\beta)sb}\big|^2 + \big|C^{\text{VLR}}_{\nu d,(\alpha\beta)sb}\big|^2\Big]
    \Bigg),
\\
   \frac{d\Gamma_{\text{incl,S}}^{\nu_\alpha \nu_\beta}}{dq^2} & = -3\frac{q^2}{m_b}\Bigg((m_\alpha ^2 + m_\beta^2 - q^2) \\
    & \quad \times \Big[\frac{1}{m_b^2}(m_b^2 + m_s^2 - q^2)\left(\left|C^{\text{SLL}}_{\nu d,\alpha\beta sb}\right|^2 + \left|C^{\text{SLR}}_{\nu d,\alpha\beta sb}\right|^2 + \left|C^{\text{SLL}}_{\nu d,\alpha\beta bs}\right|^2 + \left|C^{\text{SLR}}_{\nu d,\alpha\beta bs}\right|^2\right) \\
    & \quad + 4\frac{m_s}{m_b}\Re\big(C^{\text{SLL}}_{\nu d,\alpha\beta sb}C^{\text{SLR}*}_{\nu d,\alpha\beta sb} + C^{\text{SLL}}_{\nu d,\alpha\beta bs}C^{\text{SLR}*}_{\nu d,\alpha\beta bs}\big)\Big] \\
    & \quad + 2m_\alpha m_\beta\Big[\frac{2}{m_b^2}(m_b^2 + m_s^2 - q^2)\text{Re}\big(C^{\text{SLL}}_{\nu d,\alpha\beta sb}C^{\text{SLL}}_{\nu d,\alpha\beta bs} + C^{\text{SLR}}_{\nu d,\alpha\beta sb}C^{\text{SLR}}_{\nu d,\alpha\beta bs}\big) \\
    & \quad + 4\frac{m_s}{m_b}\Re\big(C^{\text{SLL}}_{\nu d,\alpha\beta sb}C^{\text{SLR}}_{\nu d,\alpha\beta bs} + C^{\text{SLR}}_{\nu d,\alpha\beta sb}C^{\text{SLL}}_{\nu d,\alpha\beta bs}\big)\Big]\Bigg), \\
   \frac{d\Gamma_{\text{incl,T}}^{\nu_\alpha \nu_\beta}}{dq^2} & = 
    16\frac{q^2}{m_b^3}\Bigg(\Big[3(m_\alpha ^2 + m_\beta^2 - q^2)(m_b^2 + m_s^2 - q^2) - 2\,g\big(m_\alpha ,m_\beta,m_s,\sqrt{q^2},m_b\big)\frac{1}{q^4}
    \Big] \\
    & \quad \times \Big(\left|C^{\text{TLL}}_{\nu d,\alpha\beta sb}\right|^2 + \left|C^{\text{TLL}}_{\nu d,\alpha\beta bs}\right|^2\Big) - 72m_\alpha m_\beta m_s m_b\text{Re}\big(C^{\text{TLL}}_{\nu d,\alpha\beta sb}C^{\text{TLL}}_{\nu d,\alpha\beta bs}\big)\Bigg) 
    \;.
\end{aligned}
\end{eqnarray}
Here, the function
\begin{eqnarray}
\begin{aligned}
    g(x,y,s,w,m) & = \Big(2(x^4 + y^4) - w^4 - w^2(x^2 + y^2) - 4x^2y^2\Big)\lambda(m^2,s^2,w^2) \\
    & \quad + 3w^2(x^4 + y^4 - 2x^2y^2 - w^4)(m^2 - w^2 +s^2)
\end{aligned}
\end{eqnarray}
was defined for convenience in order to shorten the expressions.
The interference terms involving the vector operators and the scalar and tensor operators read
\begin{equation}
\begin{aligned}
    \frac{d\Gamma_{\text{incl,VS}}^{\nu_\alpha \nu_\beta}}{dq^2} & = 
    -\frac{3}{m_b^2}
    \Bigg((m_\alpha  - m_\beta)\left((m_\alpha  + m_\beta)^2 - q^2\right) \\
    & \quad \times \Big[\frac{m_s}{m_b}(m_b^2 - m_s^2 + q^2\big)\Re\big(C^{\text{VLL}}_{\nu d,[\alpha\beta]sb}(C^{\text{SLL}*}_{\nu d,\alpha\beta sb} + C^{\text{SLL}}_{\nu d,\alpha\beta bs}) + C^{\text{VLR}}_{\nu d,[\alpha\beta]sb}(C^{\text{SLR}*}_{\nu d,\alpha\beta sb} + C^{\text{SLR}}_{\nu d,\alpha\beta bs})\big) \\
    & \quad + (m_b^2 - m_s^2 - q^2)\Re\big(C^{\text{VLL}}_{\nu d,[\alpha\beta]sb}(C^{\text{SLR}*}_{\nu d,\alpha\beta sb} + C^{\text{SLR}}_{\nu d,\alpha\beta bs}) + C^{\text{VLR}}_{\nu d,[\alpha\beta]sb}(C^{\text{SLL}*}_{\nu d,\alpha\beta sb} + C^{\text{SLL}}_{\nu d,\alpha\beta bs})\big)\Big] \\
    & \quad + (m_\alpha  + m_\beta)\left((m_\alpha  - m_\beta)^2 - q^2\right) \\
    & \quad \times \Big[\frac{m_s}{m_b}(m_b^2 - m_s^2 + q^2\big)\Re\big(C^{\text{VLL}}_{\nu d,(\alpha\beta)sb}(C^{\text{SLL}*}_{\nu d,\alpha\beta sb} - C^{\text{SLL}}_{\nu d,\alpha\beta bs}) + C^{\text{VLR}}_{\nu d,(\alpha\beta)sb}(C^{\text{SLR}*}_{\nu d,\alpha\beta sb} - C^{\text{SLR}}_{\nu d,\alpha\beta bs})\big) \\
    & \quad + \big(m_b^2 - m_s^2 - q^2\big)\Re\big(C^{\text{VLL}}_{\nu d,(\alpha\beta)sb}(C^{\text{SLR}*}_{\nu d,\alpha\beta sb} - C^{\text{SLR}}_{\nu d,\alpha\beta bs}) + C^{\text{VLR}}_{\nu d,(\alpha\beta)sb}(C^{\text{SLL}*}_{\nu d,\alpha\beta sb} - C^{\text{SLL}}_{\nu d,\alpha\beta bs})\big)\Big]
    \Bigg) \;, 
   \\
   \frac{d\Gamma_{\text{incl,VT}}^{\nu_\alpha \nu_\beta}}{dq^2} & = 
    \frac{36}{m_b^2}\Bigg((m_\alpha  - m_\beta)\left((m_\alpha  + m_\beta)^2 - q^2\right) \\
    & \quad \times\Big[\frac{m_s}{m_b}(m_b^2 - m_s^2 + q^2)\Re\big(C^{\text{VLL}}_{\nu d,(\alpha\beta)sb}C^{\text{TLL}*}_{\nu d,\alpha\beta sb} + C^{\text{VLR}}_{\nu d,(\alpha\beta)sb}C^{\text{TLL}}_{\nu d,\alpha\beta bs}\big) \\
    & \quad - (m_b^2 - m_s^2 - q^2)\Re\big(C^{\text{VLR}}_{\nu d,(\alpha\beta)sb}C^{\text{TLL}*}_{\nu d,\alpha\beta sb} + C^{\text{VLL}}_{\nu d,(\alpha\beta)sb}C^{\text{TLL}}_{\nu d,\alpha\beta bs}\big)\Big] \\
    & \quad + (m_\alpha  + m_\beta)\left((m_\alpha  - m_\beta)^2 - q^2\right) \\
    & \quad \times \Big[\frac{m_s}{m_b}(m_b^2 - m_s^2 + q^2)\Re\big(C^{\text{VLL}}_{\nu d,[\alpha\beta]sb}C^{\text{TLL}*}_{\nu d,\alpha\beta sb} - C^{\text{VLR}}_{\nu d,[\alpha\beta]sb}C^{\text{TLL}}_{\nu d,\alpha\beta bs}\big) \\
    & \quad - (m_b^2 - m_s^2 - q^2)\Re\big(C^{\text{VLR}}_{\nu d,[\alpha\beta]sb}C^{\text{TLL}*}_{\nu d,\alpha\beta sb} - C^{\text{VLL}}_{\nu d,[\alpha\beta]sb}C^{\text{TLL}}_{\nu d,\alpha\beta bs}\big)\Big]\Bigg)
    \;.
\end{aligned}
\end{equation}


\section{Matching to SM Effective Field Theory with Sterile Neutrinos}
\label{sec:Matching}
For completeness we present the matching to SM effective field theory (SMEFT) with sterile neutrinos. The matching conditions have been obtained by translating the existing matching results in the literature~\cite{Jenkins:2017jig,Li:2019fhz,Li:2020lba} to the operator basis we are using. 
The relevant SMEFT operators are contained in the effective Lagrangians $\mathcal{L}_{6,7}$ for operators at dimension-6 and dimension-7, respectively~\cite{Grzadkowski:2010es,Li:2020lba}
\begin{equation}
\begin{aligned}
    \mathcal{L}_6 & \supset C_H (H^\dagger H)^3 +C_{H\square} (H^\dagger H) \square (H^\dagger H) +C_{HD} (H^\dagger D^\mu H)^* (H^\dagger D_\mu H)
    \\
    &+C_{HB} H^\dagger H B_{\mu\nu}B^{\mu\nu} + C_{HW} H^\dagger H W^I_{\mu\nu} W^{I\mu\nu} + C_{HWB} H^\dagger \tau^I H W_{\mu\nu}^I B^{\mu\nu}
    \\&
    + C_{lq}^{(1)} (H^\dagger i \stackrel{\leftrightarrow}{D}_\mu H) (\bar L \gamma^\mu L)
    +C_{lq}^{(3)}(H^\dagger i \stackrel{\leftrightarrow}{D}^I_\mu H) (\bar L \tau^I\gamma^\mu L)
    +C_{ld} (\bar L \gamma_\mu L) (\bar d \gamma^\mu d)
    \\&
    +C_{QN} (\bar Q \gamma_\mu Q) (\bar N \gamma^\mu N)
    +C_{dN} (\bar d \gamma_\mu d) (\bar N\gamma^\mu N)
    \\&
    +C_{LNQd} (\bar L^\alpha N)\epsilon_{\alpha\beta} (\bar Q^\beta d)
    +C_{LdQN} (\bar L^\alpha d) \epsilon_{\alpha\beta} (\bar Q^\beta N)
    \\
    \mathcal{L}_7 & \supset 
    C_{\bar d LQLH1} \epsilon_{ij}\epsilon_{mn}(\bar d L^i)(\overline{Q^{cj}} L^m)H^n
    +C_{QNdH} (\bar QN)(\overline{N^c} d)H
    +C_{dQNH} H^\dagger (\bar d Q) (\overline{N^c}N)
    \\&
    +C_{QNLH1} \epsilon_{ij} (\bar Q \gamma_\mu Q) (\overline{N^c}\gamma^\mu L^i)H^j
    +C_{QNLH2} \epsilon_{ij} (\bar Q \gamma_\mu Q^i)(\overline{N^c}\gamma^\mu L^j)H
    \\&
    +C_{dNLH} \epsilon_{ij} (\bar d\gamma_\mu d)(\overline{N^c}\gamma^\mu L^i) H^j
    \end{aligned}
\end{equation}
where $N$ denotes right-handed neutrinos, i.e.~right-handed SM singlet fermions, $\tau^I$ denotes the Pauli spin matrices and we suppressed flavour and colour indices.
For the matching, we also 
require the modified $Z$-boson couplings~\cite{Jenkins:2017jig}
\begin{align}
    \mathcal{L}\supset -\bar g_Z Z_\mu \left[ 
    Z_{d_L} \bar d_L \gamma^\mu d_L 
    + Z_{d_R} \bar d_R \gamma^\mu d_R
    + Z_{\nu} \bar \nu_K \gamma^\mu \nu_L
    +Z_N \bar N \gamma^\mu N
    + \left( Z_{\nu N} \overline{\nu^c} \gamma^\mu N +\mathrm{h.c.}\right)
    \right]\;,
\end{align}
where $\bar g_Z$ denotes the effective gauge coupling of the $Z$ boson and depends on gauge couplings and the weak mixing angle $\bar\theta$
\begin{equation}
\begin{aligned}
    \bar g_Z  &= \frac{\bar e}{\sin \bar\theta\cos\bar\theta} \left[1+ \frac{\bar g_1^2+\bar g_2^2}{2\bar g_1 \bar g_2} v_T^2 C_{HWB}\right]\;, 
    &
    \bar e &= \bar g_2 \sin \bar\theta -\frac12 \cos\bar\theta \bar g_2 v_T^2 C_{HWB}\;,
    \\
    \cos\bar\theta & = \frac{\bar g_2}{\sqrt{\bar g_1^2+\bar g_2^2}} \left[1-\frac{C_{HWB}v_T^2}{2} \frac{\bar g_1}{\bar g_2} \frac{\bar g_2^2-\bar g_1^2}{\bar g_1^2+\bar g_2^2}\right],
    &
    \bar g_1 & = g_1(1+C_{HB} v_T^2)\;, 
    \\
    \sin\bar\theta & = \frac{\bar g_1}{\sqrt{\bar g_1^2+\bar g_2^2}} \left[1+\frac{C_{HWB}v_T^2}{2} \frac{\bar g_2}{\bar g_1} \frac{\bar g_2^2-\bar g_1^2}{\bar g_1^2+\bar g_2^2}\right],
    &
    \bar g_2 & = g_2(1+C_{HW} v_T^2) \;.
\end{aligned}
\end{equation}
The $Z$-boson couplings to the different fermion species are parameterised by
\begin{align}
   [Z_{d_L}]_{pr} & = \left(-\frac12+\frac13 \sin^2\bar\theta\right)\delta_{pr} - \frac{v_T^2}{2}\left(C_{Hq}^{(1),pr} + C_{Hq}^{(3),pr}\right)\;,
    &
    [Z_{d_R}]_{pr} & = \frac13 \sin^2\bar\theta \delta_{pr} - \frac{v_T^2}{2} C_{Hd}^{pr}\;,
    \nonumber\\
        [Z_{\nu}]_{pr} & = \frac12 \delta_{pr} - \frac{v_T^2}{2}\left(C_{Hl}^{(1),pr} - C_{Hl}^{(3),pr}\right)\;,
    &
    [Z_N]_{pr} & = -\frac{v_T^2}{2} C_{HN}^{pr}\;,
    \\\nonumber
    [Z_{\nu N}]_{pr} &= \frac{v_T^3}{4\sqrt{2}}\left(C_{NL1}^{rp}+ 2 C_{NL2}^{rp}\right)\;.
\end{align}
Following \cite{Jenkins:2017jig}, we write the renormalisable part of the SM Higgs potential as 
\begin{equation}
    V=\lambda\left(H^\dagger H - \frac{v^2}{2}\right)^2 
\end{equation}
and the 
SM Higgs doublet $H$ in unitary gauge as
\begin{align}
    H&= \frac1{\sqrt{2}} \begin{pmatrix}
    0\\
    [1+c_{H,\rm kin}]\,h + v_T
    \end{pmatrix}
    \;.
    \end{align}
The Higgs field normalisation $1+c_{H,\rm kin}$ and the Higgs VEV $v_T$ receive corrections from dimension-6 operators 
   \begin{align}
    c_{H,\rm kin} &\equiv \left(C_{H\square}-\frac14 C_{HD}\right)v^2 \;,
    &
    v_T & \equiv \left(1+\frac{3C_H v^2}{8\lambda}\right) v \;.
\end{align}
After introducing and summarising the relevant SMEFT operators and expressions, it is straightforward to present the matching of the LEFT Wilson coefficients to SMEFT.
We find for the LEFT Wilson coefficients with neutrino flavour indices $1\leq\alpha,\beta\leq3$
\begin{equation}
\begin{aligned}
    C_{\nu d,\alpha \beta pr}^{\text{VLL}} &= C^{(1),\alpha \beta p r}_{lq} - C^{(3),\alpha \beta p r}_{lq} - \frac{\bar g_Z^2}{M_Z^2} [Z_{d_L}]_{pr} [Z_{\nu}]_{\alpha\beta}, 
    &
    C_{\nu d,\alpha\beta pr}^{\text{VLR}} &= C^{\alpha \beta p r}_{ld} - \frac{\bar g_Z^2}{M_Z^2} [Z_{d_R}]_{pr} [Z_{\nu}]_{\alpha\beta}, \\
    C_{\nu d, \alpha \beta p r}^{\text{SLL}} &= -\frac{v_T}{4 \sqrt{2}} \left( C^{p \alpha r \beta}_{\bar{d} LQLH1} +  C^{p \alpha r \beta}_{\bar d LQLH1} \right), 
    &
    C_{\nu d, \alpha \beta p r}^{\text{SLR}} &= 0,\\
    C_{\nu d, \alpha \beta p r}^{\text{TLL}} &= \frac{v_T}{16 \sqrt{2}} \left( C^{p \alpha r \beta }_{\bar{d} LQLH1} - C^{p \beta r \alpha }_{\bar{d} LQLH1} \right).
\end{aligned}
\end{equation}
For Wilson coefficients pertaining only to sterile neutrinos with $\alpha,\beta\geq 4$, they are
\begin{equation}
\begin{aligned}
    C_{\nu d,\alpha\beta pr}^{\text{VLL}} &= -C^{p r \beta\alpha }_{Q N} + \frac{\bar g_Z^2}{M_Z^2} [Z_{d_L}]_{pr} [Z_{N}]_{\beta\alpha},& 
    C_{\nu d,\alpha\beta pr}^{\text{VLR}} &= -C^{p r \beta\alpha}_{d N} + \frac{\bar g_Z^2}{M_Z^2} [Z_{d_R}]_{pr} [Z_{N}]_{\beta\alpha},
    \\
    C_{\nu d, \alpha \beta p r}^{\text{SLL}} &= -\frac{v_T}{4 \sqrt{2}} \left( C^{p \alpha \beta r}_{QN dH} +  C^{p \beta \alpha r}_{QN dH} \right), 
    &
    C_{\nu d, \alpha \beta p r}^{\text{SLR}} &= \frac{v_T}{\sqrt{2}} C^{pr \alpha \beta}_{dQN H},
    \\
    C_{\nu d, \alpha \beta p r}^{\text{TLL}} &= \frac{v_T}{16 \sqrt{2}} \left( C^{p \alpha\beta r}_{QNdH} - C^{p \beta\alpha r }_{QN dH} \right)\;.
\end{aligned}
\end{equation}
The different signs and orderings of neutrino flavour indices originate from the charge conjugation in $N\equiv \nu^c$ and the symmetry properties of the bilinears
\begin{align}\label{eq:symmetry-properties}
    \overline{\psi_i^c} \Gamma\psi^{cj} &= \eta_\Gamma \overline{\psi^j} \Gamma \psi_i
    &
    C^{-1}\Gamma C &= \eta_\Gamma \Gamma^T 
    &
    \eta_\Gamma & = \begin{cases} +1 & \mathrm{for}\;\;\Gamma=1,\gamma_5,\gamma^\mu\gamma_5\\
    -1 & \mathrm{for} \;\; \Gamma = \gamma^\mu ,\sigma^{\mu\nu}, \sigma^{\mu\nu}\gamma_5
    \end{cases}
    \;.
\end{align}
Finally, SMEFT operators which contain both sterile and active neutrinos imply
\begin{equation}
\begin{aligned}
    C^{\text{VLL}}_{\nu d, \alpha \beta p r} &= \frac{v_T}{\sqrt{2}} \left( C^{pr \alpha\beta}_{QN LH1} - C^{p r \alpha \beta}_{QN LH2} \right) + \frac{\bar g_Z^2}{M_Z^2} [Z_{d_L}]_{pr} [Z_{\nu N}]_{\beta\alpha}, 
    \\
    C^{\text{VLR}}_{\nu d, \alpha \beta p r} &= \frac{v_T}{\sqrt{2}} C^{pr \alpha\beta}_{d N L H} + \frac{\bar g_Z^2}{M_Z^2} [Z_{d_R}]_{pr} [Z_{\nu N}]_{\beta\alpha},
    \\
    C^{\text{SLL}}_{\nu d, \alpha \beta p r} &= C^{\beta\alpha rp*}_{L N Qd} - \frac{1}{2}C^{\beta pr \alpha*}_{LdQ N}, 
    \\
    C^{\text{SLR}}_{\nu d, \alpha \beta pr} &= 0, 
    \\
    C^{\text{TLL}}_{\nu d, \alpha \beta pr} &= -\frac{1}{8} C^{\beta pr \alpha*}_{LdQN}
\end{aligned}
\end{equation}
when $\alpha\geq 4$ and $1\leq \beta\leq3$
 and
 \begin{equation}
\begin{aligned}
    C^{\text{VLL}}_{\nu d, \alpha \beta p r} &= \frac{v_T}{\sqrt{2}} \left( C^{rp \beta \alpha*}_{QN LH1} - C^{rp \alpha \beta*}_{QN LH2} \right) + \frac{\bar g_Z^2}{M_Z^2} [Z_{d_L}]^*_{rp} [Z_{\nu N}]^*_{\alpha\beta}, 
    \\
    C^{\text{VLR}}_{\nu d, \alpha \beta p r} &= \frac{v_T}{\sqrt{2}} C^{rp\beta \alpha*}_{d N L H} + \frac{\bar g_Z^2}{M_Z^2} [Z_{d_R}]_{rp}^* [Z_{\nu N}]^*_{\alpha\beta},
    \\
    C^{\text{SLL}}_{\nu d, \alpha \beta p r} &= C^{\alpha\beta rp*}_{L N Qd} - \frac{1}{2}C^{\alpha pr \beta*}_{LdQ N}, 
    \\
    C^{\text{SLR}}_{\nu d, \alpha \beta pr} &= 0, 
    \\
    C^{\text{TLL}}_{\nu d, \alpha \beta pr} &= -\frac{1}{8} C^{\alpha pr \beta*}_{LdQN}
\end{aligned}
\end{equation}
when $1\leq \alpha\leq 3$ and $\beta\geq4$.

\bibliography{refs}
\end{document}